\documentclass[12pt]{article}           % LaTeX 2e
\usepackage{hyperref}
\usepackage[pdftex]{graphicx}
\usepackage{amsmath,amssymb}
\usepackage{mathabx}
\usepackage{color}
\usepackage{cite}

\newcommand{\be}{\begin{equation}}
\newcommand{\ee}{\end{equation}}
\newcommand{\bea}{\begin{eqnarray}}
\newcommand{\eea}{\end{eqnarray}}

\def\p{\partial}

\newcommand{\bR}{{\mathbf{R}}}
\newcommand{\bS}{{\mathbf{S}}}

\begin{document}

\begin{titlepage}

\vspace*{1cm}
\begin{center} \Large \bf What's the Point? \\
 Hole-ography in Poincar\'e AdS
\end{center}

\begin{center}
Ricardo Esp\'indola$^{\ast\dagger}$,
Alberto G\"uijosa$^{\ast}$,
Alberto Landetta$^{\ast}$
and Juan F.~Pedraza$^{\diamond}$

\vspace{0.2cm}
$^{\ast}\,$Departamento de F\'{\i}sica de Altas Energ\'{\i}as, Instituto de Ciencias Nucleares, \\
Universidad Nacional Aut\'onoma de M\'exico,
\\ Apartado Postal 70-543, CDMX 04510, M\'exico\\
 \vspace{0.2cm}
$^{\dagger}$
Mathematical Sciences and STAG Research Centre,\\
University of Southampton,\\
Highfield, Southampton, SO17 1BJ, UK\\
\vspace{0.2cm}
$^{\diamond}$
Institute for Theoretical Physics,\\
University of Amsterdam,\\
Science Park 904, 1098 XH Amsterdam, Netherlands\\
\vspace{0.2cm}
{\tt ricardo.espindola@correo.nucleares.unam.mx, alberto@nucleares.unam.mx, landetta@ciencias.unam.mx},
{\tt jpedraza@uva.nl}
\end{center}

%\vspace*{1cm}

\begin{center}
{\bf Abstract}
\end{center}
\noindent
In the context of the AdS/CFT correspondence, we study bulk reconstruction of the Poincar\'e wedge of AdS$_3$ via hole-ography, i.e., in terms of differential entropy of the dual CFT$_2$. Previous work had considered the reconstruction of closed or open spacelike curves in global AdS, and of infinitely extended spacelike curves in Poincar\'e AdS that are subject to a periodicity condition at infinity. Working first at constant time, we find that a closed curve in Poincar\'e is described in the CFT by a family of intervals that covers the spatial axis at least twice. We also show how to reconstruct open curves, points and distances, and obtain a CFT action whose extremization leads to bulk points. We then generalize all of these results to the case of curves that vary in time, and discover that generic curves have segments that \emph{cannot} be reconstructed using the standard hole-ographic construction. This happens because, for the nonreconstructible segments, the tangent geodesics fail to be fully contained within the Poincar\'e wedge. We show that a previously discovered variant of the hole-ographic method allows us to overcome this challenge, by reorienting the geodesics touching the bulk curve to ensure that they all remain within the wedge. Our conclusion is that all spacelike curves in Poincar\'e AdS can be completely reconstructed with CFT data, and each curve has in fact an infinite number of representations within the CFT.
\vspace{0.2in}
\smallskip

\end{titlepage}

\tableofcontents

\section{Introduction and Summary}

Twenty years from the inception of the AdS/CFT correspondence \cite{malda,gkp,w}, research is still being carried out to understand how it achieves its grandest miracle: the emergence of a dynamical spacetime out of degrees of freedom living on a lower-dimensional rigid background. Over ten years ago, a crucial insight in this direction was provided by Ryu and Takayanagi \cite{rt}, who argued that areas in the bulk gravitational description are encoded as quantum entanglement in the boundary field theory. More specifically, they proposed that when the dynamics of spacetime is controlled by Einstein gravity, the area $A_{\Sigma}$
of each minimal-area codimension-two surface $\Sigma$ anchored on the boundary translates into the entanglement entropy $S$ of the spatial region in the boundary theory that is homologous to $\Sigma$, via
\begin{equation}\label{rt}
S=\frac{A_{\Sigma}}{4G_N}~.
\end{equation}
Their proposal, originally conjectural and referring only to static situations, was extended to the covariant setting in \cite{hrt} by taking $\Sigma$ to be an extremal surface, and later proved in \cite{lm,dlr}. It has been generalized beyond Einstein gravity in \cite{hms,dong,camps,castro,bdhm,flm,ew,deboer,hofman,elena,janiszewski}.
Many other notable developments have taken place, including \cite{klebanov,vr,headrick,myerssinha,chm,hhm,hm,tsunami,veronikaplateaux,fghmvr,ooguri,dongrenyi,fh,taylor,fhhprvr}. Useful reviews can be found in \cite{nrt,vrlectures,mukund}.

Another important step towards holographic reconstruction was taken in \cite{hole-ography}, working for simplicity in AdS$_3$, where the extremal codimension-two surfaces $\Sigma$ are just geodesics, and their `areas' $A_{\Sigma}$ refer to their lengths. It was discovered in that context that one can reconstruct spacelike curves $C$ that are not extremal and are not anchored on the boundary, by cleverly adding and subtracting the geodesics tangent to the bulk curve. This procedure was initially phrased in terms of the hole in the bulk carved out by the curve, and was therefore dubbed hole-ography. It entails two related insights. The first is that any given spacelike bulk curve can be represented by a specific family of spacelike intervals in the boundary theory, whose endpoints coincide with those of the  geodesics tangent to the bulk curve (in a manner that embodies the well-known UV/IR connection \cite{uvir,pp}). The second is that the length $A\equiv A_C$ of the curve can be computed in the CFT through the differential entropy $E$, a particular combination of the entanglement entropies of the corresponding intervals, whose precise definition is given below, in Eq.~(\ref{e}). The concrete relation between these two quantities takes the form inherited from (\ref{rt}), $E=A/4G_N$.

Diverse aspects of hole-ography have been explored in \cite{ewshadows,myers,veronikaresidual,cds,hmw,nutsandbolts,bartekinformation,freivogel,integralgeometry,taylorflavors,ef,keeler}. The works \cite{hole-ography,nutsandbolts} carried out the hole-ographic reconstruction of an arbitrary closed curve at constant time in \emph{global} AdS$_3$ (and also on the BTZ black hole and on the conical defect geometry).
Upon shrinking a closed curve to zero size at an arbitrary point in the bulk, a family of intervals was obtained \cite{nutsandbolts} describing a `point-curve' of vanishing length. This could then be combined with the family for a second point, to compute the distance between the two points. This framework is thus able to extract the most basic ingredients of the bulk geometry, points and distances, from the pattern of entanglement in the state of the boundary theory.

\begin{figure}[hbt]
\begin{center}
  \includegraphics[width=4cm]{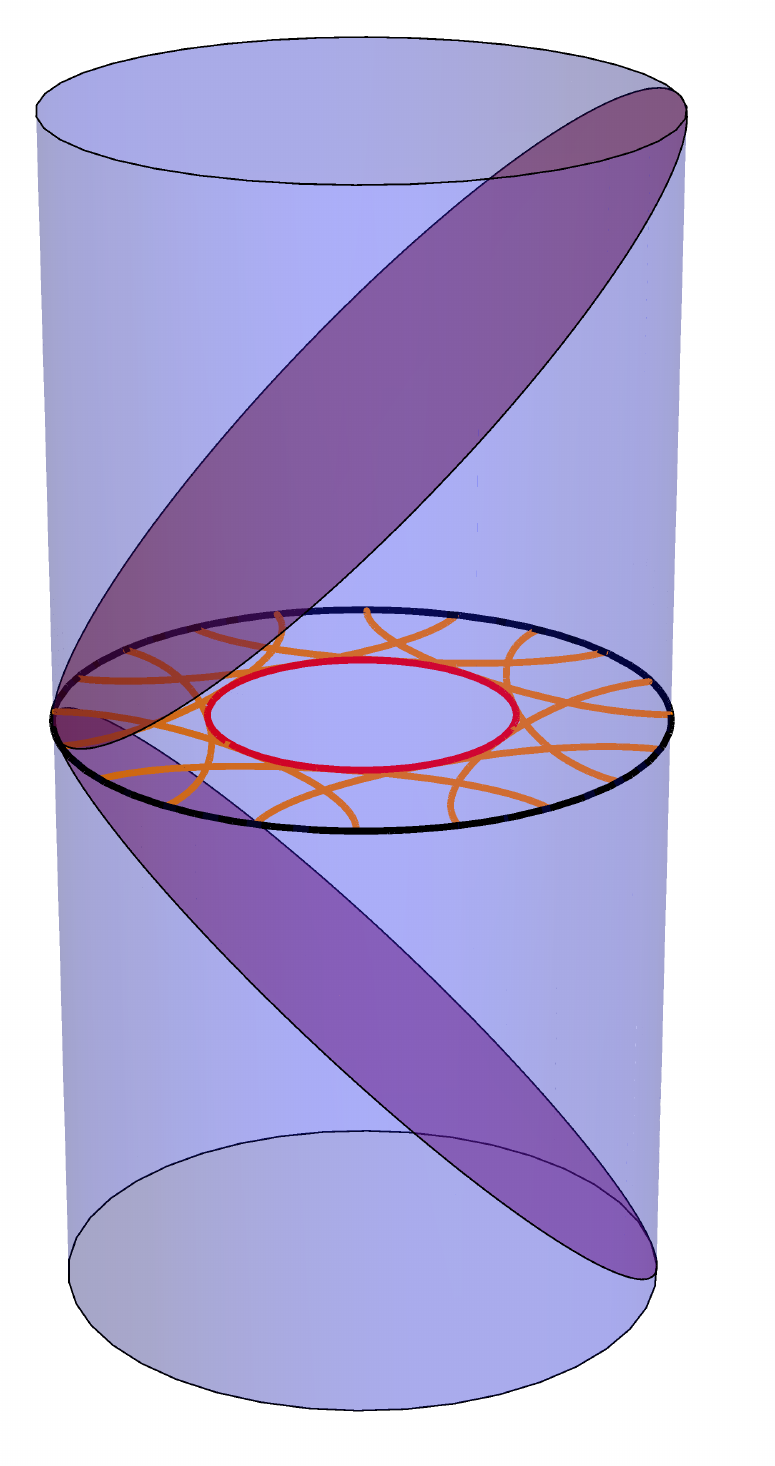}
  \hspace*{2cm}
  \includegraphics[width=4cm]{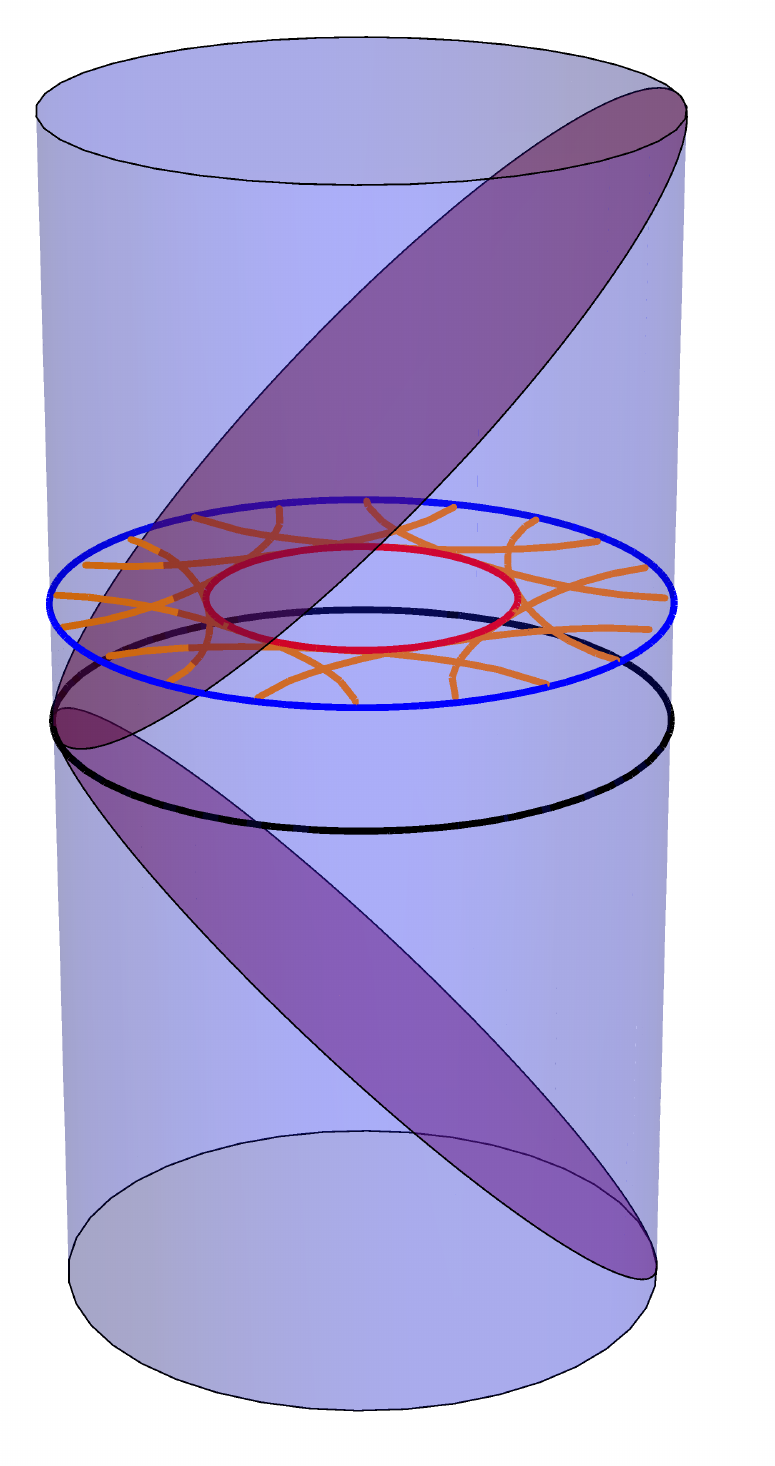}
  \setlength{\unitlength}{1cm}
\begin{picture}(0,0)
%left cylinder:
\put(-7.6,3.2){\vector(2,1){0.1}}
\put(-7.4,3.1){$x$}
\qbezier(-8.0,3.1)(-7.8,3.1)(-7.6,3.2)
\put(-8.3,4.9){\vector(-3,-1){0.1}}
\put(-7.3,5.1){$z$}
\qbezier(-7.3,5.2)(-7.9,5.1)(-8.3,4.9)
\put(-7.8,5.3){\vector(1,2){0.1}}
\put(-7.6,5.5){$t$}
\qbezier(-7.7,2)(-8.8,3.7)(-7.7,5.5)
\put(-10.7,6.0){\vector(0,1){0.5}}
\put(-10.8,6.7){$\tau$}
\put(-11.4,3.7){$\tau=0$}
\put(-9.8,6.4){\vector(1,0){0.1}}
\put(-9.6,6.2){$\theta$}
\qbezier(-10.3,6.6)(-10.0,6.4)(-9.8,6.4)
\put(-8.6,7.0){\vector(2,1){0.4}}
\put(-8.1,7.1){$\varrho$}
\put(-8.6,7.0){\circle*{0.1}}
%right cylinder:
\put(-1.3,3.2){\vector(2,1){0.1}}
\put(-1.1,3.1){$x$}
\qbezier(-1.7,3.1)(-1.5,3.1)(-1.3,3.2)
\put(-2.0,4.9){\vector(-3,-1){0.1}}
\put(-1.0,5.1){$z$}
\qbezier(-1.0,5.2)(-1.6,5.1)(-2.0,4.9)
\put(-1.5,5.3){\vector(1,2){0.1}}
\put(-1.3,5.5){$t$}
\qbezier(-1.4,2)(-2.5,3.7)(-1.4,5.5)
\put(-4.4,6.0){\vector(0,1){0.5}}
\put(-4.5,6.7){$\tau$}
\put(-3.5,6.4){\vector(1,0){0.1}}
\put(-3.3,6.2){$\theta$}
\qbezier(-4.0,6.6)(-3.7,6.4)(-3.5,6.4)
\put(-5.1,4.3){$\tau>0$}
\put(-2.3,7.0){\vector(2,1){0.4}}
\put(-1.8,7.1){$\varrho$}
\put(-2.3,7.0){\circle*{0.1}}
\end{picture}
\end{center}
\vspace*{-0.8cm}
\caption{Each of these solid cylinders is a Penrose diagram for AdS$_3$, covered in full by the global coordinates $(\varrho,\tau,\theta)$, but only in part by the Poincar\'e set $(t,x,z)$. The latter coordinates span the wedge between the AdS boundary $z=0$ ($\leftrightarrow\varrho=\pi/2$), at the surface of the cylinder, and the Poincar\'e horizon $z\to\infty$, shown as the purple disks tilted at 45 degrees. Each cylinder displays an example of a bulk curve within the Poincar\'e wedge (a circle, shown in red), together with its tangent geodesics (in orange), which in the global description allow the curve to be reconstructed hole-ographically. On the left, the curve is at fixed global time $\tau=0$ ($\leftrightarrow t=0$). In this case, complete reconstruction of the curve should be possible using data in the Minkowski CFT$_2$ dual to the Poincar\'e wedge, because all of the tangent geodesics are inside the wedge. On the right, the curve is at constant $\tau>0$ ($\leftrightarrow t\neq\,$constant), and we see that it contains a segment whose tangent geodesics exit the wedge. Even though this segment is part of Poincar\'e AdS, it \emph{cannot be reconstructed} in the Minkowski CFT$_2$ using the standard hole-ographic procedure.
\label{3dfig}}
\end{figure}
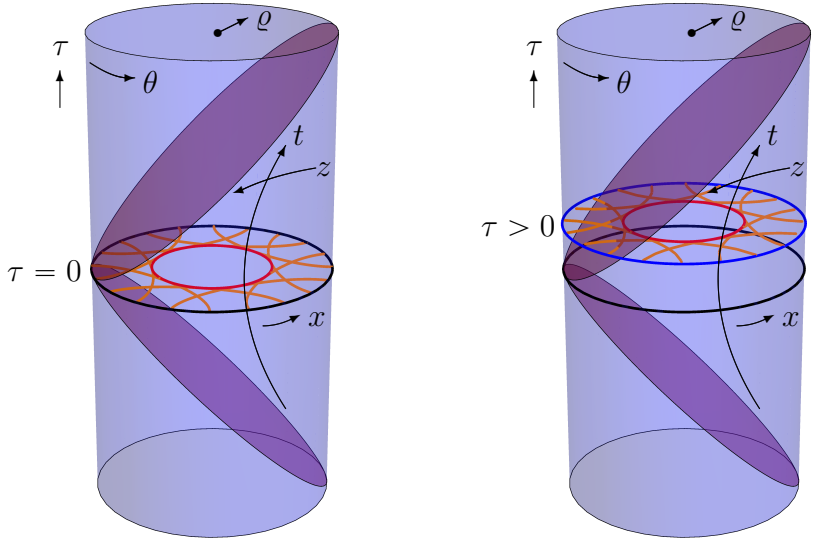

In this paper we are interested in understanding how this entire story plays out on \emph{Poincar\'e} AdS$_3$, where hole-ography faces a serious challenge.
The pure AdS geometry with coordinates $x^m\equiv(x^{\mu},z)$ and metric (\ref{poincaremetric}) is dual to the vacuum state of a CFT on 2-dimensional Minkowski spacetime, with coordinates $x^{\mu}\equiv(t,x)$.  Hole-ography in this context has been examined before, at constant time in \cite{myers} and for curves with non-trivial time-dependence in \cite{hmw}. Our motivation here is different, and its essence can be understood by looking at Fig.~\ref{3dfig}, which shows the Poincar\'e patch as a wedge within global AdS. The fact that Poincar\'e does not cover all of AdS implies that some curves within the Poincar\'e wedge can have a set of tangent geodesics whose endpoints fall outside of the wedge. Such geodesics cannot be associated with entanglement entropy in the Minkowski CFT$_2$. Their existence presents a challenge to the hole-ographic reconstruction program, because it leaves us without the means to encode in CFT language what should definitely be properties of the vacuum state.

There is one conceptual issue we should clarify. Since the global and Poincar\'e descriptions are related by a simple coordinate transformation (see Eq.~(\ref{globaltopoincare})), it might seem that the success of hole-ography in reproducing curves, points and distances in global coordinates should automatically extend to Poincar\'e. The proper length $A$ of the closed curve is certainly invariant under coordinate transformations, and naively the same would seem to be true for the entanglement entropy, which on the gravity side is also a proper length, according to the Ryu-Takayanagi prescription (\ref{rt}). Indeed, the unregulated entropy (taking the length of the geodesic all the way to the AdS boundary) is invariant, but it is also divergent, so it cannot be used directly to compute $E$. And as soon as we introduce a cutoff, we introduce coordinate dependence.
This is truly a property of regulated entanglement entropy on the field theory side: its value depends on the regularization scheme, so it is not invariant under conformal/Weyl transformations (see e.g.~\cite{entropyanomalya,entropyanomalyb,entropyanomalyc,entropyanomalyd,entropyanomalye,entropyanomalyf,entropyanomalyg}), which is what the bulk transformation from global to Poincar\'e amounts to in the CFT. As a result, equations involving $S$ cannot always be carried over directly from one set of coordinates to the other, which explains why it is important to study Poincar\'e hole-ography directly. This is what we set out to do in this paper, working first
at constant time  in Section~\ref{constanttsec}, and then at varying time in Section~\ref{covariantsec}.

In more detail, we begin by asking how to reconstruct closed curves, as opposed to the curves examined in \cite{myers,hmw}, which were infinitely extended, with a periodicity condition at infinity. A salient difference between the global and Poincar\'e settings, closely related to the geodesic incompleteness described two paragraphs above, is that in global AdS the boundary wraps all the way around the bulk. Given a closed curve, it is then easy to visualize how the sought-after family of CFT intervals will lead to geodesics that are tangent to each point on the curve.
In Poincar\'e, given a closed curve, the boundary does not wrap around it, so naively we would seem to be missing the intervals/geodesics that would be tangent to the portion of the curve that is farther away from the boundary. As explained above and seen in Fig.~\ref{3dfig}, this is allowed by the fact that Poincar\'e coordinates cover only a wedge of global AdS. But we know that the slice at Poincar\'e time $t=0$ completely coincides with the slice at global time $\tau=0$, so at least in this case, there is no possibility for geodesics to be left out. In Section \ref{closedsubsec}, our strategy will thus be to take the results of \cite{nutsandbolts} for curves at $\tau=0$ and simply perform the required change of coordinates, to obtain the corresponding Poincar\'e description. Our conclusion is that arbitrary closed curves at $t=0$ can indeed be reconstructed, but with an important novelty: the dual family of intervals must run over the $x$ axis at least \emph{twice}, for it is only on the second (or subsequent) pass(es) that we describe geodesics tangent to the more distant portion of the curve. Once we know how to do this at $t=0$, invariance of the metric (\ref{poincaremetric}) under translations in $t$ will of course allow us to reconstruct curves and points on any other fixed-$t$ slice, independently of the value of $t$.
(Translations in $\tau$, on the other hand, will give us examples of curves at variable $t$, which we examine in Section~\ref{covariantsec}.)

In Section~\ref{diffentsubsec} we show that the differential entropy $E$ gives the correct length $A$ for a generic closed curve at constant time in Poincar\'e AdS: just like in the global case examined in \cite{nutsandbolts}, we find that $E=A/4G_N$. In this particular instance, then, no subtlety arises from the coordinate transformation.
A subtlety does arise, however, when we analyze in Section \ref{opensubsec} the hole-ographic description of open curves. It was found in \cite{nutsandbolts} that in order to match the length of an open curve in global AdS, the differential entropy must be supplemented with a specific boundary function $f$, given in (\ref{fglobal}). We find that the same is true in Poincar\'e, but the relevant boundary function, Eq.~(\ref{f}), is not the direct translation of its global counterpart. Nonetheless, it does continue to be true that $f$ can be described geometrically in the bulk, and has a specific interpretation in terms of entanglement entropy in the boundary theory. This is crucial in order for open curves to be reconstructed purely with CFT data. We combine $E$ with $f$ to define a `renormalized' differential entropy $\mathcal{E}$, which directly matches the length of an arbitrary open curve, $\mathcal{E}=A/4 G_N$. A simple expression for $\mathcal{E}$ in terms of boundary data is given in (\ref{ecal}).

In Section~\ref{pointsubsec} we shrink curves down to zero size to obtain the hole-ographic description of bulk points. We find that this can be done either with closed or open curves, but in the latter case we must take the slope $\p z/\p x$ to diverge at the endpoints of the curve, in order to still be left with a non-trivial collection of geodesics in the point limit. Following \cite{nutsandbolts}, we show that the families of CFT intervals that happen to be associated with points instead of finite-size curves can be obtained by extremizing an action based on extrinsic curvature, which in terms of field theory variables takes the form (\ref{actionK2}). We then verify in Section~\ref{distancesubsec} that the distance between two arbitrary points can also be obtained from differential entropy. This can in fact be done in two different ways: using Eq.~(\ref{dworks}), which is essentially the same recipe as in \cite{nutsandbolts}, or Eq.~(\ref{elpq}), which is a generalization based on describing the points as open curves.

Moving on to the covariant case, in Section~\ref{arbitrarysubsec} we present, following \cite{hmw}, the basic formulas (\ref{geodesics})-(\ref{TXRL}) that define the intervals and geodesics associated to an arbitrary (open or closed) spacelike bulk curve, whether or not it varies in time. The corresponding differential entropy plus boundary function is written down in (\ref{ecalcovariant}), and contact is successfully made with the length $A$ of the bulk curve.

The main issue of the paper is then encountered in Section~\ref{challengesubsec}, where we show that any segment of a curve that violates condition (\ref{reconstructibility}) is nonreconstructible, in the sense that the geodesics tangent to it have at least one endpoint outside of the Poincar\'e wedge, and are consequently not associated to entanglement entropies in the CFT. Examples are given in Figs.~\ref{circletaufig}-\ref{circlecrosshorizonfig}. In Section~\ref{nullsubsec} we discover that this challenge can be overcome by making use of a variant of hole-ography formulated previously in \cite{hmw}, where one is allowed to shoot from each point on the bulk curve a geodesic aimed in a direction that differs from the tangent by a null vector satisfying (\ref{n}). We thus arrive at the central result of this paper: the statement that, {\it contrary to appearances, hole-ography can successfully reconstruct any open or closed spacelike curve within Poincar\'e-AdS$_3$, in terms of differential entropy in the CFT$_2$ on Minkowski spacetime}.

In Section~\ref{covariantpointsubsec} we study again the limit where the size of the curve vanishes, emphasizing that there are infinitely many different ways to represent any given point in terms of a family of CFT intervals. As expressed in Eq.~(\ref{lmufromxcmu}) and exemplified in Fig.~\ref{covariantpointfig}, there is one family for each distinct choice of the path traced by the center of the intervals (or equivalently, the path traced by either one of the intervals' endpoints). Generalizing the results of Section~\ref{pointsubsec}, we work out a covariant action whose extremization leads to any one of these families associated to a point. On the gravity side it is based on the normal curvature of the bulk curve, and in CFT variables it takes the form (\ref{actionKcovariant}). In the final part of the paper, Section~\ref{covariantdistancesubsec}, we show that given two bulk points, the freedom to choose a representative family from the equivalence class associated to each point allows us to easily compute the distance between the pair imitating the constant-time procedure of Section~\ref{distancesubsec}.

In Appendix~\ref{appendix} we go back to the discrete versions (\ref{diffEa})-(\ref{diffEb}) of differential entropy originally considered in \cite{hole-ography,myers}, to show that in the continuum limit they give rise to definitions that differ by a boundary term. This difference is negligible for the types of curves considered in \cite{myers,hmw}, but is important for our analysis of open curves in Sections~\ref{opensubsec} and \ref{arbitrarysubsec}. The definition (\ref{e}) of differential entropy that we use in this paper arises directly from a discrete version that differs from (\ref{diffEa}) and (\ref{diffEb}), and belongs to the one-parameter family of alternative definitions given in (\ref{diffExidiscrete}).

There are various directions for future work. Along the lines of \cite{myers,cds,hmw}, we expect our results to extend to Poincar\'e AdS in higher dimensions, under the same assumptions of symmetry for the surfaces under consideration. On a different front, Poincar\'e AdS is a particular example of an entanglement wedge \cite{densitymatrix,wall,hhlr}, with the special feature that it includes a complete global time slice, and therefore a full set of initial data for temporal evolution. A smaller entanglement wedge leaves some information out, and contains fewer complete geodesics, so it is interesting to ask whether or not it is possible again to reorient those geodesics that exit it to achieve complete hole-ographic reconstruction of any curve within the wedge. We will address this question in a separate paper \cite{wedge}.

Going beyond pure AdS, hole-ography is known to be restricted by the appearance of entanglement shadows \cite{ewshadows,freivogel} and holographic screens \cite{ef}. It may be possible to circumvent the former obstacle using entwinement, a type of entanglement between degrees of freedom in the CFT that are not spatially organized \cite{entwinement,nutsandbolts,lin,bbcdjg}. At least in the case where the gravitational description is three-dimensional, entwinement is associated with non-minimal geodesics, and it would be interesting to investigate whether the possibility of reorienting them by null vectors \cite{hmw} affords hole-ography any additional coverage. Finally, the reconstruction program has focused recently on the description of local bulk operators that are integrated over extremal surfaces, which have been shown to be dual to blocks in the CFT operator product expansion \cite{stereoscopy,guica,diamondography,blmms2,guica2,ksuw,kl,fl,insideout}. A somewhat different approach to local operators has been pursued in \cite{verlinde,mnstw,no,no2,verlinde2,gt}. One would naturally like to understand in detail how hole-ography is related to these two approaches.

\section{Hole-ography at constant Poincar\'e Time}\label{constanttsec}

\subsection{Closed curves}\label{closedsubsec}

 To fix our notation, recall that the metric of global AdS$_3$ can be written in different ways:
 \begin{eqnarray}\label{globalmetric}
 ds^2&=&-\left(1+\frac{R^2}{L^2}\right)dT^2 + \left(1+\frac{R^2}{L^2}\right)^{-1}\!dR^2+R^2 d\theta^2\nonumber\\
 &=&L^2\left(-\cosh^2\!\rho\,d\tau^2+d\rho^2+\sinh^2\!\rho\,d\theta^2\right)\nonumber\\
 &=&\frac{L^2}{\cos^2\varrho}\left(-d\tau^2+d\varrho^2+\sin^2\!\varrho\,d\theta^2\right)~,
 \end{eqnarray}
 where $L$ is the AdS radius of curvature, $T=L\tau$, and the three different choices of radial coordinate are related through $R\equiv L\sinh\rho \equiv L\tan\varrho$.
 With $\tau\in(-\infty,\infty)$, $\varrho\in[0,\pi/2)$ and $\theta\in[0,2\pi)$, the set $(\tau,\varrho,\theta)$ covers the entire anti-de Sitter spacetime. The AdS boundary is at $\varrho=\pi/2$ ($R\to\infty$). A gravitational theory on (\ref{globalmetric}) is dual to a two-dimensional CFT defined on the boundary cylinder $\bS^1\times\bR$, parametrized by $(\tau,\theta)$.

 Defining
 \begin{eqnarray}\label{globaltopoincare}
 t&=&\frac{L\sin\tau}{\cos\tau+\sin\varrho\cos\theta}~,\nonumber\\
 x&=&\frac{L\sin\theta\,\sin\varrho}{\cos\tau+\sin\varrho\cos\theta}~,\\
 z&=&\frac{L\cos\varrho}{\cos\tau+\sin\varrho\cos\theta}~,\nonumber
 \end{eqnarray}
we bring the metric to Poincar\'e form,
\begin{equation}\label{poincaremetric}
ds^2=\frac{L^2}{z^2}\left(-dt^2+dx^2+dz^2\right)~.
\end{equation}
As is well-known, with $z\in(0,\infty)$ and $t,x\in(-\infty,\infty)$, these coordinates cover only the Poincar\'e wedge of AdS, i.e., the portion
$\tau\in(-\pi,\pi)$, $\cos\theta>-\cos\tau\csc\varrho$
of global AdS (see Fig.~\ref{3dfig}). Physically, these coordinates are associated with a family of bulk observers with constant proper acceleration $a^2=-1/L^2$, and the Poincar\'e horizon at $z\to\infty$ ($\cos\theta=-\cos\tau\csc\varrho$) marks the boundary of the region with which they can interact causally. The AdS boundary is at $z=0$. The dual CFT lives on the boundary Minkowski spacetime parametrized by $(t,x)$.

 Given a curve $R(\theta)$ at fixed $T$ (fixed $\tau$) in global AdS, the associated family of tangent geodesics, or equivalently, CFT intervals, can be labeled as $\alpha(\theta_c)$, where $\theta_c$ is the angular location of the interval's center along the spatial $\bS^1$, and $\alpha$ is the interval's (half-)angle of aperture. These are given by \cite{nutsandbolts}
\begin{eqnarray}\label{globalfamily}
\tan\left(\theta-\theta_c\right)&=&\frac{L^2}{L^2+R^2}\frac{d\ln R}{d\theta}~,\\
\tan\alpha&=&\frac{L}{R}\sqrt{1+\frac{L^2}{L^2+R^2}\left(\frac{d\ln R}{d\theta}\right)^2}~.\nonumber
\end{eqnarray}
The endpoints of these geodesics/intervals are located at $\theta_{\pm}\equiv\theta_c\pm\alpha$.

 As explained in the Introduction, if we stick to the $\tau=0$ slice to begin with, we are assured that these same geodesics will provide full coverage of the bulk curve after translation to the Poincar\'e slice $t=0$. We can determine them by using (\ref{globaltopoincare}) to map the two angles $\theta_{\pm}$ to the $x$-axis. The endpoint locations corresponding to $\theta_{\pm}$ will naturally be denoted $x_{\pm}$. Halfway between these two endpoints lies the center of the interval,
\begin{equation}\label{xc0}
x_c\equiv\frac{x_+ + x_-}{2}~,
\end{equation}
and its radius is
\begin{equation}\label{ell0}
\ell\equiv\frac{x_+ - x_-}{2}~.
\end{equation}
 We will let $x_{\theta}$ denote the direct translation of the center angle $\theta_c$, which will serve then as a parameter that labels our intervals. As $\theta_c$ goes around the $\bS^1$ of the cylinder CFT, $x_{\theta}$ will run over the entire spatial axis of the Minkowski CFT. Notice that in general we expect $x_{\theta}\neq x_c$.

 Our one-parameter family of geodesics was parametrized with $\theta_c$ in the global setting, so after translation to Poincar\'e, we can naturally parametrize it with $x_{\theta}$.
 The geodesic for each  value of $x_{\theta}$ can be described with the pair $(x_-,x_+)$, or equivalently, with $(x_c,\ell)$. The latter description is sometimes more convenient.
And instead of reporting our geodesics in parametrized form, $(x_c(x_{\theta}),\ell(x_{\theta}))$, we can eliminate $x_{\theta}$
to obtain $\ell(x_c)$, which is certainly more intuitive, and directly analogous to the global expression reported in \cite{nutsandbolts} in the form $\alpha(\theta_c)$.

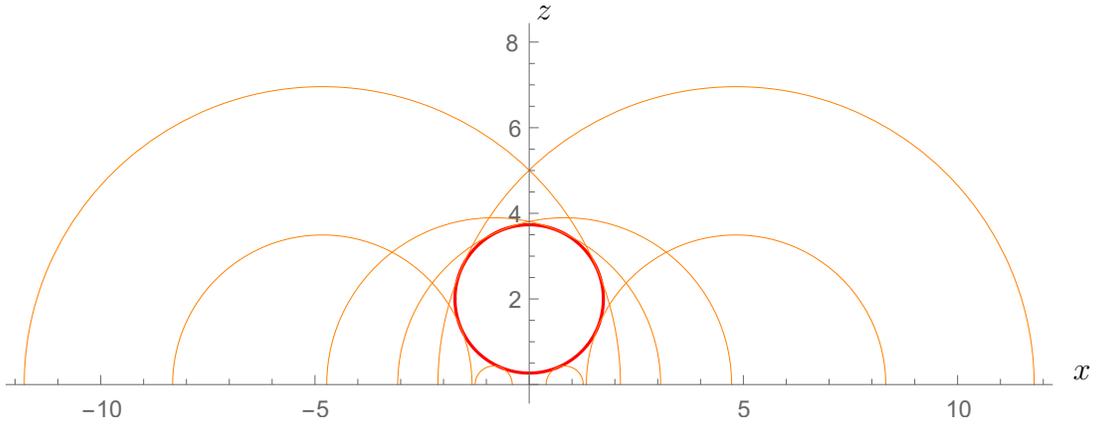
\begin{figure}[hbt]
\begin{center}
  \includegraphics[width=14cm]{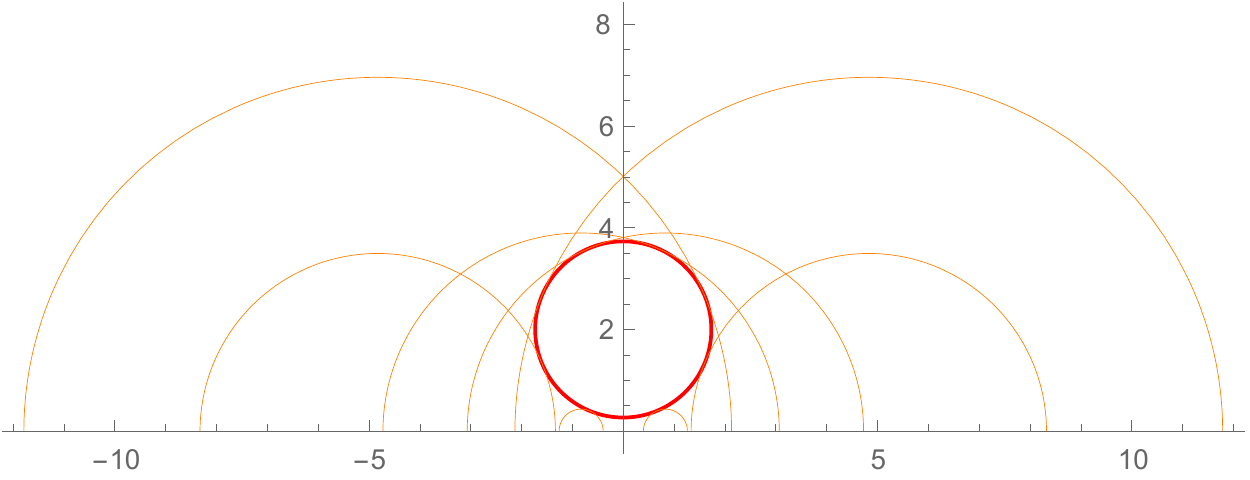}
  \setlength{\unitlength}{1cm}
\begin{picture}(0,0)
\put(-0.2,0.6){ $x$}
\put(-7.2,5.4){$z$}
\end{picture}
\end{center}
\vspace*{-0.8cm}
\caption{Circle (\ref{circlepoincare}) in Poincar\'e AdS at $t=0$, with some of its tangent geodesics, as given by (\ref{circlexpm}). In this and all subsequent plots in this paper, we set $L=1$. We have chosen the circle to be centered at $(x,z)=(0,2)$, meaning that the radius of the circle (in both global and Poincar\'e coordinates) is $R=\sqrt{3}$.
\label{circlefig}}
\end{figure}

It will be instructive to consider first the simplest concrete example of a bulk curve: a circle which in global coordinates is centered at the origin, $R=\mbox{constant}$.
It follows immediately from (\ref{globalfamily}) that the family of geodesics tangent to this circle is simply $\theta_c=\theta$, $\tan\alpha=L/R$. Using (\ref{globaltopoincare}) at $\tau=0$, we can see that the resulting bulk curve in Poincar\'e AdS is also a circle,
\begin{equation}\label{circlepoincare}
x^2+\left(z-\sqrt{L^2+R^2}\,\right)^2=R^2~.
\end{equation}
With the middle equation in (\ref{globaltopoincare}) evaluated at $\varrho=\pi/2$, we can also translate the geodesic parameters $\theta,\theta_{\pm}$. The result takes the form
\begin{equation}\label{circlexpm}
x_{\pm}=\frac{2 x_{\theta} L^2\sqrt{L^2+R^2}\pm L^2(L^2+x_{\theta}^2)}{L^2(\sqrt{L^2+R^2}+R)-x_{\theta}^2(\sqrt{L^2+R^2}-R)}~.
\end{equation}
A representative sampling of these geodesics is plotted in Fig.~\ref{circlefig}.

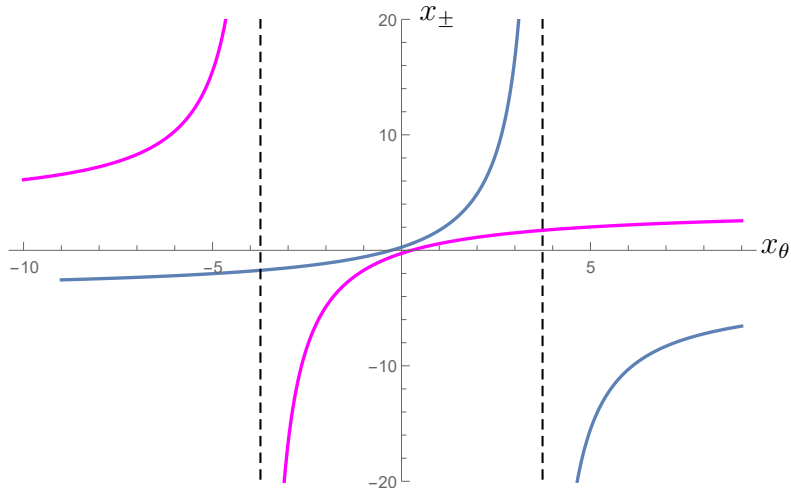
\begin{figure}[hbt]
\begin{center}
  \includegraphics[width=10cm]{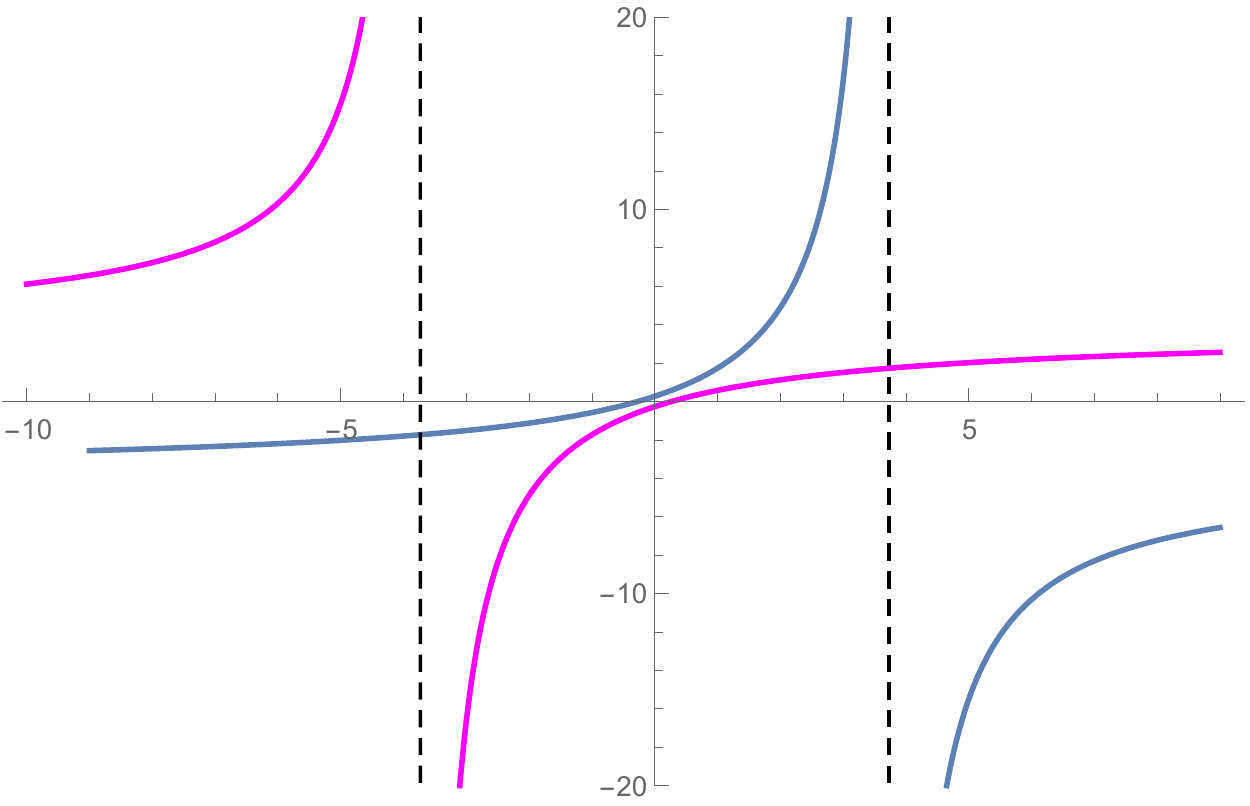}
  \setlength{\unitlength}{1cm}
\begin{picture}(0,0)
\put(-0.4,3.2){ $x_{\theta}$}
\put(-4.8,6.3){$x_{\pm}$}
\end{picture}
\end{center}
\vspace*{-0.8cm}
\caption{The blue (purple) curve shows the endpoint $x_+$ ($x_-$) of the geodesics, as given by (\ref{circlexpm}), for the the same example as in Fig.~\ref{circlefig}. The  locations where one or the other endpoint switches sign upon crossing through infinity, $x_{\theta}=\pm x_{\infty}$, are clearly visible. It is only in the middle region of the plot, $(-x_{\infty},x_{\infty})$, that the endpoints are in the canonical order $x_-<x_+$.
\label{endpointsfig}}
\end{figure}

As expected, we do find a tangent geodesic for each point on our circle.
But there is an important novelty: the denominator in (\ref{circlexpm}) vanishes at $x_{\theta}=\pm x_{\infty}$, with
\begin{equation}\label{xinfty}
x_{\infty}\equiv R+\sqrt{L^2+ R^2}~.
\end{equation}
At each of these locations, one of the endpoints changes sign.
 For $x_{\theta}\in (-x_{\infty},x_{\infty})$ we have the expected ordering $x_-<x_+$, but for other values of $x_{\theta}$ the endpoints are exchanged: as  $x_{\theta}$ increases past $x_{\infty}$, the value of $x_+$ crosses from $x\to\infty$ to $x\to-\infty$, while at $x_{\theta}=-x_{\infty}$, $x_-$ crosses in the opposite direction. The fact
that the interval radius (\ref{ell}) diverges at these crossover points implies that the corresponding geodesic is becoming vertical, and the same is true then for the bulk curve itself, i.e., $\p_{x}z\to\pm\infty$. At these points, $x(x_{\theta})$ starts to backtrack, as we pass from the lower to the upper half of the circle, or viceversa. This behavior is seen in Fig.~\ref{endpointsfig}, where we plot the endpoints as given by (\ref{circlexpm}).

The main lesson here is that, as the parameter $x_{\theta}$ ranges from $-\infty$ to $\infty$, the interval midpoint $x_c$ covers this same range \emph{twice}: once for the geodesics tangent to the lower part of our curve, which have $\ell>0$, and a second time for the geodesics tangent to the upper part, which have $\ell<0$ on account of having their endpoints reversed.

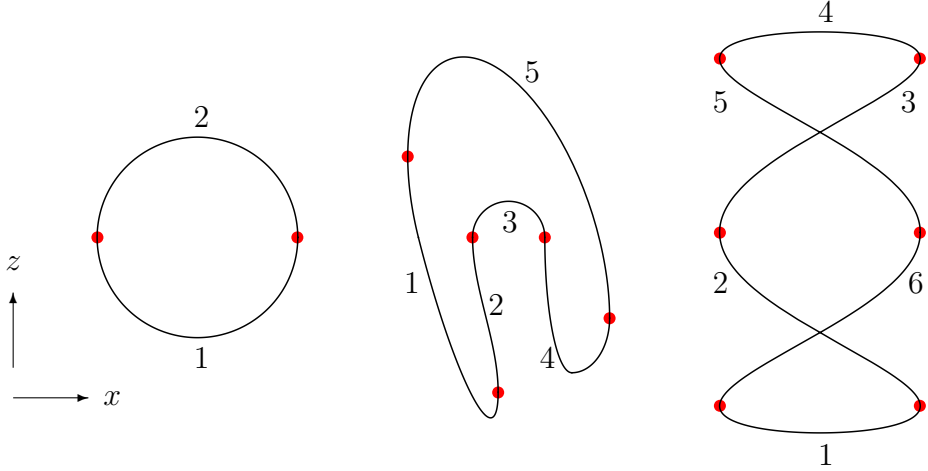
\begin{figure}[hbt]
\begin{center}
  \includegraphics[width=4cm]{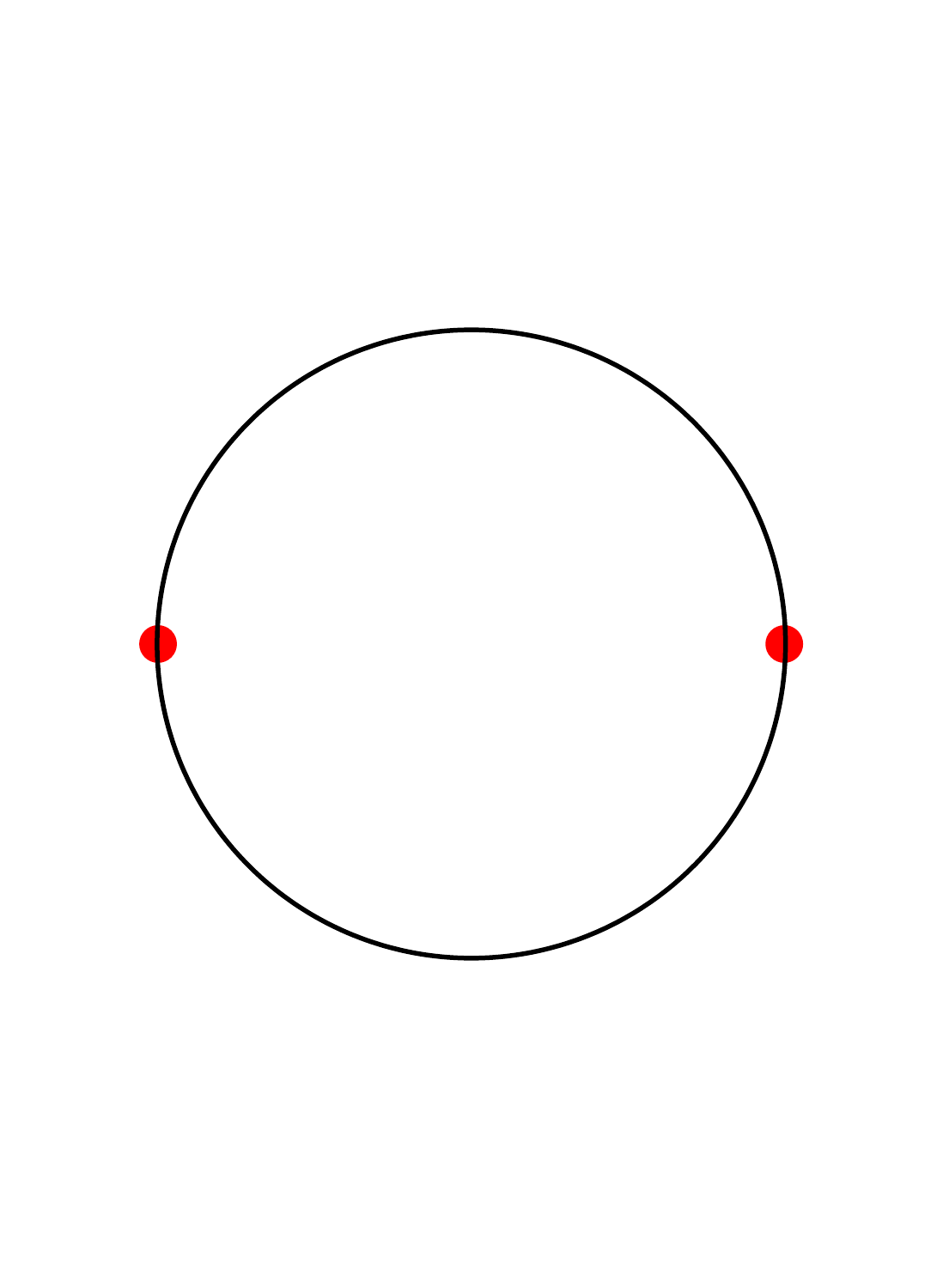}
  \includegraphics[width=4cm]{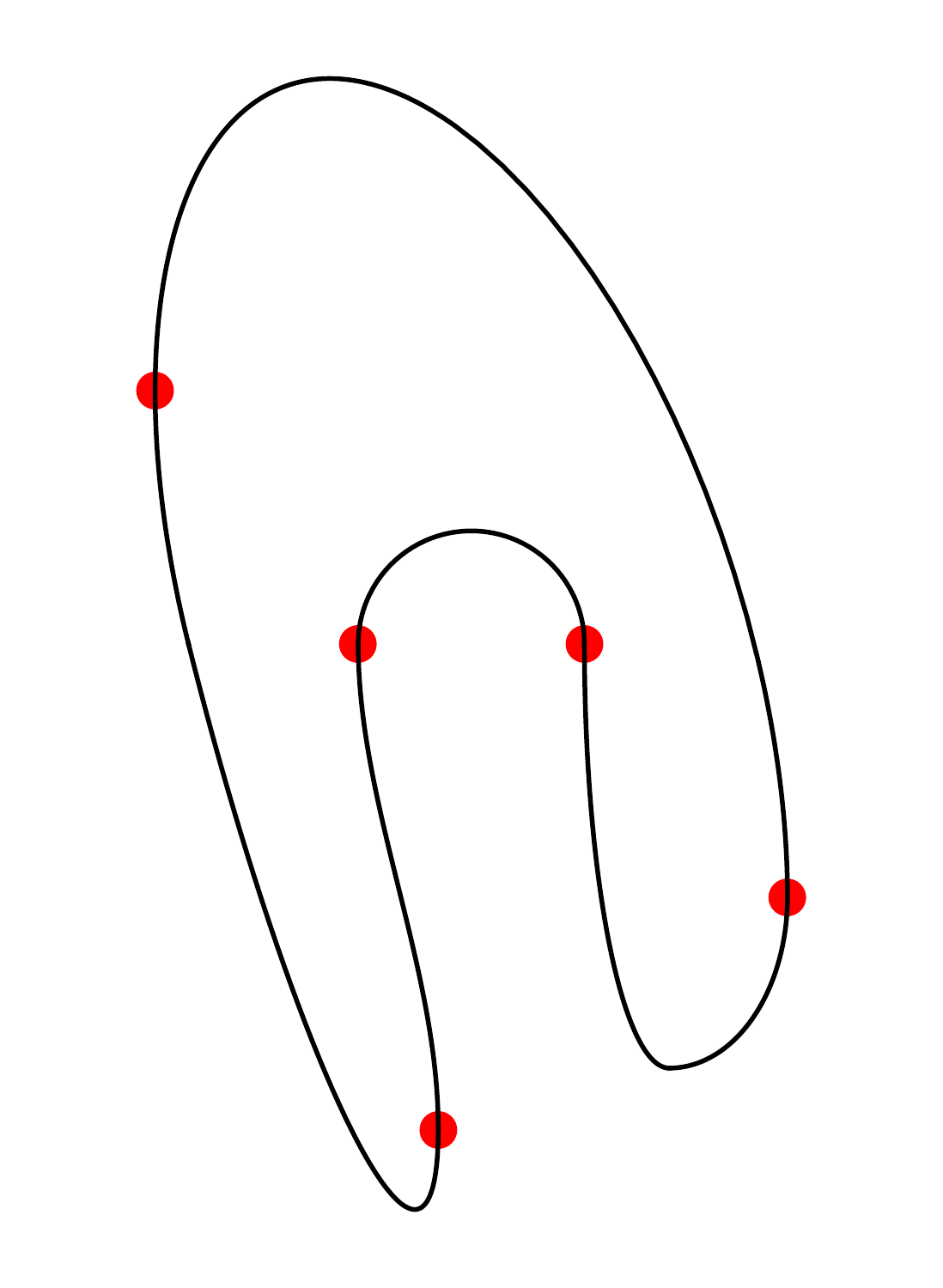}
  \includegraphics[width=4cm]{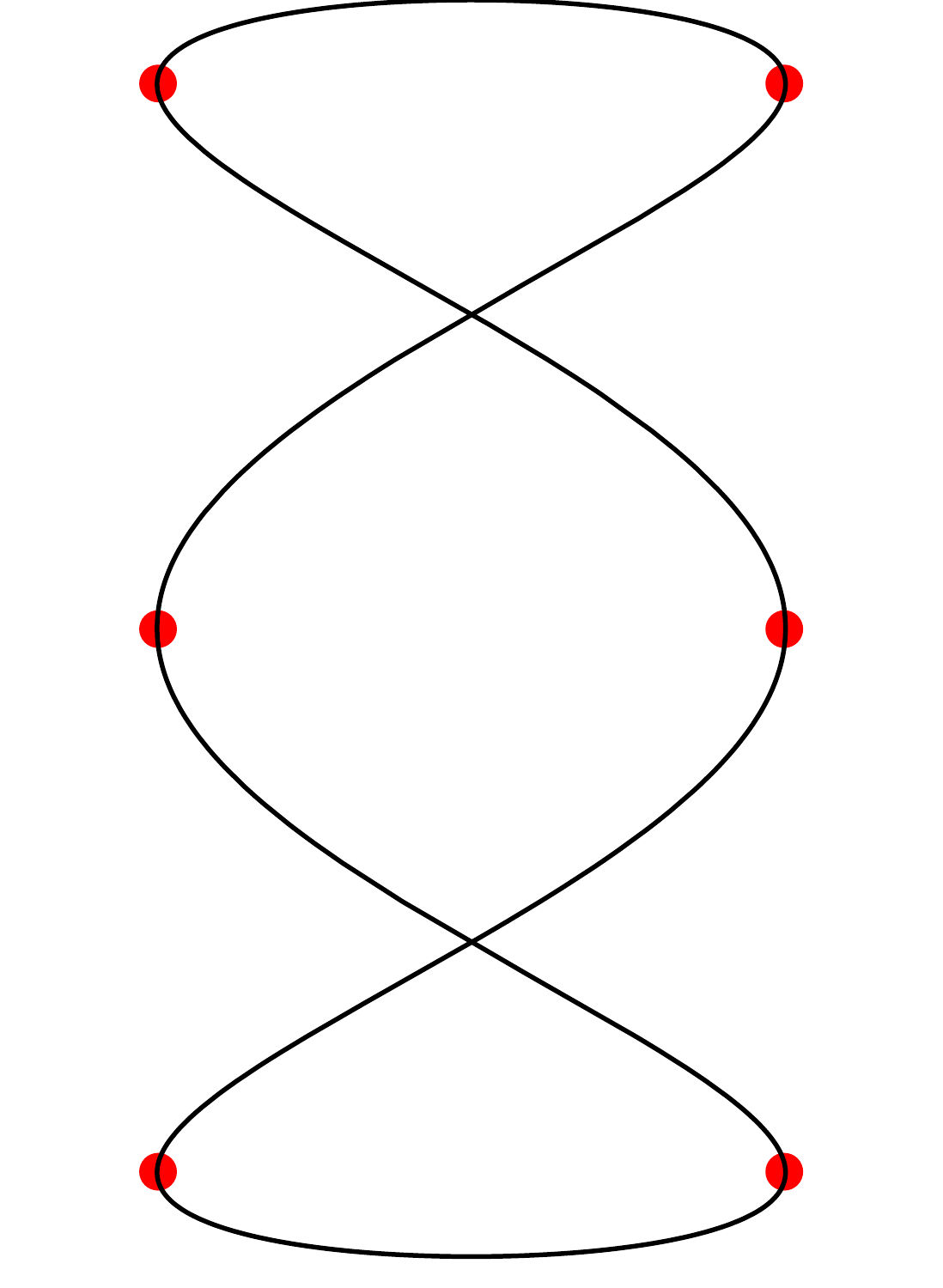}
  \setlength{\unitlength}{1cm}
\begin{picture}(0,0)
\put(-10.6,1){1}
\put(-10.6,4.2){2}
\put(-7.8,2){1}
\put(-6.68,1,7){2}
\put(-6.5,2.8){3}
\put(-6,1){4}
\put(-6.2,4.8){5}
\put(-2.3,-.3){1}
\put(-3.7,2){2}
\put(-1.2,4.4){3}
\put(-2.3,5.6){4}
\put(-3.7,4.4){5}
\put(-1.1,2){6}
\put(-11.8,0.5){$x$}
\put(-13.1,2.3){$z$}
\put(-13,1.0){\vector(0,1){1}}
\put(-13,0.6){\vector(1,0){1}}
\end{picture}
\end{center}
\vspace*{-0.6cm}
\caption{Three examples of closed curves, partitioned into $N$ segments (numbered $n=1,\ldots,N$) at the points (indicated in red) where the curves become vertical ($x'=0$), meaning the corresponding tangent geodesic has one endpoint at infinity. Through (\ref{xpm}), the sign of $x'$ on each segment determines whether the tangent geodesics have their endpoints in the canonical order ($x_-<x_+$) or not. The circle has already been discussed in the text and portrayed in Fig.~\ref{circlefig}. The second curve illustrates the fact that $N$ can be odd, because $x'$ does not necessarily flip sign in going from one segment to the next (in this example, and with the conventions described in the main text, segments $n=3,4$ are both positive). The third example illustrates the fact that the curve can self-intersect.
\label{splitfig}}
\end{figure}

This same lesson applies generally. Consider an arbitrary closed bulk curve (at constant $t$), described as $(x(\lambda),z(\lambda))$, with $\lambda$ some unspecified parameter. Since the curve is closed, the function $x(\lambda)$ must be non-monotonic, and we can find at least two values of $\lambda$ where $x'\equiv\p_{\lambda}x$ changes sign by crossing zero. At these points, the bulk curve becomes vertical, and the radius and one of the endpoints of the corresponding geodesic approach $\pm\infty$. The same would happen at points where $x'$ vanishes without changing sign. The $N\ge 2$ points where the closed curve is vertical ($x'=0$) split the curve into $N$ consecutive segments. Some examples are shown in Fig.~\ref{splitfig}. We will demand, without loss of generality, that the sign of the parameter $\lambda$ be chosen such that the point on the curve that is closest to the AdS boundary is on a segment where $x'>0$. We label this segment $n=1$, and number the remaining segments consecutively in order of increasing $\lambda$. The edges of the $n$th segment are naturally denoted $\lambda_n<\lambda_{n+1}$. As in the case of the circle (where we had $N=2$), each connected segment will be associated with a family of geodesics whose centers $x_c$ run over the entire $x$-axis. The sign of  $x'$ might or might not flip when moving from one segment to the next. We will refer to those segments where $x'>0$  ($x'<0$) as `positive' (`negative').

Either by translating the global AdS results of \cite{nutsandbolts}, or by direct computation in Poincar\'e \cite{hmw},
one finds that the geodesics tangent to our curve have endpoints at
\begin{equation}\label{xpm}
x_{\pm}(\lambda)=x(\lambda)+\frac{z(\lambda)z'(\lambda)}{x'(\lambda)}\pm\frac{z(\lambda)}{x'(\lambda)}\sqrt{x'(\lambda)^2+z'(\lambda)^2}~.
\end{equation}
Equivalently, they have midpoint
\begin{equation}\label{xc}
x_c(\lambda)=x(\lambda)+\frac{z(\lambda)z'(\lambda)}{x'(\lambda)}~,
\end{equation}
and radius
\begin{equation}\label{ell}
\ell(\lambda)=\frac{z(\lambda)}{x'(\lambda)}\sqrt{x'(\lambda)^2+z'(\lambda)^2}~.
\end{equation}
We have chosen the sign of the second denominator in (\ref{xpm}) such that the positive segments of the curve ($x'>0$) are associated with intervals whose endpoints are in the canonical order, $x_-<x_+$, while the negative parts ($x'<0$) correspond to intervals with reversed endpoints, $x_->x_+$. Through (\ref{ell}), this means that the designation as positive or negative, originally referring to an attribute of the bulk curve, also characterizes the sign of $\ell$ for the corresponding family of intervals in the CFT.
Again, the main issue here is that, to fully wrap around our closed curve, we need not one but $N\ge 2$ families of intervals whose midpoints $x_c$ run over the entire $x$-axis.

Alternatively, we can think about this as decomposing the closed curve into $N$ \emph{open} curves $z_{n}(x)$ ($n=1,\ldots,N$), which join together at the places where the slope $\partial z/\partial x$ diverges. But if we adopt this perspective, the nontrivial question is whether the information from all $N$ families of geodesics can be smoothly combined to obtain a hole-ographic description for the entire closed curve, since we know from \cite{nutsandbolts} that to obtain the length of open curves we need to add a surface term to the formula for differential entropy. We will address this question explicitly in Section \ref{opensubsec}.

Notice that (\ref{xc}) implies that $x'_c=(1+(\p_x z)^2+z\p^2_x z)x'$. This shows that the center $x_c(\lambda)$ can backtrack if $x'<0$, which happens on the negative segments that we have discussed here, or if the curve is sufficiently concave, $\p^2_x z<-(1+(\p_x z)^2)/z$. The latter possibility had been pointed out in \cite{myers,hmw,nutsandbolts}.

For use below, we note that (\ref{xc}) and (\ref{ell}) can be inverted \cite{hmw} to give the bulk curve in terms of the boundary data,
\begin{eqnarray}\label{xztilde}
x(\lambda)&=&x_c(\lambda)-\frac{\ell(\lambda)\ell'(\lambda)}{x_c'(\lambda)}~,\nonumber\\
z(\lambda)&=&\sqrt{\ell^2(\lambda)\left(1-\frac{\ell^{'2}(\lambda)}{x^{'2}_c(\lambda)}\right)}~.
\end{eqnarray}
As an additional check, these same relations can be obtained by taking the zero-mass limit $R_+\to 0$ of the expressions worked out for the static BTZ black hole \cite{btz} in Eqs.~(89)-(90) of \cite{nutsandbolts}. In this limit, the BTZ metric reduces to Poincar\'e AdS with $x\simeq x+L$.

\subsection{Differential entropy and the length of closed curves} \label{diffentsubsec}

The definition of differential entropy is most conveniently given in the form \cite{hmw}
 \begin{equation}\label{e}
 E=\int d\lambda\left.\frac{\partial S(x_L(\lambda),x_R(\bar{\lambda}))}{\partial\bar{\lambda}}\right|_{\bar{\lambda}=\lambda}~.
% =-\int d\lambda\left.\frac{\partial S(x_L(\bar{\lambda}),x_R(\lambda))}{\partial\bar{\lambda}}\right|_{\bar{\lambda}=\lambda}~.
 \end{equation}
Here $x_L$ and $x_R$ are the left and right endpoints\footnote{Notice that $x_L=x_-$ and $x_R=x_+$ only if $\ell>0$. We will return to this point below.} ($x_L\le x_R$) of a family of intervals parametrized by an arbitrary parameter $\lambda$, and $S$ denotes the corresponding entanglement entropy. The definition (\ref{e}) treats the right and left endpoints on a different footing, but as explained in \cite{hmw}, an integration by parts allows the role of $x_R$ and $x_L$ to be exchanged. The two alternative definitions differ by a boundary term, which can be neglected for the types of curves considered in \cite{hmw}, but will be important for our analysis of open curves in Section~\ref{opensubsec}. In Appendix~\ref{appendix} we show that there is in fact a one-parameter family of possible definitions of differential entropy, arising from a corresponding ambiguity (\ref{diffExidiscrete}) in the discrete version of $E$ originally considered in \cite{hole-ography,myers}.

 For convenience, from this point on we will rescale the entanglement entropy by a factor of $4G_N$, so that the Ryu-Takayanagi formula (\ref{rt}) reads $S=A_{\Sigma}$. In terms of the central charge $c$ of the CFT, reporting $S$ in units of $4G_N$ is the same as reporting it in units of $c/6L$ \cite{bh}.
For intervals at fixed time, as we are considering here, the entropy in the Minkowski space CFT$_2$  is given by (see, e.g., \cite{cc,myers})
\begin{equation}\label{s}
S(x_L,x_R)=2L\ln\left(\frac{x_R-x_L}{\epsilon}\right)~,
\end{equation}
where $\epsilon$ is a UV cutoff.\footnote{For comparison, in the case of global AdS, where the dual CFT lives on a cylinder, the entanglement entropy is (see, e.g., \cite{nutsandbolts})
$$
S(\theta_+,\theta_-)=2L\ln\left[\sin((\theta_+-\theta_-)/2\delta)\right]~.
$$
As explained in the Introduction, this equation and (\ref{s}) are not mapped into one another by mere coordinate transformation.}

In the context of holographic entanglement, the authors of \cite{myers} were the first to study curves in Poincar\'e AdS$_3$ at constant time. (Their analysis applies as well to codimension-2 surfaces in Poincar\'e AdS$_{d+1}$ with planar symmetry, i.e., translationally-invariant under $d-2$ of the $x^i$.) They restricted attention to curves that are infinitely extended along the $x$ direction, and moreover imposed periodic boundary conditions at $x\to\pm\infty$.
% (equivalent to assuming a compact spatial direction).
%Importantly, $x_{L,R}$ are assumed in \cite{hmw} to be periodic over the range of $\lambda$, so that surface terms in $\int d\lambda$ can be taken to vanish.
Under these conditions, they showed that the differential entropy (\ref{e}) for the family of intervals tangent to the curve (surface) correctly reproduces its length (area).

We will now show that the same is true for the closed curves $(x(\lambda),z(\lambda))$ that we considered in the previous subsection. Their length is given by
\begin{equation}\label{a}
A=\int d\lambda\sqrt{\gamma_{\lambda\lambda}}=\int d\lambda\,\frac{L}{z}\sqrt{x'^2+z'^2}~,
\end{equation}
where $\gamma$ is the induced metric.
%and a prime on a function denotes differentation with respect to its argument.
We want to check that this agrees with the differential entropy associated to the curve. The corresponding geodesics/intervals have endpoints located at (\ref{xpm}). For ease of reading, we will phrase our discussion for the case $N=2$ (the closed curve has only one positive and one negative segment), but the extension to $N>2$ is immediate.

For the positive part of the curve ($x'> 0$), the fact that $\ell>0$ means that the left and right endpoints are
$x_L=x_-$ and $x_R=x_+$. Using (\ref{s}), (\ref{e}) becomes
\begin{equation}\label{e2}
E=L\int d\lambda\,\frac{x_+'}{\ell}~.
\end{equation}
For the negative part, $\ell<0$ and so the endpoints are reversed, $x_L=x_+$ and $x_R=x_-$. Because of this, if we were to use (\ref{e}) as it stands, we would get some additional minus signs, and would not be able to directly obtain the total length of the curve.  But, for continuity in the family of intervals (crucial for the usefulness of differential entropy, and most clearly seen by referring back to the global AdS setup), the correct prescription is to depart from a literal reading of (\ref{e}), and keep treating $x_+$ as the \emph{right} endpoint of the interval. This ensures the appropriate cancelation of the final geodesics in the positive family against the initial geodesics in the negative family. Of course, for the logarithm in (\ref{s}) to be real, we do need to use $|x_+-x_-|$ as its argument. We are then led again to (\ref{e2}), so this single expression applies for the entire closed curve. Periodicity then guarantees, just like it did for the infinite curves considered in \cite{myers}, that surface terms can be ignored.

Let us now see explicitly the relation between differential entropy and length. Using (\ref{xpm}) and (\ref{ell}), expression (\ref{e2}) can be easily seen to take the form
\begin{eqnarray}\label{e3}
E&=&\oint d\lambda\left\{\frac{L}{z}\sqrt{x'^2+z'^2}
+L\left[\frac{\ell'}{\ell} +\frac{z''}{\sqrt{x'^2+z'^2}}-\frac{z'x''}{x'\sqrt{x'^2+z'^2}}\right]\right\}\nonumber\\
&=&A+L\oint d\lambda\,\p_{\lambda}\left[\ln\left(\frac{2|\ell|}{\epsilon}\right) +\sinh^{-1}\left(\frac{z'}{|x'|}\right)\right]~.
\end{eqnarray}
In the second line we have recognized that the first term precisely reproduces the length (\ref{a}), while the others amount to a total derivative, and do not contribute. Inside the logarithm, we have chosen a particular value of the constant of integration, involving the UV cutoff $\epsilon$. This choice will prove to be convenient in the next subsection.

\subsection{Boundary terms and the length of open curves}\label{opensubsec}

We now move on to considering (still at $t=0$) an arbitrary \emph{open} curve $(x(\lambda),z(\lambda))$. It might or might not have points where $z'/x'\to\pm\infty$, separating $N$ segments just like we discussed for closed curves (but now with $N\ge 1$).  The analysis in the preceding section directly establishes a relation between its length $A$ and the differential entropy $E$ for its associated family of intervals. This relation is again given by (\ref{e3}), with the sole difference that the integral now extends over a finite range,
\begin{eqnarray}\label{eopen}
A&=& E-L\int_{\lambda_{i}}^{\lambda_f}\ d\lambda\,
\left[\frac{\ell'}{\ell}
+\frac{z''}{\sqrt{x'^2+z'^2}}
-\frac{z'x''}{x'\sqrt{x'^2+z'^2}}\right]\nonumber\\
&=& E-f(\lambda_f)+f(\lambda_i)~.
\end{eqnarray}
In the second line we have given the name
\begin{equation}\label{f}
f(\lambda)\equiv L\ln\left(\frac{2|\ell|}{\epsilon}\right) +L\sinh^{-1}\left(\frac{z'}{|x'|}\right)~,
\end{equation}
to the (now generally non-vanishing) boundary contribution.

Let us now try to gain some understanding on the form of (\ref{f}). As we mentioned in the Introduction, the authors of \cite{nutsandbolts} showed that, when considering an \emph{open} curve in global coordinates, $R(\theta)$, with $\theta$  running from $\theta_i$ to $\theta_f$, the differential entropy
$E$ does not directly reproduce the length $A$. The two integrands differ by a total derivative. To obtain a match, one must add to $E$ a specific surface term $f(\theta_f)-f(\theta_i)$, with
\begin{equation}\label{fglobal}
f(\theta)=2L\ln\left[\frac{\sin\left(\alpha+(\theta-\theta_c)\right)}{\sin\left(\alpha-(\theta-\theta_c)\right)}\right]
=2L\ln\left[\frac{\sin\left(\theta-\theta_{-}\right)}{\sin\left(\theta_{+}-\theta\right)}\right]~,
\end{equation}
where $\alpha$ and $\theta_c$ are evaluated at the values corresponding to the bulk angle $\theta$. (Alternatively, $f$ could be expressed as a function of the boundary angle $\theta_c$.) Above their Eq.~(12), the authors of \cite{nutsandbolts} explain the geometric meaning of $f(\theta)$: it is the length of the arc of the geodesic $(\theta_c,\alpha)$ that is contained in the angular wedge between $\theta_c$ and $\theta$. Explicitly, this geodesic $R_g(\theta_g)$ is described by
\begin{equation}\label{globalgeodesic}
\tan^2(\theta_g-\theta_c)=\frac{R_g^2\tan^2\alpha-L^2}{R_g^2+L^2}~,
\end{equation}
and one can check that the length of its arc in the range of interest,
\begin{equation}\label{globalgeodesiclength}
\int_{\theta_c}^{\theta}d\theta_g\sqrt{R_g^2+\left(1+\frac{R^2_g}{L^2}\right)^{-1}\left(\frac{dR_g}{d\theta_g}\right)^2}~,
\end{equation}
indeed agrees with (\ref{fglobal}).

\emph{A priori}, it is not obvious whether a similar interpretation can be given to the Poincar\'e boundary function (\ref{f}), because, as we have explained before, entanglement entropy does not remain invariant when mapping from global to Poincar\'e AdS. In particular, the condition $\theta=\theta_c$, which makes the global boundary function (\ref{fglobal}) vanish, does not translate into $x=x_c$ or $x=x_{\theta}$.

Let us work this out for an arbitrary open curve $(x(\lambda),z(\lambda))$. The geodesic tangent to the curve at the point labeled by $\lambda$ is
\begin{equation}\label{poincaregeodesic}
z_g=\sqrt{\ell^2-(x_g-x_c)^2}~,
\end{equation}
where the radius $\ell$ and the midpoint $x_c$ are given by (\ref{xc}) and (\ref{ell}), and are therefore held fixed for the present calculation. The length of the arc of this geodesic that runs from $x$ to $x_c$ is
\begin{equation}\label{poincaregeodesiclength}
\int_{x}^{x_c}dx_g\,\frac{L}{z_g}\sqrt{1+\left(\frac{\partial z_g}{\partial x_g}\right)^2}
=-L\tanh^{-1}\left(\frac{x-x_c}{\ell}\right)
=-\frac{L}{2}\ln\left(\frac{x-x_-}{x_+-x}\right)~.
\end{equation}
Notice that, in this last form, the length (\ref{poincaregeodesiclength}) looks rather analogous to the final version of (\ref{fglobal}), except for an overall minus sign which is due to the fact that in (\ref{eopen}) we have chosen to define our $f$ with a sign opposite to that of \cite{nutsandbolts}.
Using (\ref{xpm}) and the identity $\sinh^{-1}a=\ln(a+\sqrt{1+a^2})$, this expression can be rewritten as
\begin{equation}\label{poincaregeodesiclength2}
\int_{x}^{x_c}dx_g\,\frac{L}{z_g}\sqrt{1+\left(\frac{\partial z_g}{\partial x_g}\right)^2}
=L\sinh^{-1}\left(\frac{z'}{|x'|}\right)~,
\end{equation}
which coincides with the second term of (\ref{f}).

This agreement allows us to ascribe to the term (\ref{poincaregeodesiclength2}) the entanglement interpretation developed for global AdS in Section 4.5 of \cite{nutsandbolts}. The family of intervals/geodesics associated with our open curve ends at the bulk point $(x_f,z_f)\equiv(x(\lambda_f),z(\lambda_f))$. The final member of the family is centered at $x_{c,f}\equiv x_c(\lambda_f)$, and generally $x_f\neq x_{c,f}$. We can add to the family the set of intervals whose center runs from $x_{c,f}$ to $x_f$, with radii $\ell$ chosen such that the corresponding geodesics all go through $(x_f,z_f)$, meaning that this addition does not enlarge our curve. (The added intervals belong to the family of the `point-curve' $(x_f,z_f)$, as will become clear in the next subsection.) After the addition, there is no longer any arc left for (\ref{poincaregeodesiclength}) to contribute, which means that the second term in (\ref{f}), evaluated at $\lambda_f$, represents the extra differential entropy due to the added set of intervals.  The same applies of course at the opposite endpoint of the the curve, $\lambda_i$.

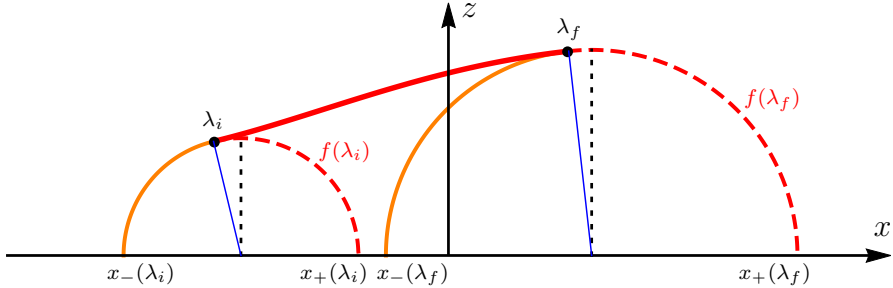
\begin{figure}[hbt]
\begin{center}
\includegraphics[angle=0,width=0.86\textwidth]{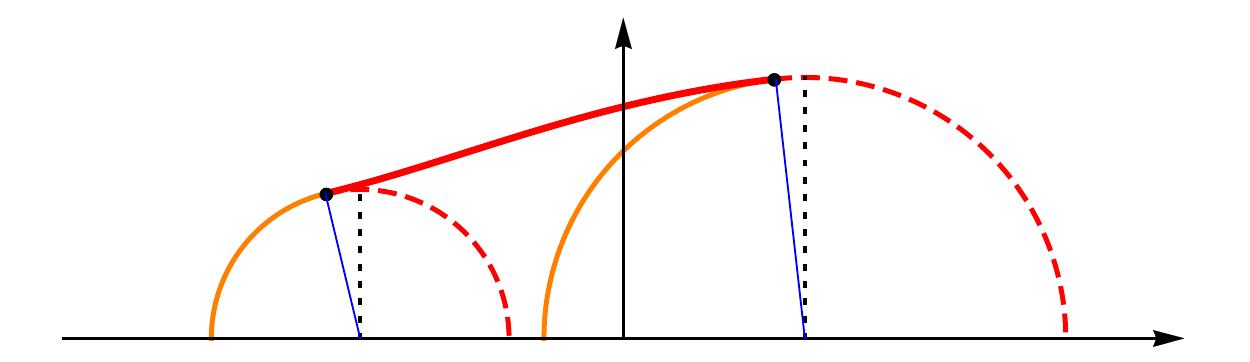}
\begin{picture}(0,0)
\put(-186,97){$z$}
\put(-30,14){$x$}
\put(-218,-2){\scriptsize{$x_-(\lambda_f)$}}
\put(-81,-2){\scriptsize{$x_+(\lambda_f)$}}
\put(-320,-2){\scriptsize{$x_-(\lambda_i)$}}
\put(-247,-2){\scriptsize{$x_+(\lambda_i)$}}
\put(-284,56){\scriptsize{$\lambda_{i}$}}
\put(-150,91){\scriptsize{$\lambda_{f}$}}
\put(-240,44){\scriptsize{{\color{red}$f(\lambda_{i})$}}}
\put(-79,64){\scriptsize{{\color{red}$f(\lambda_{f})$}}}
\end{picture}
\caption{\small Geometric interpretation of the boundary term $f(\lambda)$ in the definition (\ref{ecalrepeated}) of the renormalized differential entropy $\mathcal{E}$. The value of $f$ at each endpoint is the length of the arcs shown in dotted red.
\label{ffig}}
\end{center}
\end{figure}

Only the logarithm in (\ref{f}) remains to be interpreted. But comparing with (\ref{s}), we see that this term is half the entanglement entropy of the interval at $\lambda_f$ (or $\lambda_i$). We thus conclude that the entire formula (\ref{eopen}) for the length of our open curve admits an interpretation based on entanglement in the CFT. In the bulk description, the interpretation is very simple: the boundary function (\ref{f}) is the length of the arc of the corresponding geodesic, computed from the edge of our curve, at $x$, all the way to the right endpoint of the geodesic, $x_+$. Or, more precisely, to the regularized version of this endpoint,
\begin{equation}\label{xpepsilon}
x^\epsilon_+\equiv x_+-\frac{\epsilon^2}{2\ell}~,
\end{equation}
where the geodesic reaches the UV cutoff $z=\epsilon$. This geometric interpretation is illustrated in Fig.~\ref{ffig}.

For an open curve whose associated geodesics cover the entire $x$-axis, such as the positive or negative semicircle that we analyzed in Section~\ref{closedsubsec}, (\ref{f}) is logarithmically divergent (because both $\ell$ and $\partial z/\p x$ diverge at the endpoints $\lambda_{i,f}$). In that case, it is more convenient to reexpress $f$ as the integral over $\lambda$ of a total $\lambda$-derivative, so that it can be subtracted directly from the \emph{integrand} of $E$ in (\ref{e}), to get a finite result. This takes us back to the total-derivative terms in the top line of (\ref{e3}), which can be rewritten in the the form
\begin{eqnarray}\label{fpoincareint}
f(\lambda_f)&=&L\int^{\lambda_f}d\lambda\left(\frac{\ell'}{\ell} +\frac{z''}{\sqrt{x'^2+z'^2}}-\frac{z'x''}{x'\sqrt{x'^2+z'^2}}\right)\nonumber\\
{}&=&L\int^{\lambda_f}d\lambda \left(\frac{\ell'}{\ell}+\frac{\ell'x_c''-\ell''x_c'}{\ell'^2 -x_c'^2}\right)~.
\end{eqnarray}
In the second line we have used (\ref{xztilde}) to express $f$ purely in terms of boundary data. Combining (\ref{fpoincareint}) with (\ref{e2}), we can define a `renormalized' differential entropy
\begin{equation}\label{ecal}
\mathcal{E}[\ell ]\equiv E[\ell] - f(\lambda_f)+ f(\lambda_i)
=L\int_{\lambda_{i}}^{\lambda_{f}}d\lambda\left(\frac{x_c'}{\ell}+\frac{\ell'x_c''-\ell''x_c'}{x_c'^2-\ell'^2}\right)~.
\end{equation}
{}From our previous analysis, this directly reproduces the length of an arbitrary open curve,
\begin{equation}\label{eamatching}
A=\mathcal{E}~.
\end{equation}

As an example, consider the circle (\ref{circlepoincare}), shown in Fig.~\ref{circlefig}. In the language of Section~\ref{closedsubsec}, its lower half is a positive segment ($x'>0$) and is labeled $n=1$, whereas its upper half is a negative segment, denoted $n=2$. These two semicircles are open curves described by
\begin{equation}\label{circlezul}
z_{1,2}(x)=\sqrt{L^2+R^2}\mp\sqrt{R^2-x^2}~,
\end{equation}
with $x$ ranging between $-R$ and $R$.
Their length is
\begin{equation}\label{semicirclelengths}
A_{1,2}=\pi R\mp 2R\tan^{-1}(R/L)~.
\end{equation}
Notice that the length of the two semicircles is different, even though they do add up to the correct total, $A=2\pi R$. This is due to the $z$-dependence of the metric.
Using (\ref{ecal}), we find
\begin{equation}\label{circleeamatch}
\mathcal{E}_{1,2}=\pm A_{1,2}.
\end{equation}
The reversal of sign for the negative semicircle is as expected from the convention adopted in the previous subsection and not implemented when writing down (\ref{semicirclelengths}): for a negative segment, $\lambda$ should run in the direction of decreasing $x$. It is with this orientation that the full circle is traced by the original parameter $\theta$ or $x_{\theta}$. Indeed, if we take this sign into account, we find that upon combining the two semicircles the contribution of the boundary function cancels, and we have
\begin{equation}\label{circlefull}
\mathcal{E}_1-\mathcal{E}_2=E_1-E_2=A_1+A_2=A~.
\end{equation}

\subsection{Points} \label{pointsubsec}

Now that we have a formula that computes the lengths of arbitrary closed or open bulk curves in terms of boundary entanglement entropies, we can shrink these curves as in \cite{nutsandbolts}, to obtain points. To describe a given point, we have two options. One is to start with a closed curve, which we will take for simplicity to have only one postive and one negative portion (e.g., a circle). Closed curves have the advantage of not needing any boundary terms, but require a family of intervals/geodesics covering the $x$-axis at least twice. The other option is to start with an open (positive or negative) curve (e.g., a semicircle) whose slope $\partial z/\partial x$ diverges at the edges, so that (via (\ref{xc}) and (\ref{ell})) the corresponding intervals/geodesics cover the entire $x$-axis.\footnote{If we started instead with an open bulk curve whose slope is not divergent at the endpoints, then the range of $x$ covered by the corresponding CFT intervals would be finite, and when we shrink the size of the curve we would end up with nothing.} In this case we do not have to deal with the double-valuedness of $x_c$, but the price we pay is that we must include the boundary contribution (\ref{f}).

Either way, upon shrinking the size of the curve all the way down to zero, we obtain the family of intervals $(x_c(\lambda),\ell(\lambda))$ (equivalently, $x_\pm(\lambda)$) whose associated geodesics all pass through the desired bulk point $(x,z)$. These intervals can of course be determined directly from (\ref{poincaregeodesic}),
\begin{equation}\label{ell2}
\ell=\pm\sqrt{(x-x_c)^2+z^2}~,
\end{equation}
where the family with the upper (lower) sign is needed to describe a positive (negative) open `point-curve', and both families are needed to assemble a closed point-curve. If we wanted to, we could by convention always pick the positive branch of the square root in (\ref{ell2}), which would amount to changing our notation to always insist on having $x_+\ge x_-$. But when putting together the positive and negative segments to construct a closed point-curve, we would still need to use the appropriate signs, as discussed in the previous two subsections. Eq.~(\ref{ell2}) can be rewritten in terms of the intervals' endpoints as
\begin{equation}\label{geodesicRL1}
(x_+ - x)(x-x_{-})=z^2~.
\end{equation}

It is interesting to ask what the special property is that allows the particular set of CFT intervals $\ell(x_c)$ in (\ref{ell2}) to be identified as describing a bulk point in AdS. This is important if we are attempting to reconstruct the bulk starting just from the boundary theory. By taking the first and second derivative of (\ref{ell2}), we can see that our point-curves are solutions to the equation of motion
\begin{equation}\label{pointeom}
\ell \ell''+\ell'^2-1=0~.
\end{equation}
This is then the analog of Eq.~(21) in \cite{nutsandbolts}. As explained there, it is natural to obtain a second-order differential equation, since there must be two integration constants, associated with the coordinates of the bulk point, $(x,z)$. Incidentally, we might wonder why, to single out a point, we are prescribing here an infinite family of geodesics that pass through it, when it should suffice to specify just \emph{two} such geodesics to locate the point where they intersect. Indeed, given only two intersecting geodesics (equivalently, two overlapping intervals in the CFT), we know the radii at the given midpoints, $\ell(x_{c,1})$ and $\ell(x_{c,2})$, and these two data pick out a unique solution to (\ref{pointeom}), i.e., a unique family that covers the entire spatial axis and includes both of the geodesics that we started with. What we gain by thinking of the entire family instead of the original pair is that we can analyze the point-curve in parallel with any other bulk curve, and in particular verify that it has vanishing length by computing its differential entropy.

Following \cite{nutsandbolts}, we expect the equation of motion (\ref{pointeom}) to  follow from an action principle based on extremizing the extrinsic curvature of closed curves. The idea is the following: in negatively curved spaces, the Gauss-Bonnet theorem states that
\begin{equation}\label{gbth}
\oint_C d\lambda\sqrt{\gamma}\,K=2\pi-\int_{\Sigma} d\Sigma\,\mathcal{R}\geq2\pi\,,
\end{equation}
for any closed curve $C$ such that $C=\partial\Sigma$, where $d\lambda\sqrt{\gamma}$ is the length element along the curve, $K$ is the extrinsic curvature and $\mathcal{R}$ is the Ricci scalar on the surface $\Sigma$ bounded by the loop. Evidently, if the loop shrinks to a point the second integral vanishes, and the inequality is saturated. Thus, we can find bulk points by extremizing the left-hand side of (\ref{gbth}).

The extrinsic curvature is computed from
\begin{equation}\label{extrinsiccurvature}
K_{mn}=\frac{1}{2}\left(n^p\partial_p g_{mn}+g_{pn}\partial_m n^{p}+g_{pm}\partial_n n^{p}\right)\,,
\end{equation}
where $n_m$ is a normal unit vector and $\gamma_{mn}=g_{mn}+n_m n_n$ is the induced metric on the curve.
%{If we just want to assemble the LHS of (\ref{gbth}), we probably don't need the full tensor (\ref{extrinsiccurvature}). Also, it'd be best if we can use a language and notation here that runs in parallel with what is needed for the covariant case, Eq.~(\ref{extrinsiccurvatures}).}
The scalar extrinsic curvature is computed by contracting $K_{mn}$ with $\gamma^{mn}$.

For an arbitrary (time-independent) closed curve, our proposed action
$I\equiv\int d\lambda\,\mathcal{L}$, with Lagrangian $\mathcal{L}\equiv\sqrt{\gamma}\,K$, is found to take the form
\begin{equation}\label{actionK}
I=\int d\lambda\,\frac{x'(\lambda)^3-z(\lambda) z'(\lambda) x''(\lambda)+x'(\lambda) \left(z'(\lambda)^2+z(\lambda) z''(\lambda)\right)}{z(\lambda)\left(x'(\lambda)^2+z'(\lambda)^2\right)}~.
\end{equation}
As we can see, this action contains second-order derivatives. Nonetheless, the Euler-Lagrange equations,
\begin{equation}
\frac{d}{d\lambda^2}\frac{\partial\mathcal{L}}{\partial z''}-\frac{d}{d\lambda}\frac{\partial\mathcal{L}}{\partial z'}+\frac{\partial\mathcal{L}}{\partial z}=0\,,\qquad \frac{d}{d\lambda^2}\frac{\partial\mathcal{L}}{\partial x''}-\frac{d}{d\lambda}\frac{\partial\mathcal{L}}{\partial x'}+\frac{\partial\mathcal{L}}{\partial x}=0\,,
\end{equation}
simplify drastically, leading to $x'(\lambda)=0$ and $z'(\lambda)=0$, respectively. The solution defines the bulk point $(x,z)$, which serves as a consistency check of the functional (\ref{actionK}).

In terms of boundary data, we can rewrite (\ref{actionK}) as
\begin{equation}\label{actionK2}
I= 2\int d\lambda \frac{\sqrt{x_+'(\lambda)x_-'(\lambda)}}{x_+(\lambda)-x_-(\lambda)}= \int d\lambda \frac{\sqrt{x_c'(\lambda)^2-\ell'(\lambda)^2}}{\ell(\lambda)}\,.
\end{equation}
In the second form, the Lagrangian is independent of $x_{c}$, so there is an associated conserved momentum,
\begin{equation}
\frac{d}{d\lambda}\frac{\partial \mathcal{L}}{\partial x_c'}=0\qquad\Rightarrow\qquad \frac{\partial \mathcal{L}}{\partial x_c'}=\frac{x_{c}'}{\ell\sqrt{x_c'^2-\ell'^2}}=\Pi\,.
\end{equation}
Solving for $x_{c}'(\lambda)$,
\begin{equation}\label{xpluseq}
x_{c}'(\lambda)=\pm\frac{\Pi \ell(\lambda)\ell'(\lambda)}{\sqrt{\Pi^2 \ell(\lambda)^2-1}}\,,
\end{equation}
and plugging it back into (\ref{actionK2}) we obtain
\begin{equation}\label{actionK3}
I=  \int \frac{d\lambda}{\ell(\lambda)} \sqrt{\frac{\ell'(\lambda)^2}{\Pi^2 \ell(\lambda)^2-1}}\,.
\end{equation}
The equation for $\ell$ derived from (\ref{actionK3}) is trivially satisfied, so we can focus on (\ref{xpluseq}) only. We can get rid of $\lambda$ by writing (\ref{xpluseq}) as
\begin{equation}
\frac{d x_{c}}{d \ell}=\pm\frac{\Pi \ell}{\sqrt{\Pi^2 \ell^2-1}}\,,
\end{equation}
which has solution
\begin{equation}
x_{c}=\pm\sqrt{\ell^2-\Pi^{-2}}+\zeta\,.
\end{equation}
If we identify the integration constants as $\Pi=z^{-1}$ and $\zeta=x$ we recover equation (\ref{ell2}), as expected. Consistent with this, if in (\ref{actionK2}) we choose $\lambda=x_c$ and then extremize, we indeed recover the equation of motion (\ref{pointeom}).

\subsection{Distances} \label{distancesubsec}

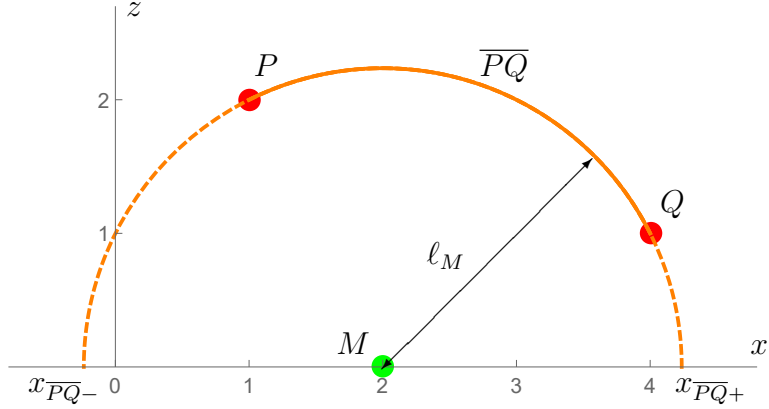
\begin{figure}[hbt]
\begin{center}
  \includegraphics[width=10cm]{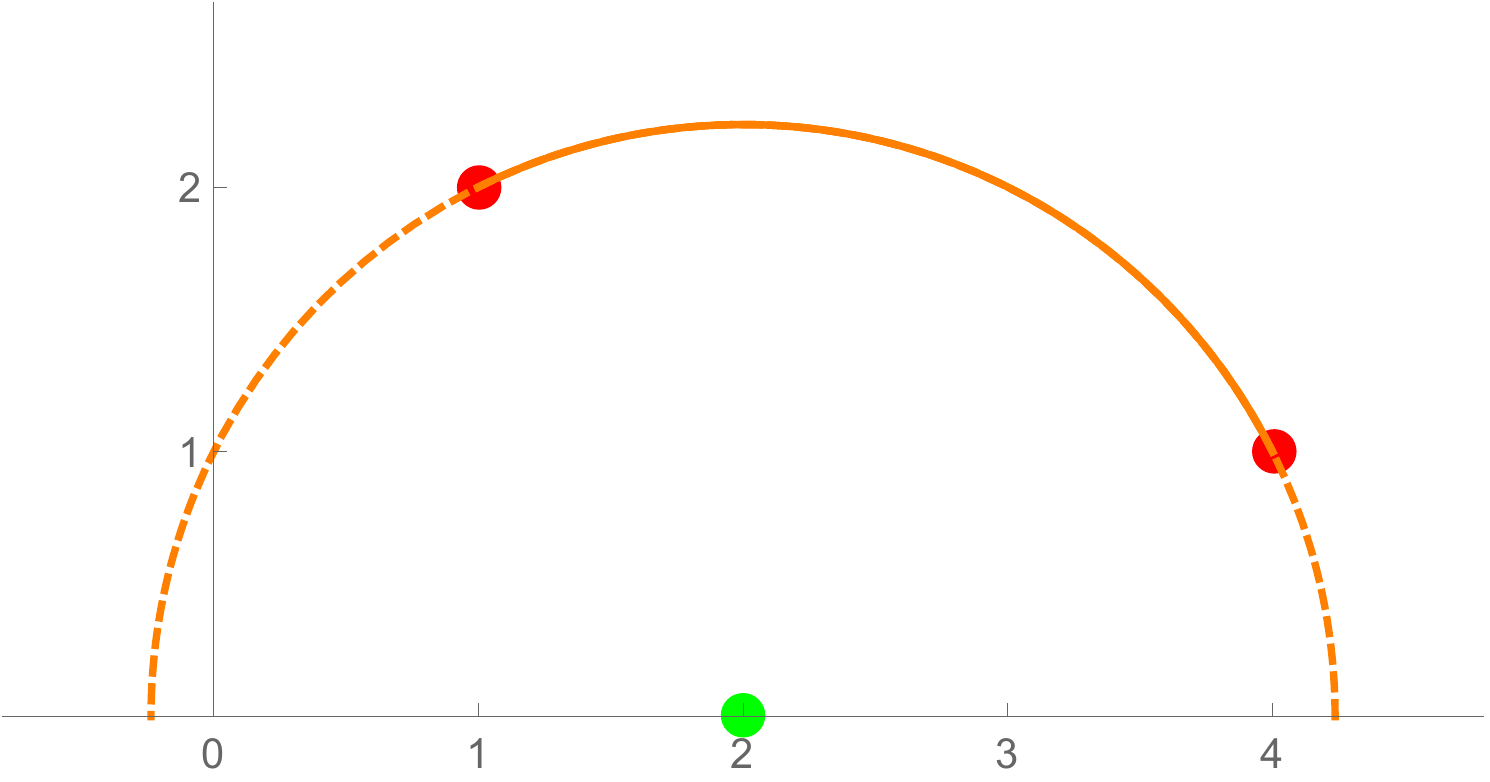}
  \setlength{\unitlength}{1cm}
\begin{picture}(0,0)
\put(-0.4,0.6){$x$}
\put(-8.7,5.1){$z$}
\put(-7.0,4.3){$P$}
\put(-1.6,2.5){$Q$}
\put(-5.9,0.6){$M$}
\put(-4.7,1.8){$\ell_M$}
\put(-4.0,4.3){$\overline{PQ}$}
\put(-1.4,0.1){$x_{\overline{PQ}+}$}
\put(-10.0,0.1){$x_{\overline{PQ}-}$}
\put(-3.9,1.8){\vector(-1,-1){1.4}}
\put(-3.9,1.8){\vector(1,1){1.4}}
\end{picture}
\end{center}
\vspace*{-0.5cm}
\caption{Setup discussed in the main text, with two bulk points $P$ and $Q$, shown in red, and the geodesic $\overline{PQ}$ that goes through them, shown in orange. The center of this geodesic is at $M$, shown in green, its radius is denoted by $\ell_M$, and its left and right endpoints are labeled $x_{\overline{PQ}\mp}$. The proper length of the arc running from $P$ to $Q$ (solid orange) defines the distance $d(P,Q)$, given explicitly in (\ref{distancepq}).
\label{pqgeodesicfig}}
\end{figure}

 We will now study how to compute the distance between two bulk points $P$ and $Q$, in terms of differential entropy. Let $P$ have coordinates $(x_P,z_P)$, and similarly for $Q$. In this subsection we choose $\lambda=x_c$, and therefore denote the families of intervals in the CFT dual to our bulk points by $\ell_P(x_c)$ and $\ell_Q(x_c)$.
For concreteness, we will take $Q$ to be to the right of $P$, $x_Q\ge x_P$.
The geodesic that connects the two points, which we will denote $\overline{PQ}$, is centered at the point $M$ on the boundary that is `equidistant' from $P$ and $Q$, in the sense that $\ell_P(x_M)=\ell_Q(x_M)$. The setup is illustrated in Fig.~\ref{pqgeodesicfig}. Explicitly,
\begin{equation}\label{xm}
x_M= \frac{x_Q+x_P}{2}+\frac{z^2_Q-z^2_P}{2(x_Q-x_P)}~,
\end{equation}
and the radius of $\overline{PQ}$ is
\begin{equation}\label{pqradius}
\ell_{M}=\frac{1}{2}\sqrt{(x_P-x_Q)^2
+2(z_P^2+z_Q^2)
+\frac{(z_P^2-z_Q^2)^2}{(x_P-x_Q)^2}}~.
\end{equation}
The distance between $P$ and $Q$ is given by the arclength along this geodesic. Using (\ref{poincaregeodesic}), this can be written as
\begin{eqnarray}\label{distancepq}
d(P,Q)
&=&\int\limits_{x_P}^{x_Q} dx_g\,\frac{L}{z_g}\sqrt{1+\left(\frac{\p z_g}{\p x_g}\right)^2}\nonumber\\
{}&=&\frac{L}{2} \left( \ln\left( \frac{x_Q-x_{\overline{PQ}-}}{x_{\overline{PQ}+}-x_Q}\right)
-\ln\left( \frac{x_P-x_{\overline{PQ}-}}{x_{\overline{PQ}+}-x_P}\right)\right) ,
\end{eqnarray}
where $z_g(x)$ is the parametrization of $\overline{PQ}$, and $x_{\overline{PQ}\pm}\equiv x_M\pm\ell_{M}$ refer to the left/right endpoints (at the AdS boundary) of the geodesic. Equation (\ref{poincaregeodesiclength}), which we used in our analysis of the boundary function $f$, is a special case of (\ref{distancepq}), with $x_P=x_c$ and $x_Q=x$.

Expression (\ref{distancepq}) is what we want to  reproduce using differential entropy. Notice that this formula computes the \emph{signed} length between the two points, and satisfies $d(P,Q)=-d(Q,P)$.
We also note in passing that from (\ref{geodesicRL1}) we know that
\begin{equation}
(x-x_{\overline{PQ}-})(x_{\overline{PQ}+}-x)=z^2~,
\end{equation}
for any point on the geodesic centered at $x_M$, and using this we can rewrite the distance between $P$ and $Q$ in the simplified form
\begin{equation}\label{distancepq2}
d(P,Q) = L \ln \left( \frac{x_{\overline{PQ}+}-x_P}{x_{\overline{PQ}+}-x_Q} \right)~.
\end{equation}

In defining $\ell_P(x_c)$ and $\ell_Q(x_c)$, if we regard each point as a vanishingly small \emph{open} curve, then we pick only one sign in (\ref{ell2}), and $x_c$ runs over the real axis once. Lengths in that case are determined using the `renormalized' differential entropy (\ref{ecal}), which includes the contribution of the boundary function (\ref{f}). {}From (\ref{eopen}), we know that for an arbitrary open curve $E=A+f(x_{c,f})-f(x_{c,i})$,
which implies that $E=f(+\infty)-f(-\infty)$ for open point-curves (which have $A=0$). Additionally, in the paragraph above (\ref{xpepsilon}) we learned that the boundary function $f(x_c)$ has a simple geometric interpretation: as seen in Fig.~\ref{ffig}, it is the distance between the edge of the curve which that geodesic is tangent to, $(x(x_c),z(x_c))$ and the regularized right endpoint of the geodesic centered at $x_c$, $(x_+^{\epsilon},\epsilon)$.

Given these results, a strategy naturally suggests itself.  To be able to extract information about the geodesic $\overline{PQ}$, we should compute the differential entropy not for the complete family $\ell_P(x_c)$, but for a truncation of
 it to the range $x_c \in(-\infty,x_{M}]$, so that the final interval in the family is precisely the one associated with $\overline{PQ}$. We will denote the truncated version by $\hat{\ell}_P(x_c)$, which we take to vanish for $x_c>x_M$.
The corresponding differential entropy will be denoted with the same symbol, $\Hat{E}_P\equiv E[\hat{\ell}_P]$. {}From what explained in the previous paragraph, we know that
\begin{equation}\label{ephat}
\Hat{E}_P=-d(\overline{PQ}+^{\epsilon},P)-f_P(-\infty)~,
\end{equation}
where $\overline{PQ}+^{\epsilon}$ refers to the regularized right endpoint, located at $x_{\overline{PQ}+}^{\epsilon}$. To obtain the distance $d(P,Q)$, we can combine this with a version of $\ell_Q(x_c)$ that is truncated to the complementary range $x_c \in (x_M, + \infty)$, so that $\overline{PQ}$ is now associated with the initial interval of the family. We will denote this truncation by $\check{\ell}_Q(x_c)$. (In this notation, $\ell_P(x_c)=\hat{\ell}_P(x_c)+\check{\ell}_P(x_c)$, and likewise for $\ell_Q$.) The corresponding differential entropy is
\begin{equation}\label{eqchek}
\Check{E}_Q=f_Q(+\infty)+d(\overline{PQ}+^{\epsilon},Q)~.
\end{equation}
Defining the combined family
\begin{equation}\label{lpq}
\ell_{PQ}(x_c)\equiv\hat{\ell}_P(x_c)+\check{\ell}_Q(x_c)~,
\end{equation}
we find that its differential entropy is
\begin{equation}\label{elpq}
E[\ell_{PQ}]=\Hat{E}_P+\Check{E}_Q=d(P,Q)-f_P(-\infty)+f_Q(+\infty)~.
\end{equation}
This serves as a formula for the desired distance between the two points in terms of entanglement entropy, save for the uncomfortable fact that the two remaining $f$ terms (which can also be expressed as distances) are both divergent: $f_P(-\infty)=\ln((x_P^2+z_P^2)/\epsilon^2)$ and $f_Q(+\infty)=\ln(4I^2/\epsilon^2)$, with $\epsilon\to 0$ and $I\to\infty$. Along the way, we have arrived in (\ref{lpq}) at precisely the same combined family $\ell_{PQ}(x_c)$ that was constructed in \cite{nutsandbolts}, in the alternative form
\begin{equation}\label{ellmin}
\ell_{PQ}(x_c)\equiv\min(\ell_P(x_c),\ell_Q(x_c))~.
\end{equation}

To avoid having to deal with the divergences arising from the boundary function (\ref{f}), we can consider the points $P,Q$ as vanishingly small \emph{closed} curves. There is then no boundary contribution, at the cost of $x_c$ covering the real axis $N\ge 2$ times, as we saw in Section \ref{closedsubsec}. For concreteness, we focus here on the case with $N=2$. In the terminology and notation of Section \ref{closedsubsec}, we can decompose this type of closed curve into one positive and one negative segment, $\ell^{(n)}_{P}(x_c)$, with $n=1,2$ and $x_c \in(-\infty,\infty)$ in each segment. Since we are dealing with a point, these two are in fact the same families of intervals/geodesics, and differ only in orientation. The positive and negative segments are obtained by choosing opposite signs in  (\ref{ell2}), so
$\ell^{(1)}_{P}=-\ell^{(2)}_{P}$ (and likewise for $Q$). For the $n=1$ portion of the curves, where $\ell_P,\ell_Q>0$, we form the same combination as in (\ref{lpq}),
\begin{equation}\label{lpq1}
\ell^{(1)}_{PQ}(x_c)\equiv\hat{\ell}^{(1)}_P(x_c)+\check{\ell}^{(1)}_Q(x_c)~.
\end{equation}
For the $k=2$ portion, where $\ell_P,\ell_Q<0$, we exchange $P$ and $Q$,
\begin{equation}\label{lpq2}
\ell^{(2)}_{PQ}(x_c)\equiv\hat{\ell}^{(2)}_Q(x_c)+\check{\ell}^{(2)}_P(x_c)~.
\end{equation}
This exchange will be seen to be necessary in the calculation that follows, and is also consistent with the definition (\ref{ellmin}) given in \cite{nutsandbolts}.

With these definitions, the differential entropy for the positive ($n=1$) portion of the combined family (\ref{lpq1}) takes the form
\begin{align}\label{eclosedlower}
 E[\ell^{(1)}_{PQ}] & = \Hat{E}_P^{(1)} + \Check{E}_Q^{(1)}
 \nonumber\\
 &=  L \Big( \int_{-\infty}^{x_M}\dfrac{dx_c}{\hat{\ell}^{(1)}_P(x_c)} \Big(1 + \partial_{x_c}\hat{\ell}^{(1)}_P(x_c)  \Big) +
  \int_{x_M}^{\infty}\dfrac{dx_c}{\check{\ell}^{(1)}_Q(x_c)} \Big(1 + \partial_{x_c}\check{\ell}^{(1)}_Q(x_c)  \Big)\Big)\nonumber\\
  &=L\Big( \ln \Big(\frac{x_{\overline{PQ}+}- x_P}{x_{\overline{PQ}+}- x_Q} \Big)
  - \ln \Big(\frac{x^2_P+z_P^2}{x^2_Q+z_Q^2} \Big)\Big)~.
\end{align}
Here we have used the fact that $\int dx_c/\ell$ for $\ell>0$ can be written in the form
\begin{equation}\label{int0}
\int \frac{dx_c}{\pm \sqrt{(x-x_c)^2+z^2}} = \ln \left[ \pm (x_c-x) + \ell \right]~,
\end{equation}
with the upper choice of sign.
To extract the second logarithm in the result (\ref{eclosedlower}), it is necessary to regularize the $x_c\to\pm\infty$ endpoint of the integrals as $x_c=\pm 1/\delta$, with $\delta\to 0$ in the end.

For the negative ($n=2$) portion, the differential entropy of the combined family (\ref{lpq2}) takes the form
 \begin{align}\label{eclosedupper}
 E[\ell^{(2)}_{PQ}] & =\Hat{E}_Q^{(2)} + \Check{E}_P^{(2)}
 \nonumber\\
 &= L \Big( \int_{-\infty}^{x_M}\dfrac{dx_c}{\hat{\ell}^{(2)}_Q(x_c)} \Big(1 + \partial_{x_c}\hat{\ell}^{(2)}_Q(x_c)  \Big) +
  \int_{x_M}^{\infty}\dfrac{dx_c}{\check{\ell}^{(2)}_P(x_c)} \Big(1 + \partial_{x_c}\check{\ell}^{(2)}_P(x_c)  \Big)\Big)\nonumber\\
  & = L \Big( \int_{-\infty}^{x_M}\dfrac{dx_c}{-\hat{\ell}^{(1)}_Q(x_c)} \Big(1 - \partial_{x_c}\hat{\ell}^{(1)}_Q(x_c)  \Big) +
  \int_{x_M}^{\infty}\dfrac{dx_c}{-\check{\ell}^{(1)}_P(x_c)} \Big(1 - \partial_{x_c}\check{\ell}^{(1)}_P(x_c)  \Big)\Big)\nonumber\\
  & =L \Big( \ln \Big(\frac{x_Q - x_{\overline{PQ}-}}{x_P - x_{\overline{PQ}-}} \Big)
  + \ln \Big(\frac{x^2_P+z_P^2}{x^2_Q+z_Q^2} \Big)\Big)~.
 \end{align}
 Here we have used (\ref{int0}) with the lower choice of sign.
Adding up (\ref{eclosedlower}) and (\ref{eclosedupper}), dividing by two and comparing with equation (\ref{distancepq}), we arrive at
\begin{equation}\label{dworks}
d(P,Q)=\frac{1}{2}E[\ell_{PQ}(x_c)]~.
\end{equation}
This has exactly the same form as the formula deduced for global AdS in \cite{nutsandbolts}. We conclude then that distances in Poincar\'e AdS can be computed using entanglement entropy in the CFT, through either (\ref{elpq}) or (\ref{dworks}).

\section{Covariant Hole-ography} \label{covariantsec}

\subsection{Arbitrary curves}\label{arbitrarysubsec}

Moving on to the time-dependent case, consider an arbitrary spacelike bulk curve
\begin{equation}\label{bulkcurve}
C^m(\lambda)=\left( t(\lambda),x(\lambda),z(\lambda)\right)~,
\end{equation}
 parametrized by some parameter $\lambda$. For each value of $\lambda$, there is a spacelike geodesic tangent to the curve, with endpoints at
 $x^{\mu}_{\pm}(\lambda)\equiv (t_{\pm}(\lambda),x_{\pm}(\lambda))$.
 If we boost to the frame, labeled $*$, where both endpoints are simultaneous (i.e., $t^{*}_+=t^{*}_-$), the geodesic will be a semicircle, centered at $x^{*\mu}_c$, the boosted version of
 \begin{equation}\label{xcmu}
 x^{\mu}_c(\lambda)\equiv \frac{1}{2}\left(x^{\mu}_+(\lambda)+x^{\mu}_-(\lambda)\right)~,
 \end{equation}
 and with radius\footnote{Notice that the symbol $\ell$ here denotes the \emph{unsigned} norm of the radius vector $\ell^{\mu}$. In the static case of the previous section, what we had defined as $\ell$ in (\ref{ell}) did carry a sign, and is precisely what will be henceforth denoted as $\ell^x$ (now that we generically have $\ell^t\neq 0$).}
 \begin{equation}\label{ellmu}
 \ell(\lambda)\equiv
 %\sigma
 \sqrt{\ell^{\mu}\ell_{\mu}}~,\qquad
 \ell^{\mu}(\lambda)\equiv\frac{1}{2}\left(x^{\mu}_+(\lambda) - x^{\mu}_-(\lambda)\right)~.
 %\qquad \sigma\equiv\mbox{sgn}(x^1_+-x^1_-)~.
 \end{equation}
 After boosting back to the original frame, the entire family of tangent geodesics can be shown to take the form \cite{hmw}
 \begin{eqnarray}\label{geodesics}
 \Gamma^m(s,\lambda) &=&\Big(\ t(\lambda) + \frac{ z(\lambda) z'(\lambda) t'(\lambda)}{ x'(\lambda)^2-t'(\lambda)^2} - \frac{t'(\lambda)  \ell(\lambda)}{\sqrt{x'(\lambda)^2-t'(\lambda)^2}} \cos s~,
 \\
& &  \quad x(\lambda)
 + \frac{ z(\lambda) z'(\lambda) x'(\lambda)}{ x'(\lambda)^2-t'(\lambda)^2} - \frac{x'(\lambda)  \ell(\lambda)}{\sqrt{x'(\lambda)^2-t'(\lambda)^2}} \cos s~,   \ell(\lambda)\sin s \Big)~,\nonumber
 \end{eqnarray}
where $0\le s\le\pi$ is a parameter running along each geodesic, and
\begin{equation}
 \label{r*}
 \ell(\lambda)= z(\lambda) \sqrt{1 + \frac{z'(\lambda)^2}{x'(\lambda)^2-t'(\lambda)^2}}~.
\end{equation}
The geodesic endpoints
$\Gamma^{\mu}({\pi\atop 0},\lambda)=x^{\mu}_{\pm}(\lambda)=(t_{\pm}, x_{\pm})$
 are given by
 \begin{eqnarray} \label{TXRL}
  t_{\pm}(\lambda) &=& t(\lambda) + \frac{ z(\lambda) z'(\lambda) t'(\lambda)}{ x'(\lambda)^2-t'(\lambda)^2} \pm \frac{t'(\lambda)  \ell(\lambda)}{\sqrt{x'(\lambda)^2-t'(\lambda)^2}}~,\\
  x_{\pm}(\lambda) &=&x(\lambda) + \frac{ z(\lambda) z'(\lambda) x'(\lambda)}{ x'(\lambda)^2-t'(\lambda)^2} \pm \frac{x'(\lambda)  \ell(\lambda)}{\sqrt{x'(\lambda)^2-t'(\lambda)^2}}~.\nonumber
 \end{eqnarray}
Using (\ref{geodesics}), we can check that, for any fixed value of $\lambda$, all points on the geodesic (given by all values of $s$) lie as expected on a boosted version of the semicircle (\ref{ell2}),
\begin{equation}\label{geodesicT}
-(t-t_c)^2+(x-x_c)^2 + z^2= \ell^2~,
\end{equation}
or equivalently, of (\ref{geodesicRL1}),
\begin{equation}\label{geodesicRL2}
(x_+-x)^\mu(x-x_-)_\mu=z^2~.
\end{equation}
Expressions (\ref{r*})-(\ref{TXRL}) can be inverted using the boundary-to-bulk relations provided in Section~4.4 of \cite{hmw}. This leads to
\begin{eqnarray}\label{covariantinvert}
  x^{\mu}(\lambda) &=& x^{\mu}_c(\lambda)-\ell^{\mu}(\lambda)\chi(\lambda)~, \\
  z(\lambda) &=& \ell(\lambda)\sqrt{1-\chi(\lambda)^2}~,\nonumber\\
  \chi(\lambda) &\equiv& \frac{\ell^x(\lambda)t_c'(\lambda)-\ell^{t}(\lambda)x_c'(\lambda)}
  {\ell^{x}(\lambda)\ell^{t\prime}(\lambda)-\ell^{t}(\lambda)\ell^{x\prime}(\lambda)}~. \nonumber
\end{eqnarray}

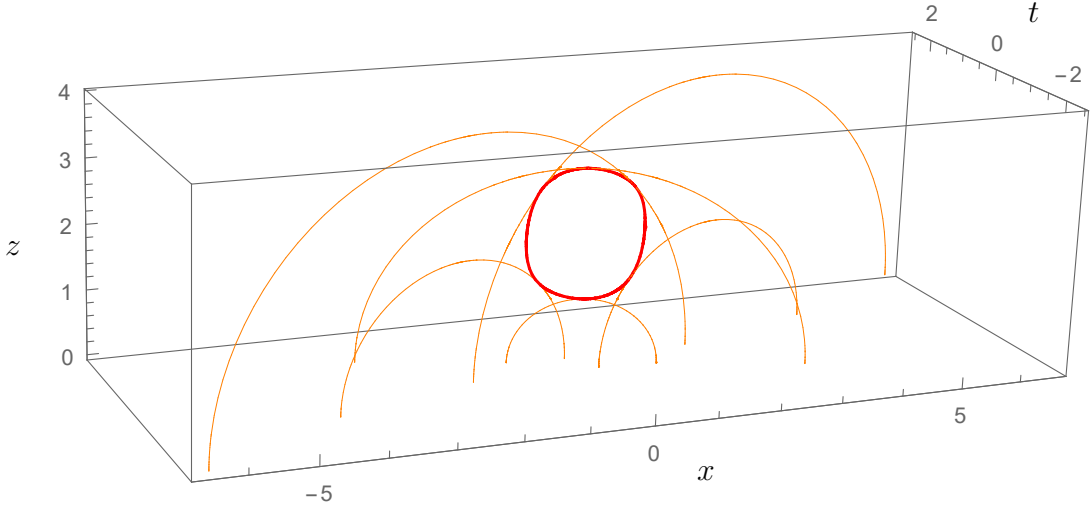
\begin{figure}[htb]
\begin{center}
  \includegraphics[angle=0,width=0.9\textwidth]{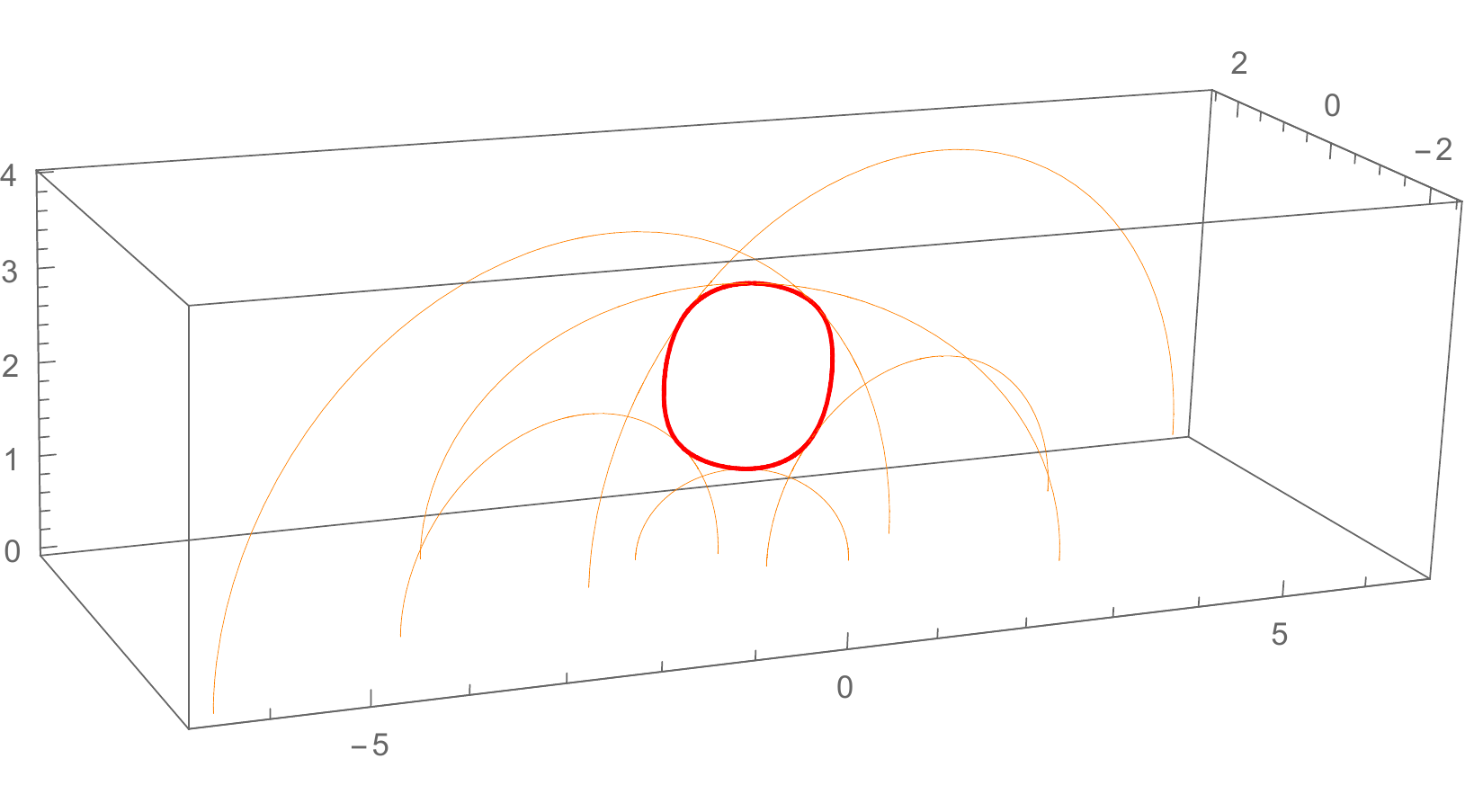}
\setlength{\unitlength}{1cm}
\begin{picture}(0,0)
\put(-5.5,0.8){$x$}
\put(-14.7,3.8){$z$}
\put(-1.1,6.9){$t$}
\end{picture}
\vspace*{-0.8cm}
\caption{\small The red curve is a plot of the undulating circle (\ref{circleundulating}), with $(\bar{t},\bar{x},\bar{z})=(0,0,2)$, $r=1$, $a=1/9$ and $n=3$. Some of the geodesics tangent to the circle are depicted in orange, for $\lambda/2\pi=2/16,4/16,6/16,10/16,12/16,14/16$.
\label{circleundulatingfig}}
\end{center}
\end{figure}

As a concrete example, we consider a circle that undulates in time, centered at $(\bar{t},\bar{x},\bar{z})$, with radius $r$ and undulation amplitude $a$:
\begin{eqnarray}\label{circleundulating}
t(\lambda)&=&\bar{t}-a\cos n\lambda~,\\
x(\lambda)&=&\bar{x}-r\cos\lambda~,\nonumber\\
z(\lambda)&=&\bar{z}-r\sin\lambda~.\nonumber
\end{eqnarray}
where $n\in\mathbb{Z}$. For the curve to be spacelike everywhere, we must demand that
\begin{equation}
x'(\lambda)^2+z'(\lambda)^2-t'(\lambda)^2=  r^2-a^2 n^2\sin^2 n \lambda>0~.
\end{equation}
This constraint is satisfied for all $\lambda\in[0,2\pi)$ as long as
\begin{equation}\label{constrainttime}
|a|<\frac{r}{|n|}~.
\end{equation}
A particular example satisfying (\ref{constrainttime}) is shown in Fig.~\ref{circleundulatingfig}.

 Going back to the general analysis, the entanglement entropy of each interval in the CFT can be computed in the boosted frame, where it is given by (\ref{s}), and then carried over to the original coordinates,
 \begin{equation}
 S (x^{\mu}_{-},x^{\mu}_{+}) = 2L \ln \left( \frac{\big|x^{\mu}_{+} - x^{\mu}_{-}\big|}{\epsilon} \right)
 = L \ln \left( \frac{(x_{+} - x_{-})^2 - (t_{+}-t_{-})^2}{\epsilon^2} \right)
 = L \ln \left( \frac{4\ell^2}{\epsilon^2} \right)~.
 \end{equation}
The differential entropy (\ref{e}) then takes the form
\begin{eqnarray}\label{Ecovariant}
E &=& L \int_{\lambda_i}^{\lambda_f} d\lambda\, \frac{\ell\cdot x'_+}{\ell^2} \nonumber\\
&=&L\int_{\lambda_i}^{\lambda_f} d\lambda\,\frac{1}{\ell^2}\left(\ell\cdot\ell'+\ell\cdot x'_c\right) \nonumber\\
&=&\frac{L}{2} \ln \left.\frac{4\ell^2}{\epsilon^2}\right|_{\lambda_i}^{\lambda_f}
+L \int_{\lambda_i}^{\lambda_f}d\lambda\,\frac{\ell\cdot x'_c}{\ell^2}~,
\end{eqnarray}
which correctly reproduces (\ref{e2}) in the case of constant time.

The term that remains within the integral in (\ref{Ecovariant})
can be processed by means of expressions (\ref{r*})-(\ref{TXRL}), to obtain
\begin{eqnarray}\label{ecovariant2}
E &=&\frac{L}{2} \ln \left.\frac{4\ell^2}{\epsilon^2}\right|_{\lambda_i}^{\lambda_f}
+ L\int_{\lambda_i}^{\lambda_f} d\lambda
\left(\frac{1}{z(\lambda)} \sqrt{x'(\lambda)^2 + z'(\lambda)^2 -t'(\lambda)^2}\right.   \\
{}&{}&
+\left. \frac{z''(\lambda)}{\sqrt{x'(\lambda)^2 + z'(\lambda)^2 -t'(\lambda)^2}}
+ \frac{z'(\lambda)}{x'(\lambda)^2-t'(\lambda)^2} \frac{t'(\lambda) t''(\lambda) - x'(\lambda) x''(\lambda)}{\sqrt{x'(\lambda)^2 + z'(\lambda)^2 -t'(\lambda)^2}} \right)~.\nonumber
\end{eqnarray}
 At the end of the first line we recognize the term that yields the length $A$ of the curve. The terms in the second line are the $\lambda$-derivative of
\begin{equation}\label{hmwtotalderivative}
L\sinh^{-1}\left(\dfrac{z'(\lambda)}{\sqrt{x'(\lambda)^{2}-t'(\lambda)^{2}}}\right)~.
\end{equation}
For closed bulk curves, the total-derivative terms drop out, and we find that $A=E$, as expected. For open curves, we arrive instead at a generalization of (\ref{eopen})-(\ref{f}), $A=E-f(\lambda_f)+f(\lambda_i)$, where now
\begin{equation}\label{ftime}
f(\lambda)=\frac{L}{2} \ln\frac{4\ell^2}{\epsilon^2}+L\sinh^{-1}\left(\dfrac{z'(\lambda)}{\sqrt{x'(\lambda)^{2}-t'(\lambda)^{2}}}\right)~.
\end{equation}

The boundary function (\ref{ftime}) can be easily seen to have the same geometric interpretation as in the case of constant time. In particular, its second term matches the arclength between $x^{\mu}_c$ and $x^{\mu}$ along the geodesic tangent to the curve at the given point $\lambda$,
allowing us to rewrite
 \begin{equation}\label{fcovariant}
 f(\lambda) = \frac{L}{2} \ln\frac{4\ell^2}{\epsilon^2} + \frac{L}{2} \ln \left( \frac{x_+  -x}{x - x_-} \right)~.
 \end{equation}
This can equivalently be expressed as a contribution to the $\lambda$-integrand. Our final expression for the `renormalized' differential entropy is then found to be
\begin{eqnarray}\label{ecalcovariant}
\mathcal{E} &\equiv& E - f(\lambda_f)+f(\lambda_i) \\
{}&=& E
- L \int_{\lambda_i}^{\lambda_f} \!d\lambda\, \Big( \frac{1}{2} \frac{\partial_\lambda \ell^2}{\ell^2}
+ \frac{\ell_x(x' - x'_c) + (x_c - x) \ell'_x  }{\ell_x^2-(x-x_c)^2}\Big)~.\nonumber
\end{eqnarray}
 Using  (\ref{r*})-(\ref{TXRL}), we can verify that indeed $A=\mathcal{E}$.
Expression (\ref{ecalcovariant}) can be written purely in terms of CFT data by means of (\ref{covariantinvert}).

\subsection{A challenge to hole-ography in Poincar\'e AdS} \label{challengesubsec}

As explained in the Introduction, the fact that Poincar\'e coordinates $(t,x,z)$ defined in (\ref{globaltopoincare}) cover only a wedge within the full AdS$_3$ spacetime (\ref{globalmetric}) implies that, when considering a generic $t$-dependent spacelike bulk curve (\ref{bulkcurve}), some of its tangent geodesics will \emph{not} be fully contained within the Poincar\'e wedge. See Fig.~\ref{3dfig}. This presents a challenge to hole-ographic reconstruction, because when it happens, we are unable to encode the length of the curve into CFT data using differential entropy.

To see exactly where the problem resides, recall that, given any two points $x^{\mu}_{-}$ and $x^{\mu}_{+}$ on the boundary of AdS that are spacelike separated, there does exist a bulk geodesic that connects them, and it has the shape of a boosted semicircle, Eq.~(\ref{geodesicT}). The projection of this geodesic onto the AdS boundary is simply a straight line connecting the two points. If the geodesic happens to be tangent to some bulk curve, then clearly the boundary projection of the vector tangent to the curve will lie on the same straight line, and will therefore be spacelike. It follows from this that \emph{at a given point $\lambda$, a bulk curve in Poincar\'e AdS has a tangent geodesic that reaches the boundary if and only if the boundary projection of its tangent vector at that point is spacelike,}
\begin{equation}\label{reconstructibility}
-t'(\lambda)^2+x'(\lambda)^2>0~.
\end{equation}
Indeed, we see explicitly in (\ref{TXRL}) that the endpoint positions $x^{\mu}_{\pm}$  are real only when this condition is satisfied. This, then, is our criterion for reconstructibility of the bulk curve. Importantly, it differs from the condition for the bulk curve itself to be spacelike, $ -t'^2+x'^2+z'^2>0$, and can therefore be violated.

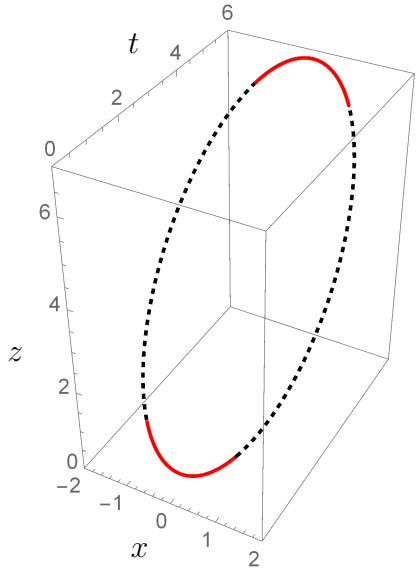
\begin{figure}[hbt]
\begin{center}
  \includegraphics[width=5cm]{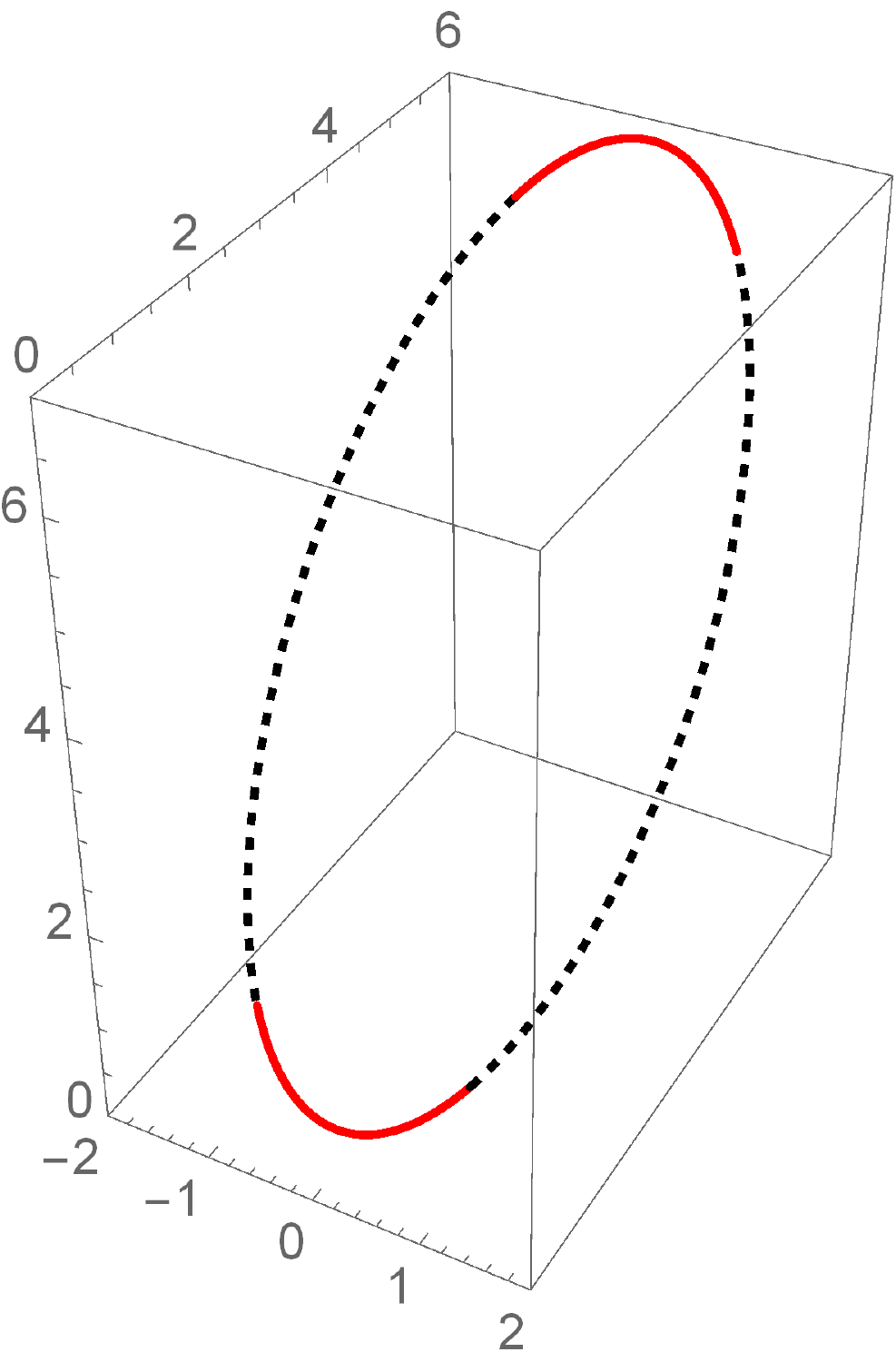}
  \setlength{\unitlength}{1cm}
\begin{picture}(0,0)
\put(-4.2,0.2){ $x$}
\put(-5.7,2.8){$z$}
\put(-4.1,6.9){$t$}
\end{picture}
\end{center}
\vspace*{-0.8cm}
\caption{An example of a closed spacelike curve at varying Poincar\'e time $t$, given by (\ref{circletau}) with $\tau=\pi/5,\varrho=\pi/4$. The top and bottom, shown in solid red, have tangent geodesics lying fully within the Poincar\'e wedge. This is not true for the segments on the sides, shown in dashed black, which violate the condition (\ref{reconstructibility}) and are therefore nonreconstructible. As described in the main text, the tilted oval seen here is the Poincar\'e counterpart of the global circle shown in the right image of Fig.~\ref{3dfig}.
\label{circletaufig}}
\end{figure}

As a concrete example, consider the closed curve that is obtained by mapping to Poincar\'e AdS the same circle at fixed global time that we discussed in Section~\ref{closedsubsec}, $\varrho(\theta)=\mbox{constant}$ (recall that $R\equiv L\tan\varrho$),  but now displaced to $\tau\neq 0$. Using (\ref{globaltopoincare}) and choosing $\lambda=\theta$, this is
\begin{eqnarray}\label{circletau}
t(\lambda)&=&\frac{L\sin\tau}{\cos\tau+\sin\varrho\cos\lambda}\nonumber\quad,\\
x(\lambda)&=&\frac{L\sin\lambda\,\sin\varrho}{\cos\tau+\sin\varrho\cos\lambda}\quad,\\
z(\lambda)&=&\frac{L\cos\varrho}{\cos\tau+\sin\varrho\cos\lambda}\nonumber\quad.
\end{eqnarray}
The two constants $\tau,\varrho$ are parameters that specify our choice of curve. For the curve to be fully contained within the Poincar\'e wedge, we must have
$|\tau|+\varrho<\pi/2$.
Notice from (\ref{circletau}) that $t\propto z$. As shown in Fig.~\ref{circletaufig}, this curve is an oval
tilted in the $t$ direction.

In the global description it is evident that the entire curve (\ref{circletau}) is spacelike. In the Poincar\'e description, there is a region of it that violates the reconstructibility condition (\ref{reconstructibility}). The edge of this region is located at the points where
$x'(\lambda)^{2}-t'(\lambda)^{2}=0$. Solving this equation, we find the four points
\begin{eqnarray}\label{circlenonreconstrutible}
\lambda_{1}&=&\arctan\left[\sin(\tau-\varrho),\cos(\tau-\varrho)\right]~,\nonumber\\
\lambda_{2}&=&\arctan\left[-\sin(\tau+\varrho),\cos(\tau+\varrho)\right]~,\nonumber\\
\lambda_{3}&=&\arctan\left[-\sin(\tau+\varrho),-\cos(\tau+\varrho)\right]~,\\
\lambda_{4}&=&\arctan\left[\sin(\tau-\varrho),-\cos(\tau-\varrho)\right]~.\nonumber
\end{eqnarray}
The notation here picks out a quadrant for the inverse tangent function: $\arctan[s,c]$ means an angle whose sine and cosine are respectively $s$ and $c$.
We thus find two nonreconstructible segments, $(\lambda_{1},\lambda_{2})$ and $(\lambda_{3},\lambda_{4})$, which as shown in Fig.~\ref{circletaufig} are located on the sides of the circle. Since the top of the curve is closest to the horizon, it might seem surprising that it is reconstructible, but the geodesics tangent to points in that region do fit inside the Poincar\'e wedge. This can also be verified directly for the circle in the original global coordinates, shown in Fig.~\ref{3dfig}.

\subsection{Resolution via `null vector alignment'} \label{nullsubsec}

 In the previous subsection we have seen that there are spacelike bulk curves in Poincar\'e AdS with segments that are nonreconstructible, in the sense that they violate  condition (\ref{reconstructibility}), and are therefore tangent to geodesics that are not fully contained within the Poincar\'e wedge. Such geodesics are not associated with entanglement entropy in the dual CFT defined on Minkowski spacetime, so we are left wondering whether there is some way to encode such bulk curves in the field theory language. For this we must find some way to select a family of intervals in the CFT, whose entropies manage to capture the information about the nonreconstructible segments in spite of not being associated to geodesics that are tangent to them.

 Fortunately, a prescription that gives us the necessary margin for maneuvering in this direction was discovered in \cite{hmw}. The authors of that work showed that the standard formula for differential entropy, Eq.~(\ref{e}), correctly computes the length $A$ of a bulk curve even if we  choose a non-standard family of intervals/geodesics, obtained by reorienting the tangent vector to the curve, $u^m\equiv(t',x',z')$, according to $u\to U\equiv u+n$, where $n$ is a null vector orthogonal to $u$, i.e.,
 \begin{equation}\label{n}
 n\cdot n=0~,\qquad n\cdot u=0~.
 \end{equation}
 As long as these two conditions are satisfied, $n$ can be any differentiable function of $\lambda$.

 If from each point $x^m(\lambda)$ on the curve we shoot a geodesic along $U^m(\lambda)$ instead of $u^m(\lambda)$, we select a family of intervals in the CFT whose endpoints are given by (\ref{TXRL}) with the replacement $u\to U$. Running through the steps leading to (\ref{ecovariant2}), it is straightforward to arrive at
 \begin{eqnarray}\label{en}
E&=&\frac{L}{2}\left.\ln \frac{4\ell^2}{\epsilon^2}\right|^{\lambda_f}_{\lambda_i}
+L\int d\lambda \left(\frac{1}{z(\lambda)}\sqrt{U_{x}^{2}(\lambda)+U_{z}^{2}(\lambda)-U_{t}^{2}(\lambda)}\right. \nonumber\\
{}&{}&\qquad\qquad\qquad\qquad\qquad+\frac{z''(\lambda)+n_{z}'(\lambda)}{\sqrt{U_{x}^{2}(\lambda)+U_{z}^{2}(\lambda)-U_{t}^{2}(\lambda)}} \\
{}&{}&
\mkern-36mu +\left.\frac{z'(\lambda)+n_{z}(\lambda)}{U_{x}^{2}(\lambda)-U_{t}^{2}(\lambda)}
\frac{t'(\lambda)t''(\lambda)-x'(\lambda)x''(\lambda)
+n_{z}(\lambda)z''(\lambda)+z'(\lambda)n'_{z}(\lambda)+n_{z}(\lambda)n'_{z}(\lambda)}{\sqrt{U_{x}^{2}(\lambda)
+U_{z}^{2}(\lambda)-U_{t}^{2}(\lambda)}}\right)~,\nonumber
\end{eqnarray}
where just like before $\ell^2\equiv\ell_x^2-\ell_t^2$, but now the components of $\ell^{\mu}$ depend on our choice of $n(\lambda)$.
The terms in the second and third line are the $\lambda$-derivative of
 \begin{equation}
\sinh^{-1}\left(\frac{U_{z}(\lambda)}{\sqrt{U_{x}^{2}(\lambda)-U_{t}^{2}(\lambda)}}\right)\quad,
\end{equation}
and together with the logarithm can therefore be ignored for the type of curves considered in \cite{hmw}, which are infinitely extended and have periodic boundary conditions at $x\to\pm\infty$. In that case, then, all that is left is the final term in the top line of (\ref{en}). Since
 conditions (\ref{n}) guarantee that $U\cdot U=u\cdot u$, we recognize this term as the length of the bulk curve, thereby verifying that $A=E$, as claimed by \cite{hmw}.

The authors of \cite{hmw} referred to the replacement $u\to U$ as `null vector alignment', as opposed to the standard `tangent vector alignment'. They employed the freedom afforded by the choice of $n(\lambda)$ to show that an arbitrary differentiable family of spacelike intervals $(x_-^{\mu}(\lambda),x_+^{\mu}(\lambda))$ in the CFT can always be used to construct at least one (and usually two) bulk curve(s), whose differential entropy agrees with its length. This boundary-to-bulk construction runs in the opposite direction to the  bulk-to-boundary procedure we had discussed heretofore, where one starts with a bulk curve and uses its tangent geodesics to obtain a family of intervals in the CFT. For curves at constant time, there is no essential difference between these two directions, but in the covariant case it is in general necessary to employ null vector alignment when proceeding in the boundary-to-bulk direction.

The result $E=A$ for arbitrary $n(\lambda)$ evidently extends immediately from \cite{hmw} to the arbitrary closed curves considered in this paper. In the case of open curves, it is generalized to $A=E-f(\lambda_f)+f(\lambda_i)\equiv\mathcal{E}$, where the $n$-dependent boundary function is given by
\begin{equation}\label{fn}
f(\lambda)=\frac{L}{2} \ln\frac{4\ell^2}{\epsilon^2}+L\sinh^{-1}\left(\dfrac{U_z(\lambda)}{\sqrt{U^2_x(\lambda)-U^2_t(\lambda)}}\right)~.
\end{equation}

The important takeaway from all of this is that, from the global AdS perspective, there are in fact infinitely many choices for the family of CFT intervals that reconstructs a given bulk curve. More specifically, there is one choice for each function $n(\lambda)$, and since this null vector is subject to the two constraints (\ref{n}), on AdS$_3$ this amounts to the freedom of choosing one of its components ($d-1$ components on AdS$_{d+1}$). In Poincar\'e coordinates, given any choice of $n^z$ we can solve (\ref{n}) to find the remaining components of $n$,
\begin{eqnarray}\label{nfromnz}
n^t&=&\frac{n^z u^t u^z\pm|n^z u^x|\sqrt{-u_t^2+u_x^2+u_z^2}}{u_t^2-u_x^2}~,\\
n^x&=&-\frac{n^z u^z}{u^x}+\frac{n^z u_t^2 u^z\pm u^t |n^z u^x|\sqrt{-u_t^2+u_x^2+u_z^2}}{u^x(u_t^2-u_x^2)}~,\nonumber
\end{eqnarray}
where the two choices of sign are correlated. Equivalently, we can choose $n^t$ arbitrarily and from it determine
\begin{eqnarray}\label{nfromnt}
n^x&=&\frac{n^t u^t u^x\pm|n^t u^z|\sqrt{-u_t^2+u_x^2+u_z^2}}{u_x^2+u_z^2}~,\\
n^z&=&-\frac{n^t u^t}{u^z}-\frac{n^t u_x^2 u^t \pm u^x |n^t u^z|\sqrt{-u_t^2+u_x^2+u_z^2}}{u^z(u_x^2+u_z^2)}~.\nonumber
\end{eqnarray}

We would like to establish whether this freedom allows us to address the problem encountered in the previous subsection.
Consider a spacelike ($u\cdot u > 0 $) bulk curve that has a region where (\ref{reconstructibility}) is violated, i.e.,
$-u_{t}^{2}+u_{x}^{2}=u\cdot u-u_{z}^{2}\le 0$. In this region, tangent vector alignment yields geodesics that are not fully contained within the Poincar\'e wedge. Invoking null vector alignment instead, we can use geodesics along $U(\lambda)=u(\lambda)+n(\lambda)$. To achieve reconstructibility with these new geodesics, we must demand that
\begin{equation}\label{nreconstructibility}
-U_{t}^{2}+U_{x}^2>0\quad\leftrightarrow\quad (u^z+n^z)^2<(z^2/L^2)u\cdot u\quad.
\end{equation}
The inequality on the right follows from the fact that $U\cdot U=u\cdot u$.
For each $\lambda$, (\ref{nreconstructibility}) is a single inequality imposed on the completely free component $n^z$, so there are infinitely many solutions.
Two concrete examples are:
\begin{itemize}
\item $U^z=0$. Plugging $n^{z}=-u^{z}$ into (\ref{nfromnz}), we find a specific choice of $n^m(\lambda)$ which evidently satisfies the right inequality in (\ref{nreconstructibility}).
In this case, all geodesics in the family touch the bulk curve at their point furthest away from the boundary.

\item $U^t=0$. Taking $n^{t}=-u^{t}$ and using (\ref{nfromnt}), we find another choice of $n^m(\lambda)$ that evidently satisfies the left inequality in (\ref{nreconstructibility}). In this case, we only use geodesics at constant time, even though the value of $t$ is in general different for each geodesic.
\end{itemize}

\begin{figure}[hbt]
\begin{center}
  \includegraphics[width=12cm]{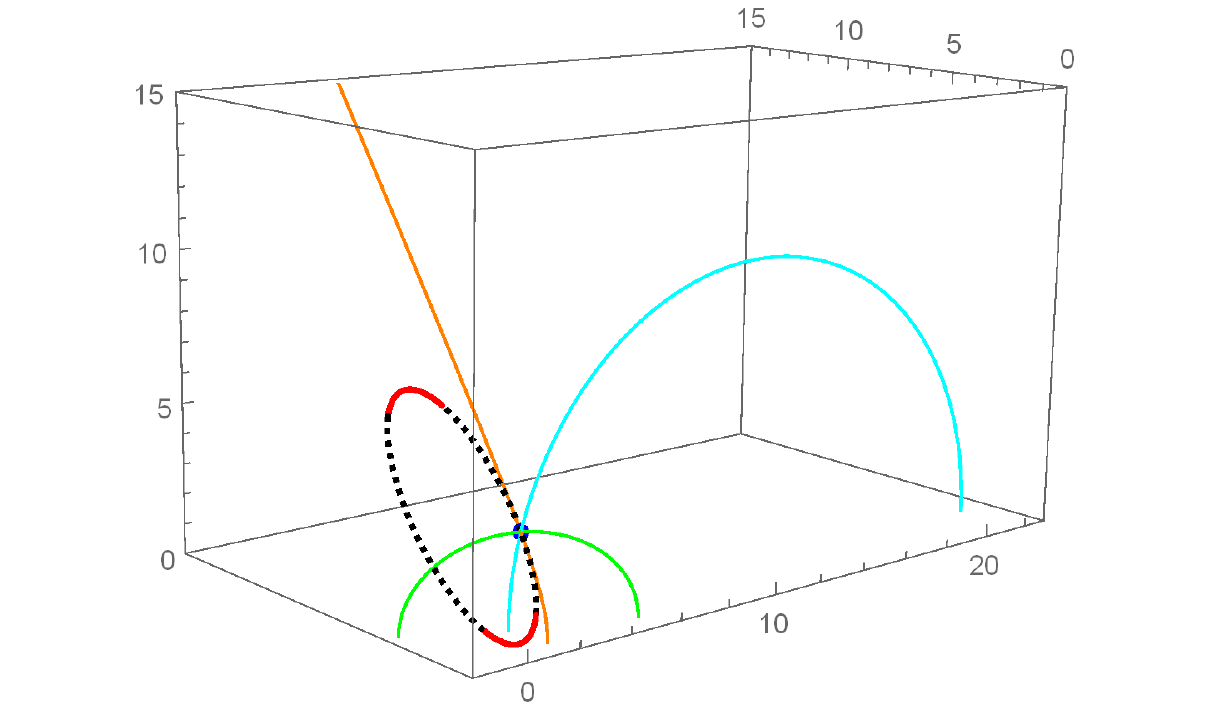}
\setlength{\unitlength}{1cm}
\begin{picture}(0,0)
\put(-9.9,0.3){\vector(0,1){0.5}}
\put(-9.9,0.3){\vector(3,1){0.4}}
\put(-9.9,0.3){\vector(-2,1){0.4}}
\put(-10.5,0.5){$t$}
\put(-10.0,0.9){$z$}
\put(-9.5,0.5){$x$}
\end{picture}
\end{center}
\vspace*{-0.8cm}
\caption{We see here the same tilted oval (\ref{circletau}) as in Fig.~\ref{circletaufig}, from a greater distance and a different perspective. The point $\lambda=4\pi/5$ in the nonreconstructible region is marked, and the geodesic tangent to the oval at that point is shown in orange. One of its endpoints exits the Poincar\'e wedge through the horizon, $z\to\infty$, so it cannot be associated with entanglement entropy in the CFT. Nonetheless, null vector alignment $u\to U=u+n$ allows us to reorient this geodesic so that both of its endpoints land on the boundary of Poincar\'e AdS, $z=0$. Among the infinitely many different ways in which this can be done, we illustrate the two examples described in the main text: the green geodesic has $U^z=0$, and the cyan geodesic has $U^t=0$. With either of these choices, we are able to translate the given point into CFT language.
\label{circlereconstructfig}}
\end{figure}
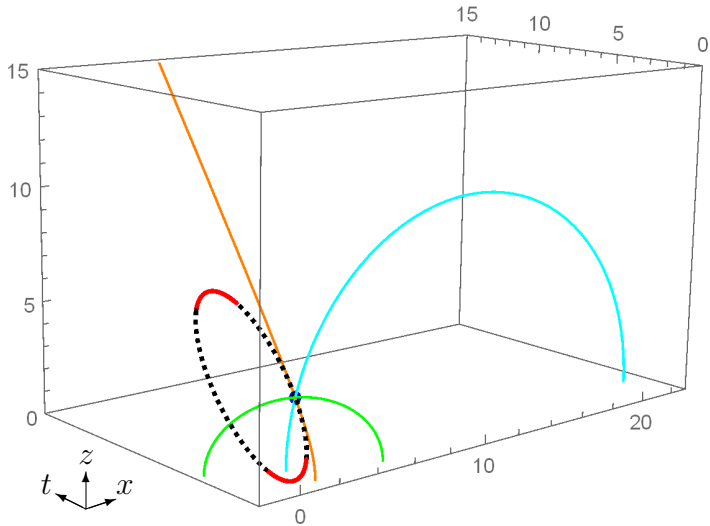

To understand how this works in practice, let us go back to the example of the tilted oval that we had in (\ref{circletau}). In global coordinates this is simply a circle of radius $R=L\tan\varrho$ at fixed $\tau$, so its total length is $A=2\pi R$. The points where $-u_t^2+u_x^2$ changes sign are the $\lambda_i$ defined in (\ref{circlenonreconstrutible}), and split the oval into four segments, as shown  in Fig.~\ref{circletaufig}. The two segments $(\lambda_{1},\lambda_{2})$ and $(\lambda_{3},\lambda_{4})$, shown in dashed black in the figure, are nonreconstructible with tangent vector alignment. We now know that they can be described using null vector alignment instead. In Figure~\ref{circlereconstructfig} we see how this is possible: for a point in the nonreconstructible region, the addition of a null vector allows us to reorient the geodesic touching the curve in such a way that both of its endpoints reach the boundary of the Poincar\'e wedge.
Notice that, if we employ some $n(\lambda)\neq 0$ only for the two nonreconstructible segments, then even though our entire curve is closed, the contribution of the boundary function (\ref{fn}) will generally not cancel between adjacent segments, because it depends on $n$. So $E_1+E_2+E_3+E_4\neq A$ in general, but what we have shown  for arbitrary curves implies that
$\mathcal{E}_1+\mathcal{E}_2+\mathcal{E}_3+\mathcal{E}_4= A$. Alternatively, we can use null vector alignment for the entire oval, with some choice of $n(\lambda)$ that is smooth across the points $\lambda_i$ (e.g., $U^z=0$ or $U^t=0$). In this case the boundary function does drop out, and we have $E=A$, independently of the choice of $n(\lambda)$.

\begin{figure}[hbt]
\begin{center}
  \includegraphics[width=6cm]{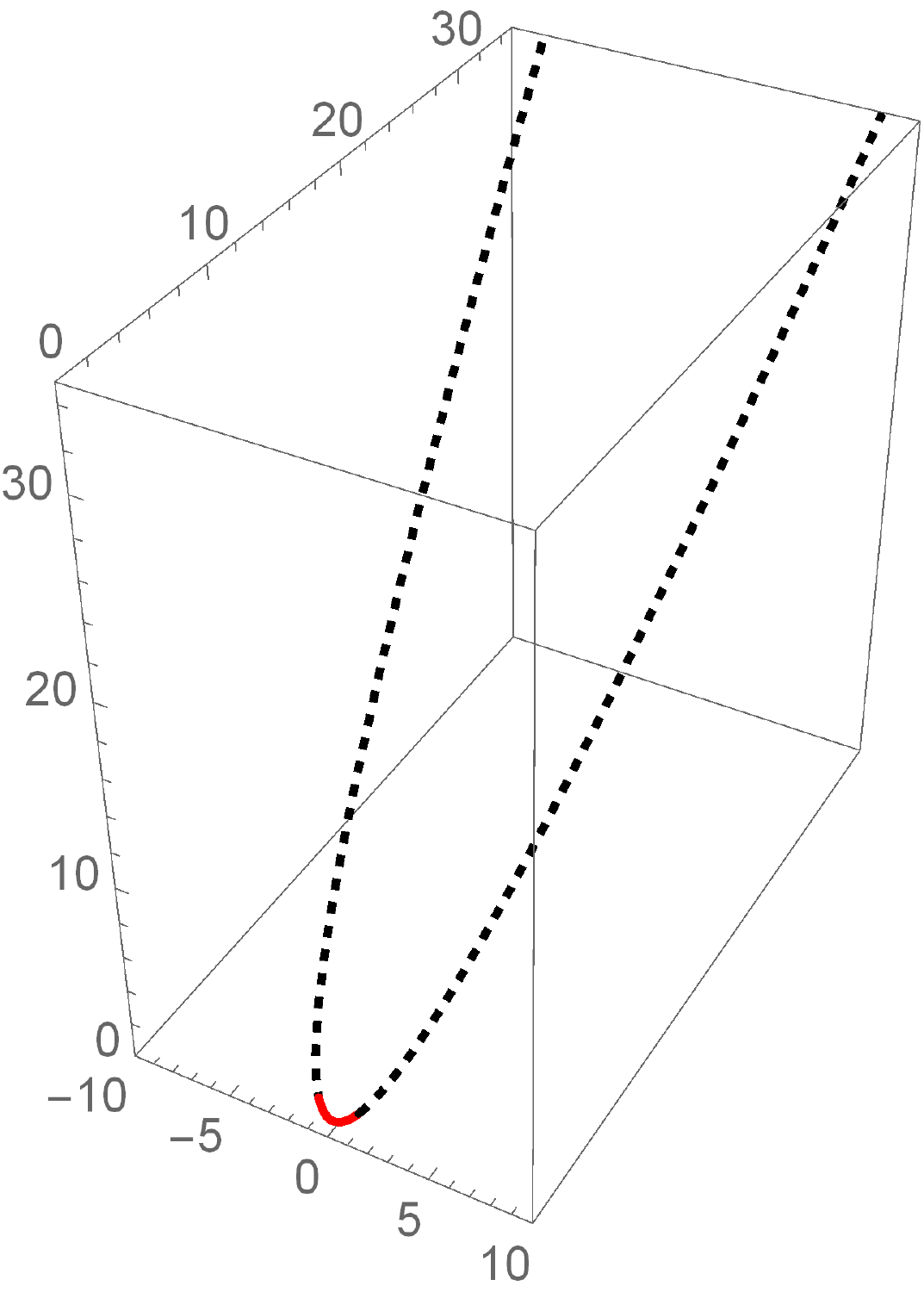}
  \includegraphics[width=7cm]{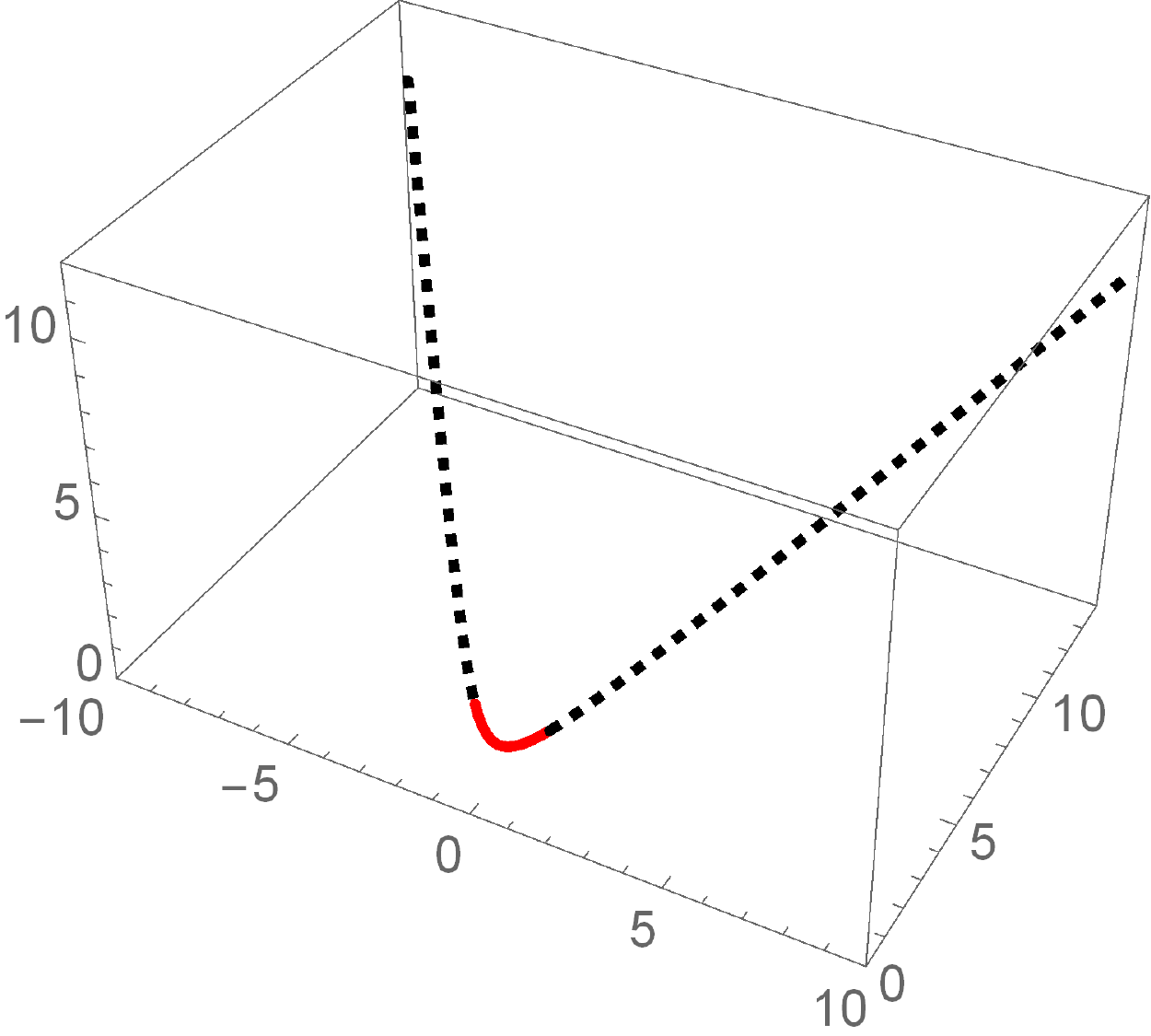}
\setlength{\unitlength}{1cm}
\begin{picture}(0,0)
%\put(-4.2,0.2){ $x$}
%\put(-5.7,2.8){$z$}
%\put(1.1,0.9){$t$}
\put(-8.1,1.1){\vector(-1,4){0.03}}
\put(-8,0.6){\line(-1,6){0.08}}
\put(-8,0.6){\vector(1,1){0.4}}
\put(-8.0,0.6){\vector(2,-1){0.4}}
\put(-7.7,0.2){ $x$}
\put(-8.2,1.3){$z$}
\put(-7.5,1.0){$t$}
\end{picture}
\end{center}
\vspace*{-0.8cm}
\caption{Open spacelike curves at varying Poincar\'e time $t$, with both endpoints reaching the Poincar\'e horizon. The curve on the left is given by (\ref{circletau}) with $\tau=\pi/4,\varrho=\pi/4$. In global coordinates it corresponds to a circle that barely fits within the Poincar\'e wedge, and touches the horizon at a single point. The curve on the right has $\tau=\pi/2,\varrho=\pi/4$, and is a global circle that partly lies behind the Poincar\'e horizon. For both curves the bottom segment, shown in solid red, is reconstructible with tangent vector alignment, but the sides, shown in dashed black, violate condition (\ref{reconstructibility}) and require null vector alignment to be reconstructed.
\label{circlecrosshorizonfig}}
\end{figure}
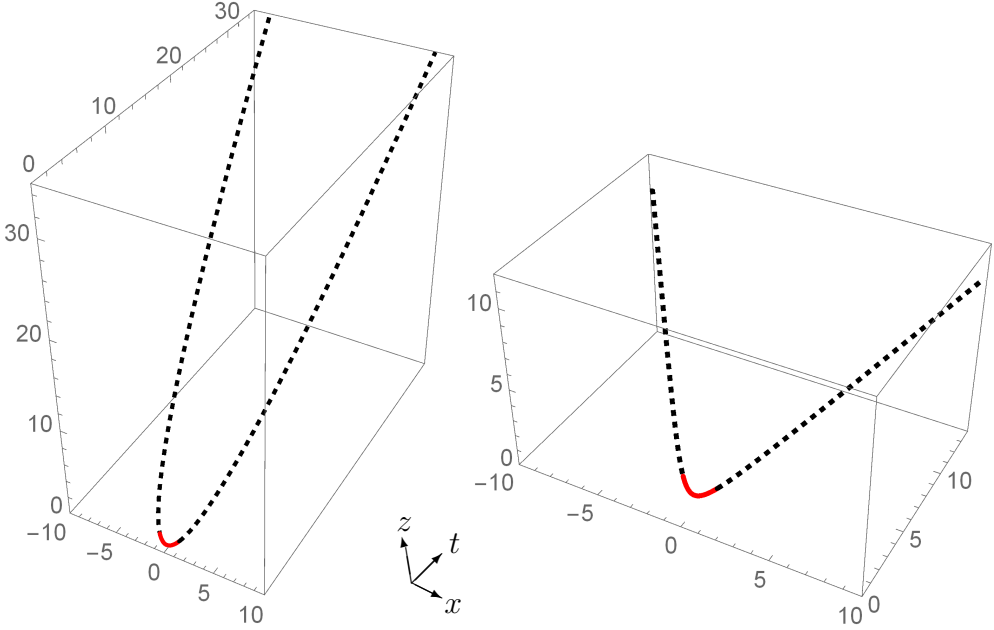

It is also natural to wonder what happens in the case of a curve that is closed in global coordinates but is not fully contained within the Poincar\'e wedge. In Poincar\'e coordinates this translates into an open curve with both of its endpoints at the Poincar\'e horizon (at $t\to\pm\infty$, $x\to\pm\infty$). One question is whether we might be able to reconstruct the portion of the curve beyond the horizon, using null vector alignment to shoot geodesics into the Poincar\'e wedge. This is quickly seen to be impossible, because on AdS there is a unique geodesic associated with each pair of boundary points, and all geodesics with both endpoints on the boundary of the Poincar\'e patch are known to lie entirely within the patch. There is no option then but to treat this case as an open curve. We know that any nonreconstructible segments of it will be accessible via null vector alignment.
%Depending on the shape of the curve and the choice of $n(\lambda)$, it might be the case that the boundary function (\ref{fn}) diverges on the horizon.
Two examples of this type of curve are shown in Fig.~\ref{circlecrosshorizonfig}.

\subsection{Points}\label{covariantpointsubsec}

Now that we know how to encode an arbitrary closed or open bulk curve, we can again reason as in \cite{nutsandbolts} and shrink these curves down to arbitrary bulk points.  Each resulting `point-curve' will be associated with a family of intervals/geodesics with endpoints $x^{\mu}_{\pm}(\lambda)$, or equivalently, with center and radius vectors $x^{\mu}_c(\lambda)$ and $\ell^{\mu}(\lambda)$. If the curve is open, then as in Section~\ref{pointsubsec} we must demand that it is vertical at its beginning and end, in order for the family of intervals not to disappear in the point limit. If the curve is closed, as in Section~\ref{closedsubsec} we will obtain a family that crosses from $x\to\infty$ to $x\to-\infty$ some number $N\ge 2$ of times before smoothly coming back to itself.

Importantly, there are infinitely many different families that describe the same bulk point, because there are infinitely many choices for the shape of the curve that we shrink to any given point. For any such choice, and for any choice of $n^{\mu}(\lambda)$ if we decide to use null vector alignment as in the previous subsection, after reducing to zero size we will simply get some family of geodesics that pass through the desired bulk point. The family described in Section~\ref{pointsubsec}, where all intervals/geodesics are on the same time slice as the bulk point, is just one particular example. Evidently we could also use geodesics on any boosted time slice. More generally, we get one family of intervals/geodesics for each choice of curve on the boundary of AdS that is spacelike separated from the given bulk point, by taking the center vectors $x^{\mu}_c(\lambda)$ (or the right or left endpoint) of the intervals to lie on the chosen boundary curve.

 Just like in the constant-time case, when our curve shrinks down to a point, the generic equations (\ref{TXRL}) degenerate and are not directly useful, because all derivatives vanish. Nonetheless, it is easy to work out the required description.
  Consider a bulk point $P$, whose coordinates are denoted $x^m_P\equiv(t_P,x_P,z_P)$. According to (\ref{geodesicT}), the center and radius vectors of each geodesic passing through $P$
satisfy
\begin{equation}\label{geodesicT2}
-(t_P-t_c)^2+(x_P-x_c)^2 + z_P^2= -\ell_t^2+\ell_x^2~.
\end{equation}
 Since this geodesic is just a boosted semicircle, there exists a frame, denoted *, where the entire geodesic lies at constant time, implying in particular that the same boost that sets $t_P^*-t_c^*=0$ also sets $\ell_t^*=0$. {}This requires
\begin{equation}\label{boostrelation}
\frac{\ell_t}{\ell_x} = \frac{t_P - t_c}{x_P-x_c}~.
\end{equation}
{}From (\ref{geodesicT2}) and (\ref{boostrelation}) we can deduce the explicit expression that determines our family of intervals for each choice of center curve $x^{\mu}_c(\lambda)$,
\begin{equation}\label{lmufromxcmu}
\ell^{\mu}(\lambda)=\pm\left(x^{\mu}_P-x^{\mu}_c(\lambda)\right)\sqrt{\frac{-(t_P-t_c(\lambda))^2 + (x_P-x_c(\lambda))^2 + z_P^2}{-(t_P-t_c(\lambda))^2 + (x_P-x_c(\lambda))^2}}~.
\end{equation}
The choice of overall sign determines the orientation of the interval/geodesic, and as in Sections~\ref{closedsubsec} and \ref{pointsubsec}, if we describe the point as a shrunk closed curve the sign changes when we pass from the positive to the negative part of the curve. Knowing (\ref{lmufromxcmu}), the complete set of geodesics is given by
\begin{equation}\label{geodesicspoint}
\Gamma^m(s,\lambda)=\left(x^{\mu}_c(\lambda)-\ell^{\mu}(\lambda)\cos s, \ell(\lambda)\sin s\right)~,
\end{equation}
in analogy with (\ref{geodesics}).

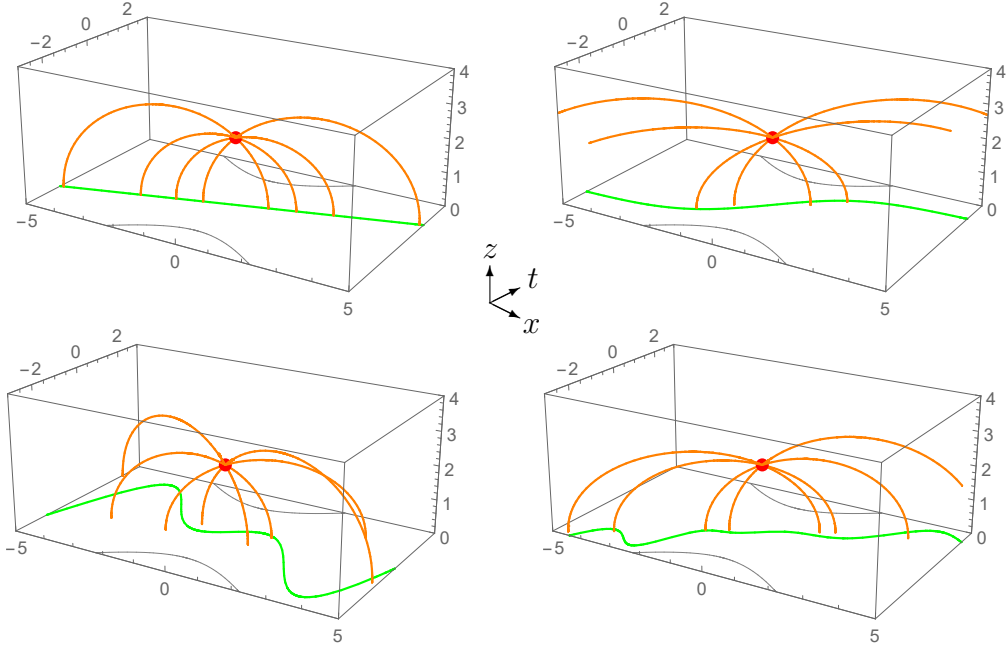
\begin{figure}[hbt]
\begin{center}
  \includegraphics[width=7cm]{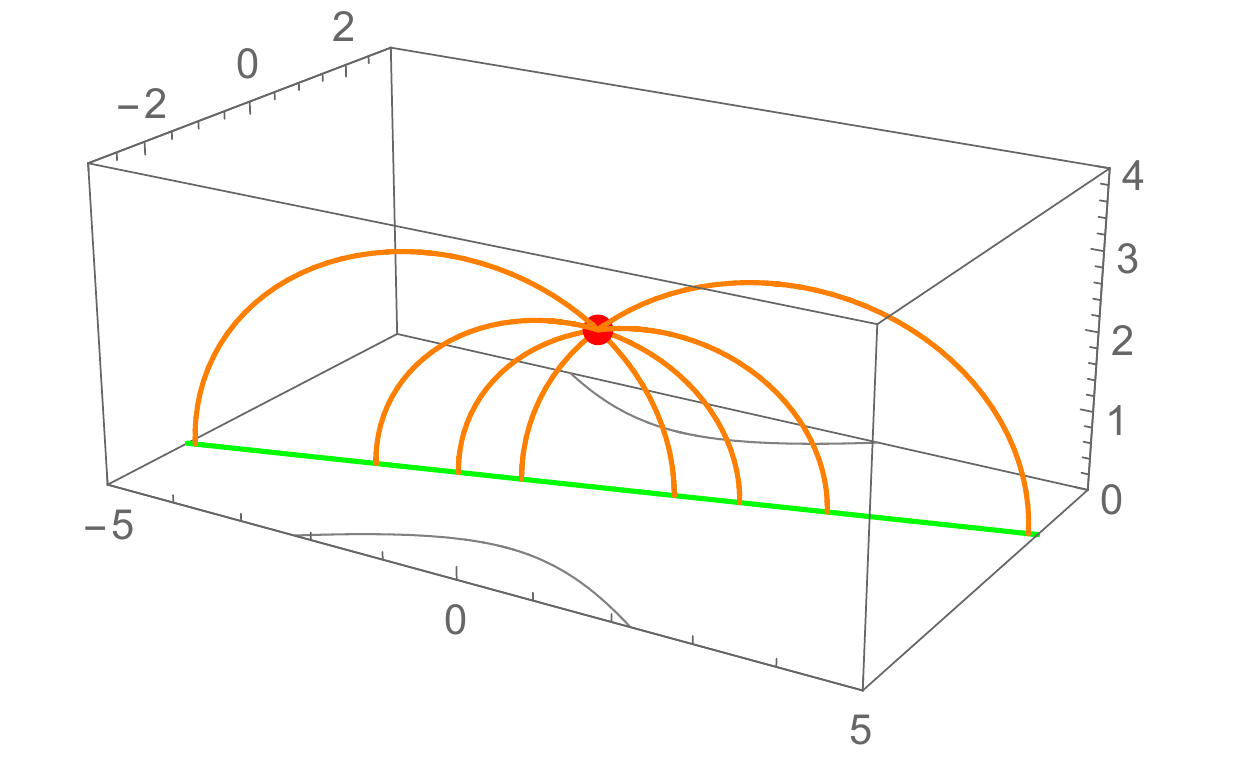}
  \includegraphics[width=7cm]{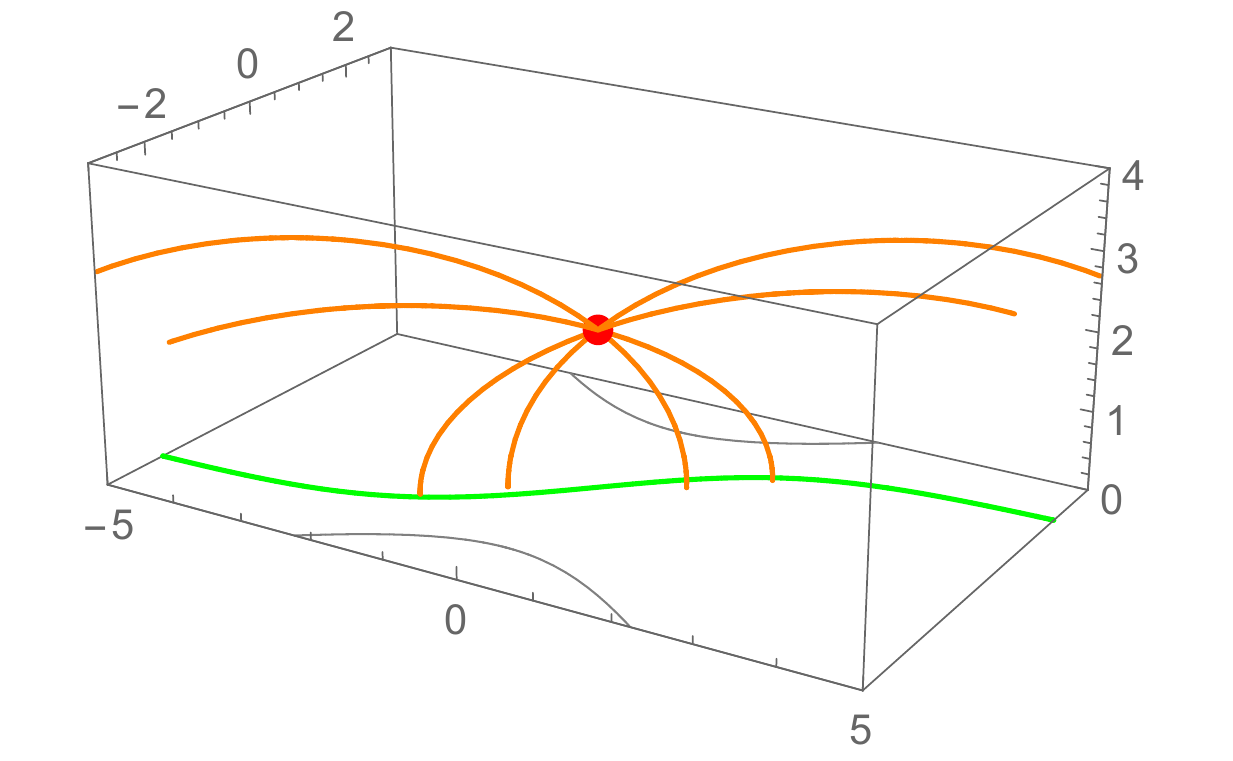}
  \includegraphics[width=7cm]{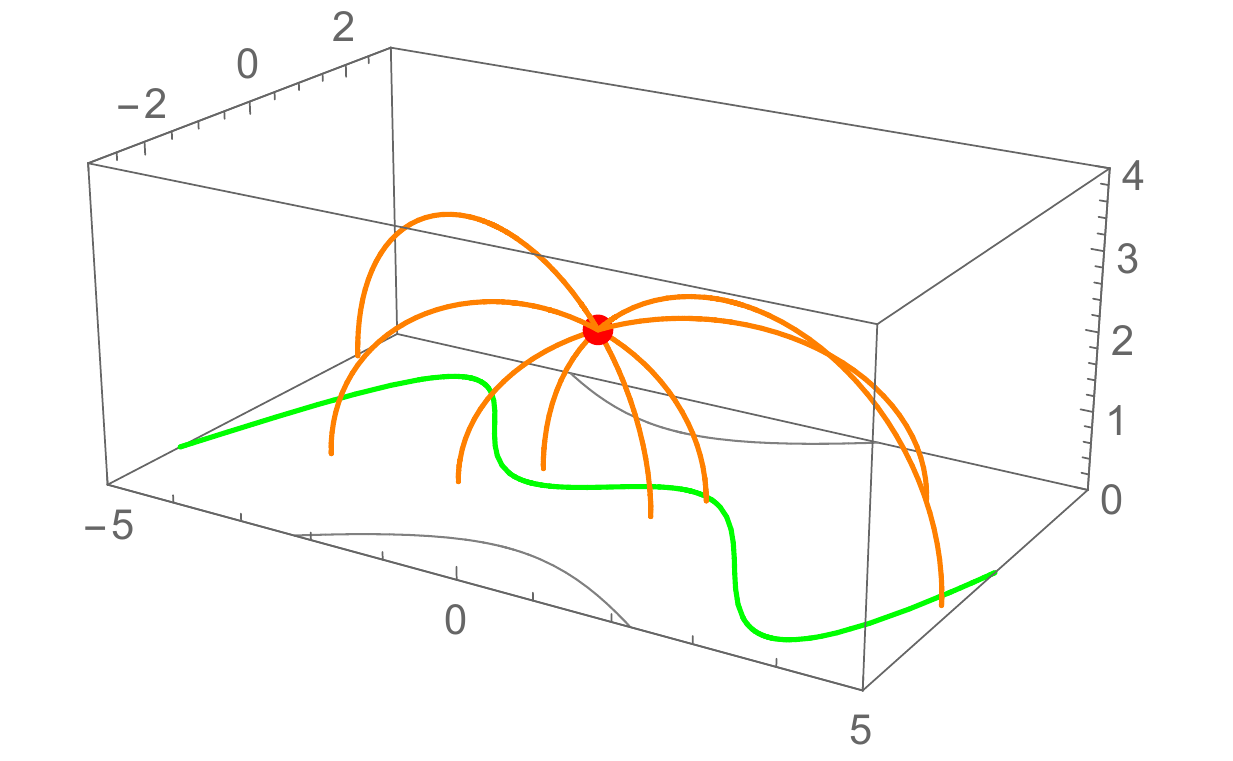}
  \includegraphics[width=7cm]{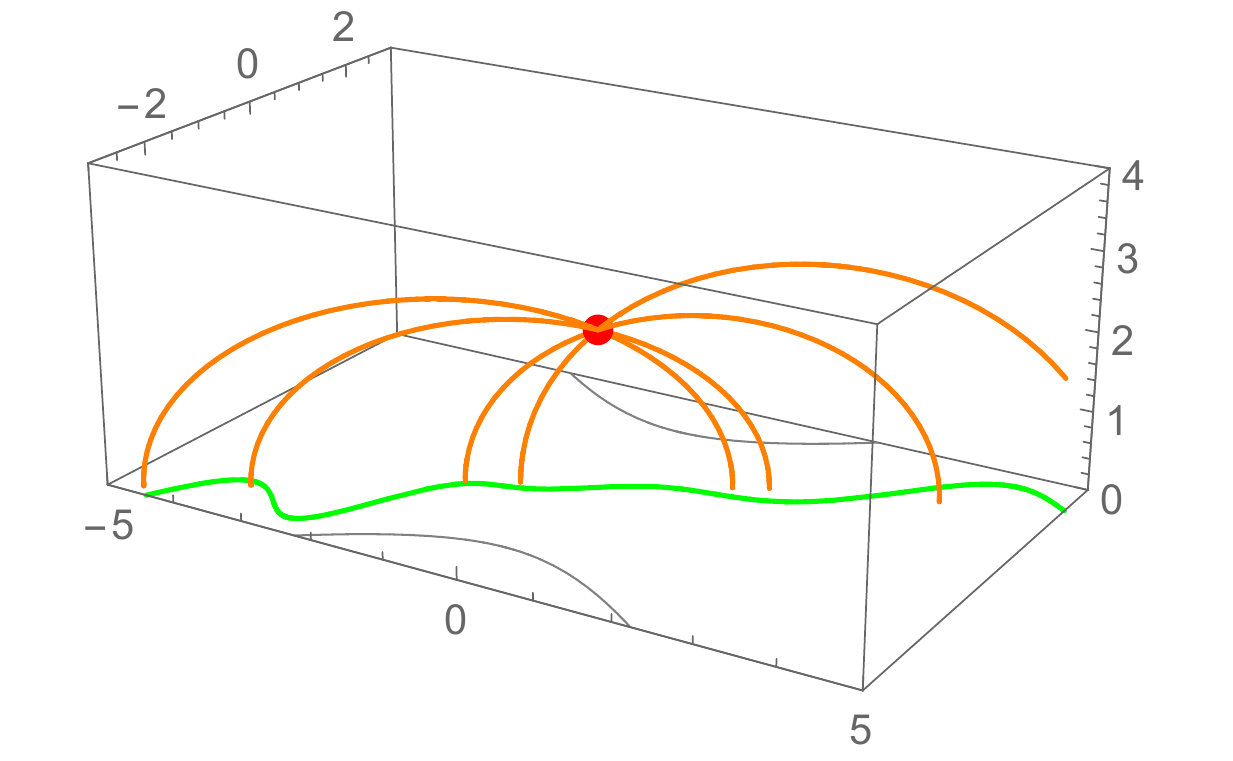}
\setlength{\unitlength}{1cm}
\begin{picture}(0,0)
\put(-7.5,4.6){\vector(0,3){0.5}}
\put(-7.5,4.6){\vector(2,1){0.4}}
\put(-7.5,4.6){\vector(2,-1){0.4}}
\put(-7.2,4.2){ $x$}
\put(-7.6,5.2){$z$}
\put(-7.0,4.8){$t$}
\end{picture}
\end{center}
\vspace*{-0.8cm}
\caption{Four different ways to describe the same bulk point $x_P^m=(0,0,2)$, marked in red, and a small sample of the corresponding geodesics, in orange. Shown in green on the AdS boundary is the curve traced by the center of the intervals in the CFT, $x^{\mu}_c(\lambda)$, from which the entire set of geodesics follows via (\ref{lmufromxcmu})-(\ref{geodesicspoint}). The gray dotted curves delimit the region on the boundary that is spacelike separated from $x^m_P$, from which any chosen green curve must not exit. Starting from the top left, our choice of center curve is $x^{\mu}_c(\lambda)=(0.3\lambda,\lambda)$,  $(2\tanh(\lambda/2),\lambda)$, $(0.7\lambda\cos(\lambda-0.2),\lambda)$ and $(0.8\cos(\sin\lambda),\lambda+\sin^2(\lambda/3))$, respectively.}
\label{covariantpointfig}
\end{figure}

If we take $t_c=t_P$ in (\ref{lmufromxcmu}), we correctly recover Eq.~(\ref{ell2}), describing a semicircle at constant time. In Fig.~\ref{covariantpointfig} we plot four different choices of center curve $x^{\mu}_c(\lambda)$ for the same given bulk point, and a small sample of the intervals/geodesics they give rise to.

The distinguishing feature of the family of CFT intervals described by (\ref{lmufromxcmu}) is that when we substitute it in the formula for differential entropy (\ref{ecalcovariant}), the complete integrand vanishes, as expected from the association with a bulk curve of vanishing length. As in Section \ref{pointsubsec}, we would like to search for a variational principle that selects families of this type.
The natural idea here is to try to generalize the extrinsic curvature argument to time-dependent situations.

 As in Section \ref{arbitrarysubsec}, consider an arbitrary spacelike curve $C^m(\lambda)$ in AdS$_3$, with tangent vector $u^{m}(\lambda)=(t'(\lambda),x'(\lambda),z'(\lambda))$. We can define the `acceleration' vector $a(\lambda)$ as the covariant derivative of $u(\lambda)$, normalized with respect to its magnitude:
\begin{equation}\label{vecadef}
a^{m}(\lambda)\equiv\frac{1}{\sqrt{u(\lambda)\cdot u(\lambda)}}\left(\frac{du^{m}(\lambda)}{d\lambda}+\Gamma^{m}_{nl}u^{n}(\lambda)u^{l}(\lambda)\right)~,
\end{equation}
where $\Gamma^{m}_{nl}$ are the usual Christoffel symbols.
The curvature of $C(\lambda)$ is defined as the norm of $a$,
\begin{equation}\label{curvaturedef}
\kappa=\sqrt{a(\lambda)\cdot a(\lambda)}~.
\end{equation}
Evidently, for a spacetime geodesic the curvature (\ref{curvaturedef}) is exactly zero. In the general case, $\kappa$ serves as a measure of how much the given curve differs from a geodesic. We can decompose (\ref{vecadef}) as the sum of two orthogonal contributions,
\begin{equation}
a^{m}(\lambda)=a^{m}_{\parallel}(\lambda)+a^{m}_{\perp}(\lambda)~,
\end{equation}
where
\begin{equation}\label{aparaperp}
a^{m}_{\parallel}(\lambda)\equiv\frac{u(\lambda)\cdot a (\lambda)}{u(\lambda) \cdot u (\lambda)}u^{m}(\lambda)~,
\qquad
a^{m}_{\perp}(\lambda)\equiv a^{m}(\lambda)-\frac{u(\lambda)\cdot a(\lambda)}{u(\lambda) \cdot u(\lambda)}u^{m}(\lambda)
\end{equation}
are the components of $a$ parallel and perpendicular to $u$. The norms of these components are called the geodesic curvature and the normal curvature, respectively,
\begin{equation}\label{extrinsiccurvatures}
\kappa_{\parallel}\equiv\sqrt{a_{\parallel}(\lambda)\cdot a_{\parallel}(\lambda)}~,\qquad
\kappa_{\perp}=\sqrt{a_{\perp}(\lambda)\cdot a_{\perp}\lambda)}~,
\end{equation}
and obviously satisfy $\kappa^2=\kappa_{\parallel}^2+\kappa_{\perp}^2$.

When a closed loop shrinks down to a point, its normal curvature diverges, so we expect that by expressing $\kappa_{\perp}(\lambda)$ as a function of $x^{\mu}_{\pm}$ and extremizing we can recover the boundary definition of a bulk point.
In the time-independent case, $t'(\lambda)=0$, we find after some algebra that
\begin{eqnarray}
\kappa_{\parallel}^2&=&\frac{\left(x'(\lambda)^2 z'(\lambda)+z'(\lambda)^3-z(\lambda) x'(\lambda) x''(\lambda)-z(\lambda ) z'(\lambda) z''(\lambda )\right)^2}{z(\lambda )^2 \left(x'(\lambda )^2+z'(\lambda )^2\right)^2}\,,\\
\kappa_{\perp}^2&=&\frac{\left(x'(\lambda)^3-z(\lambda) z'(\lambda) x''(\lambda)+x'(\lambda) \left(z'(\lambda)^2+z(\lambda) z''(\lambda)\right)\right)^2}{z(\lambda)^2\left(x'(\lambda)^2+z'(\lambda)^2\right)^2}\,.\nonumber
\end{eqnarray}
This expression for $\kappa_{\perp}(\lambda)$ coincides with the Lagrangian $\mathcal{L}$ defined in terms of extrinsic curvature above (\ref{actionK}). Thus, in static configurations extremizing normal curvature, as we are proposing here, is in fact the same as extremizing extrinsic curvature as in \cite{nutsandbolts}.
%Isn't there some standard relation between the normal curvature and the trace of the extrinsic curvature?

The generalization for time-dependent case is straightforward. In this case we find
\begin{eqnarray}\label{kappan}
\kappa_{\perp}^2=\frac{1}{z(\lambda)^2\left(x'(\lambda)^2+z'(\lambda)^2-t'(\lambda)^2\right)^2}
\bigg[
2\left(x'(\lambda)^2-t'(\lambda)^2\right)^2 \left(z'(\lambda)^2+z(\lambda) z''(\lambda )\right)&&\nonumber\\
+x'(\lambda)^6- t'(\lambda)^6+\left(x''(\lambda )^2-t''(\lambda )^2\right)z(\lambda )^2 z'(\lambda )^2&&\nonumber\\
-\left(x'(\lambda)^2-t'(\lambda)^2\right)\left(3x'(\lambda)^2t'(\lambda)^2-\left(z'(\lambda)^2+z(\lambda) z''(\lambda)\right)^2\right)
&&\nonumber\\
-2\left(x'(\lambda)^2-t'(\lambda)^2+z'(\lambda)^2+z(\lambda) z''(\lambda)\right)\left(x'(\lambda)x''(\lambda)-t'(\lambda)t''(\lambda)\right) z(\lambda) z'(\lambda)
&&\nonumber\\
-\left(x'(\lambda) t''(\lambda)-t'(\lambda) x''(\lambda)\right)^2z(\lambda)^2
\bigg]\,.&&
\end{eqnarray}
Again, this functional depends on second derivatives so it in general leads to fourth-order differential equations. As a consistency check, however, we have verified that the point-like ansatz $x^{\mu}(\lambda)=(t_P,x_P,z_P)$ is indeed a solution of these equations.

With some work, we can rewrite the normal curvature (\ref{kappan}) of our bulk curve as a function of the endpoints $x^{\mu}_{\pm}(\lambda)$ of the corresponding CFT intervals, given by (\ref{TXRL}). Taking the result as our Lagrangian, we arrive at
\begin{eqnarray}\label{actionKcovariant}
I&\equiv&\int d\lambda\,\mathcal{L}
=2\int d\lambda\,\sqrt{\frac{x_+'(\lambda)x_-'(\lambda)+t_+'(\lambda)t_-'(\lambda)}{(x_+(\lambda)-x_-(\lambda))^2+(t_+(\lambda)-t_-(\lambda))^2}}\\
&=&\int d\lambda\,\sqrt{\frac{x_c'(\lambda)^2-\ell_x'(\lambda)^2+t_c'(\lambda)^2-\ell_t'(\lambda)^2}{\ell_x(\lambda)^2+\ell_t(\lambda)^2}}~.\nonumber
\end{eqnarray}
For constant time, we correctly recover our previous action (\ref{actionK2}). Similar to what we had in that case, we see in the last line of (\ref{actionKcovariant}) that the action is independent of $x_c^{\mu}(\lambda)$, so the conjugate momenta $\Pi_{\mu}\equiv\p\mathcal{L}/\p x_c'^{\mu}$ are constants of motion. These conditions determine a particular choice of center curve, $t_c=(\Pi_t/\Pi_x) x_c +\mbox{constant}$ (corresponding to fixed time in a boosted frame), with a specific parametrization $x_c(\lambda)$.

\subsection{Distances}\label{covariantdistancesubsec}

In the previous subsection we have learned that any given bulk point $P$ is described not by a unique family of intervals in the CFT, $\{x_c^{\mu}(\lambda),\ell^{\mu}(\lambda)\}_P$, but by an entire equivalence class of such families, which we will denote $\mathcal{F}_P\equiv\left[\{x_c^{\mu}(\lambda),\ell^{\mu}(\lambda)\}_P\right]$. Each family in this class can be selected by specifying a center curve $x_c^{\mu}(\lambda)$ within the region of the AdS boundary that is spacelike separated from $P$, and then using (\ref{lmufromxcmu}) to obtain the corresponding radius vectors $\ell^{\mu}(\lambda)$. Some examples that illustrate the range of options were portrayed in Fig.~\ref{covariantpointfig}.

Given two bulk points
 $P$ and $Q$, by carrying out a boost with parameter $\beta=(t_P-t_Q)/(x_P-x_Q)$ to the frame where they are simultaneous, and then boosting back to the original frame, we can deduce that the geodesic $\overline{PQ}$ that connects them is centered at
 \begin{eqnarray}\label{pqcenter}
 t_{M}&=&\frac{2t_P(x_P-x_Q)^2-(t_P-t_Q)\left(t_P^2-t_Q^2+(x_P-x_Q)^2-z_P^2+z_Q^2\right)}{2(-(t_P-t_Q)^2+(x_P-x_Q)^2)}~,\nonumber\\
x_{M}&=&\frac{(x_P-x_Q)\left(x_P^2-x_Q^2+z_P^2-z_Q^2\right)-(t_P-t_Q)^2(x_P+x_Q)}{2(-(t_P-t_Q)^2+(x_P-x_Q)^2)}~,
 \end{eqnarray}
and has radius vector
\begin{eqnarray}\label{pqradius}
\ell^t_{M}&=&\frac{(t_P-t_Q)\mbox{sgn}(x_P-x_Q)\sqrt{(-(t_P-t_Q)^2+(x_P-x_Q)^2+z_P^2+z_Q^2)^2-4z_P^2 z_Q^2}}{2(-(t_P-t_Q)^2+(x_P-x_Q)^2)}~,
\nonumber\\
\ell^x_{M}&=&\frac{\sqrt{(x_P-x_Q)^2}\sqrt{(-(t_P-t_Q)^2+(x_P-x_Q)^2+z_P^2+z_Q^2)^2-4z_P^2 z_Q^2}}{2(-(t_P-t_Q)^2+(x_P-x_Q)^2)}~.
\end{eqnarray}
The distance between the two points is given by the arclength along this geodesic,
\begin{eqnarray}\label{covariantdistancepq}
d(P,Q)&=&L\ln\left(
\frac{-(t_P-t_Q)^2-z_P^2+z_Q^2+(x_P-x_Q)(x_P-x_Q-\Delta)}
{(t_P-t_Q)^2-z_P^2+z_Q^2-(x_P-x_Q)(x_P-x_Q+\Delta)}\right)~,\nonumber\\
\Delta&\equiv&\sqrt{\frac{(-(t_P-t_Q)^2+(x_P-x_Q)^2+z_P^2+z_Q^2)^2-4z_P^2 z_Q^2}{(x_P-x_Q)^2}}~.
\end{eqnarray}
This is the boosted version of (\ref{distancepq}) or (\ref{distancepq2}).

To reproduce (\ref{covariantdistancepq}) in terms of differential entropy imitating the procedure in Section~\ref{distancesubsec}, we begin with the equivalence classes of families of intervals for the two points, $\mathcal{F}_P$ and $\mathcal{F}_Q$, and select a representative from each class that happens to include the geodesic $\overline{PQ}$. If in addition we narrow down our selection by demanding that these two representatives have the same center curve $x_c^{\mu}(\lambda)$, the situation becomes directly analogous to what we had before. We can define the truncated families $\hat{\ell}^{\mu}_P(\lambda)$ and $\check{\ell}^{\mu}_P(\lambda)$ (and likewise for $Q$) by including intervals up to or starting from $x_{M}^{\mu}(\lambda)$. With these we can again form the combination
$\ell^{\mu}_{PQ}(\lambda)\equiv \hat{\ell}^{\mu}_{P}(\lambda)+\hat{\ell}^{\mu}_{Q}(\lambda)$, and compute its differential entropy. Evidently the simplest choice is to take $x_c^{\mu}(\lambda)$ along the boosted time slice that contains $P$, $Q$ and $\overline{PQ}$. In that case our calculation in Section~\ref{distancesubsec} applies directly. E.g., for closed point-curves we find again that
\begin{equation}\label{covariantdworks}
d(P,Q)=\frac{1}{2}E[\ell^{\mu}_{PQ}(\lambda)]~.
\end{equation}
We expect this relation to hold also for other choices of $x_c^{\mu}(\lambda)$, but we will not attempt to prove that here. The important conclusion is that there does exist a procedure to compute bulk distances from CFT data.

\section*{Acknowledgements}

It is a pleasure to thank Mariano Chernicoff, Bartek Czech, Sagar Lokhande, Hirosi Ooguri, Mukund Rangamani and Marika Taylor for useful conversations, and Sagar Lokhande for comments on the manuscript.
The work of RE, AG and AL was partially supported by Mexico's National Council of Science and Technology (CONACyT) grant 238734 and DGAPA-UNAM grant IN107115. JFP was supported by the Netherlands
Organization for Scientific Research (NWO) under the VENI scheme.

\appendix

\section{Discrete Versions of Differential Entropy}\label{appendix}

Given a family of $K$ successively overlapping intervals $I_k$  that cover a time slice of the CFT$_2$,
the original discrete definition of differential entropy (inspired by strong subadditivity) is \cite{hole-ography}
\be\label{diffEa}
E^{(1)}\equiv\sum_{k=1}^{K}\left[S(I_k)-S(I_k\cap I_{k+1})\right]~.
\ee
An `averaged' version of this was considered in \cite{myers},
\be\label{diffEb}
E^{(2)}\equiv\sum_{k=1}^{K}\left[S(I_k)-\frac{1}{2}S(I_{k-1}\cap I_{k})-\frac{1}{2}S(I_k\cap I_{k+1})\right]\,.
\ee
For closed curves,
%(or open curves with periodic boundary conditions)
it is understood that $I_{K+1}\equiv I_1$. The fact that each interval has neighbors on both sides implies that $E^{(1)}=E^{(2)}$. In the following we will show that in the continuum limit the two definitions actually differ by a boundary term, which is relevant when the curve in consideration is open.

The ingredients that we will need, which we transcribe here for convenience, are the entanglement entropy (\ref{s}) for the interval associated with point $\lambda$ on the curve,
\be
S(\lambda)=2L \log\left(\frac{|x_+(\lambda)-x_-(\lambda)|}{\epsilon}\right)=2L \log\left(\frac{2|\ell(\lambda)|}{\epsilon}\right)\,,
\ee
and the expressions (\ref{xpm}) for the endpoints of this interval in terms of bulk data,
\be
x_\pm(\lambda)=x(\lambda)+\frac{z(\lambda)z'(\lambda)}{x'(\lambda)}\pm\frac{z(\lambda)}{x'(\lambda)}\sqrt{x'(\lambda)^2+z'(\lambda)^2}\,,
\ee
or, equivalently, $x_\pm(\lambda)=x_c(\lambda)\pm\ell(\lambda)$, where according to (\ref{xc})-(\ref{ell}),
\be
x_c(\lambda)=x(\lambda)+\frac{z(\lambda)z'(\lambda)}{x'(\lambda)}\,,\qquad \ell(\lambda)=\frac{z(\lambda)}{x'(\lambda)}\sqrt{x'(\lambda)^2+z'(\lambda)^2}\,.
\ee
For concreteness, we will consider a `positive' curve, meaning that $x_+(\lambda)>x_-(\lambda)$, $x'(\lambda)>0$ and $\ell(\lambda)>0$.

Without loss of generality, we will assume that $\lambda\in[0,1]$, and discretize this domain as follows:
\be
\lambda_k=\frac{k-1}{K-1}=\left\{0,\frac{1}{K-1},\frac{2}{K-1},\ldots,1\right\}\,,\qquad\text{for}\qquad k=\{1,2,3,\ldots,K\}\,.
\ee
Clearly, the difference between two consecutive $\lambda$'s approaches zero in the continuum limit,
\be
\delta\lambda\equiv\lambda_{k+1}-\lambda_{k}=\frac{1}{K-1}\to 0\qquad\text{as}\quad K\to\infty.
\ee

\begin{figure}[hbt]
\begin{center}
\includegraphics[angle=0,width=0.86\textwidth]{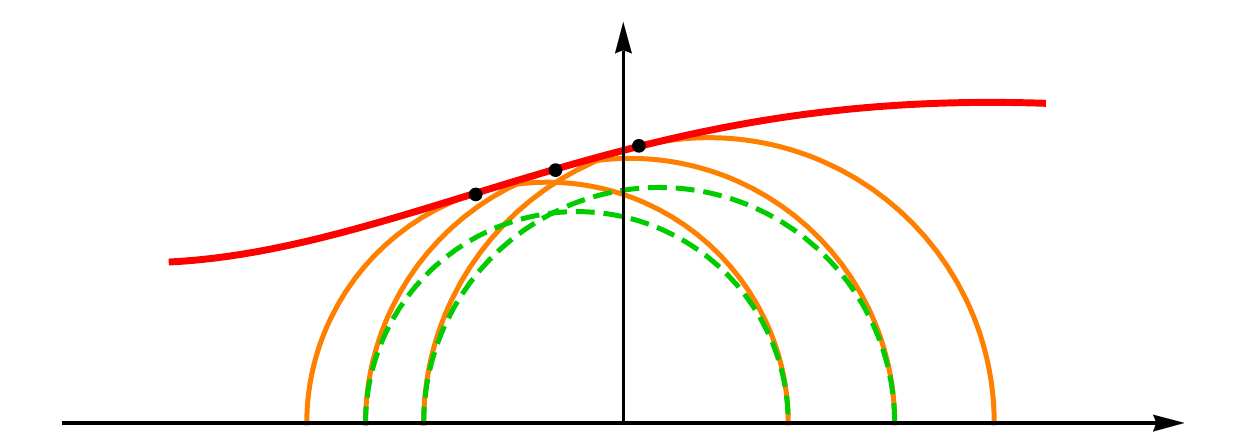}
\begin{picture}(0,0)
\put(-186,121){$z$}
\put(-30,15){$x$}
\put(-148,-2){\scriptsize{$x_+^{k-1}$}}
\put(-116,-2){\scriptsize{$x_+^{k}$}}
\put(-87,-2){\scriptsize{$x_+^{k+1}$}}
\put(-292,-2){\scriptsize{$x_-^{k-1}$}}
\put(-274,-2){\scriptsize{$x_-^{k}$}}
\put(-257,-2){\scriptsize{$x_-^{k+1}$}}
\put(-240,83){\scriptsize{$\lambda_{k-1}$}}
\put(-215,89){\scriptsize{$\lambda_{k}$}}
\put(-190,97){\scriptsize{$\lambda_{k+1}$}}
\qbezier(-320,62)(-300,63)(-280,70)
\put(-280,70){\vector(3,1){0.2}}
\put(-304,68){\scriptsize{$\lambda$}}
\end{picture}
\caption{\small Discrete reconstruction of differential entropy. To each point of the (red) curve $(x(\lambda_k),z(\lambda_k))$ we can associate
a geodesic (depicted in solid orange) whose endpoints reach the AdS boundary at the points $x_{\pm}^{k}$, specifying a boundary interval $I_k$. The discrete versions of differential entropy (\ref{diffEa}) and (\ref{diffEb}) involve a particular combination of the entanglement entropy of such intervals $S(I_k)$, as well as the entanglement entropy of the intersections of $I_k$ with their immediate neighbours, $S(I_k\cap I_{k+1})$ and $S(I_{k-1}\cap I_{k})$ (depicted in dashed green). A one-parameter generalization of the discrete version of differential entropy is given by (\ref{diffExi}). It includes the two original versions, and differs in general by a boundary term. \label{fig:disc}}
\end{center}
\end{figure}
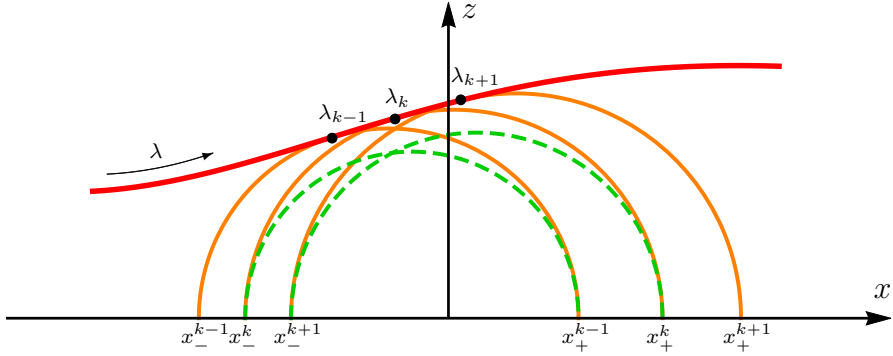

Evaluating the bulk/boundary data in these discrete values of the parameter $\lambda$ we get $x_{\pm}^k\equiv x_{\pm}(\lambda_k)$ (see Figure \ref{fig:disc}), and similarly for $x_c(\lambda)$ and $\ell(\lambda)$. Thus, in their discrete versions we have $x^k_\pm=x^k_c\pm\ell_k$. We are interested in taking the continuum limit, so it will be convenient to define the following quantities, which we truncate at linear order in $\delta\lambda$:
\be
x_{\pm}^{k\pm1}=x_{\pm}^{k}\pm x'\,\!\!_{\pm}^{\,k} \delta\lambda\,,\qquad\text{with}\qquad x'\,\!\!_{\pm}^{\,k}=\frac{d x_{\pm}(\lambda)}{d\lambda}\bigg|_{\lambda=\lambda_k}\,.
\ee

Similarly we can write $x_c^{k\pm1}=x_c^{k}\pm x'\,\!\!_{c}^{\,k} \delta\lambda$ and $\ell_{k\pm1}=\ell_{k}\pm \ell'\,\!\!_{k} \delta\lambda$.
Consider now the following boundary intervals, and their corresponding entanglement entropies:
\begin{itemize}
  \item $I_k$: Its length is given by $x_+^k-x_-^k=2\ell_k$, therefore
  \be
  S(I_k)=2L \log\left(\frac{2 \ell_k}{\epsilon}\right)\,.
  \ee
  \item $I_k\cap I_{k+1}$: Its length is given by $x_+^k-x_-^{k+1}=2\ell_k-(x'\,\!\!_{c}^{\,k}-\ell'\,\!\!_{k})\delta\lambda$, therefore
  \be
  S(I_k\cap I_{k+1})=2L \log\left(\frac{2\ell_k-(x'\,\!\!_{c}^{\,k}-\ell'\,\!\!_{k})\delta\lambda}{\epsilon}\right)=2L \log\left(\frac{2 \ell_k}{\epsilon}\right)-\frac{L(x'\,\!\!_{c}^{\,k}-\ell'\,\!\!_{k})}{\ell_k}\delta\lambda\,.
  \ee
  \item $I_{k-1}\cap I_{k}$: Its length is given by $x_+^{k-1}-x_-^{k}=2\ell_k-(x'\,\!\!_{c}^{\,k}+\ell'\,\!\!_{k})\delta\lambda$, therefore
    \be
  S(I_{k-1}\cap I_{k})=2L \log\left(\frac{2\ell_k-(x'\,\!\!_{c}^{\,k}+\ell'\,\!\!_{k})\delta\lambda}{\epsilon}\right)=2L \log\left(\frac{2 \ell_k}{\epsilon}\right)-\frac{L(x'\,\!\!_{c}^{\,k}+\ell'\,\!\!_{k})}{\ell_k}\delta\lambda\,.
  \ee
\end{itemize}

Putting this together, we arrive to the following formulas for the two discrete versions of differential entropy (\ref{diffEa})-(\ref{diffEb}):
\be\label{e1continuum}
E^{(1)}=L\sum_{k=1}^{K}\frac{x'\,\!\!_{c}^{\,k}-\ell'\,\!\!_{k}}{\ell_k}\delta\lambda\to L\int d\lambda\,\frac{x'_-}{\ell}\,,
\ee
\be\label{e2continuum}
E^{(2)}=L\sum_{k=1}^{K}\frac{x'\,\!\!_{c}^{\,k}}{\ell_k}\delta\lambda\to L\int d\lambda\,\frac{x'_c}{\ell}\,,
\ee
which differ by a boundary term,
\be
E^{(2)}-E^{(1)}=L \int d\lambda\,\frac{\ell'}{\ell}=L\int d\lambda\, \partial_\lambda \ln\left(\frac{2\ell}{\epsilon}\right)=\frac{1}{2}S(\lambda_f)-\frac{1}{2}S(\lambda_i)\,.
\ee

Finally, notice that as a generalization of (\ref{diffEa}) and (\ref{diffEb}) one can write down a one-parameter family of discrete differential entropies,
\be\label{diffExidiscrete}
E^{(\xi)}\equiv\sum_{k=1}^{K}\left[S(I_k)-\frac{1}{2}(\xi-1) S(I_{k-1}\cap I_{k})-\frac{1}{2}(3-\xi)S(I_k\cap I_{k+1})\right]\,,
\ee
which are all in agreement for closed curves. In the continuum limit one obtains
\be\label{diffExi}
E^{(\xi)}\to L\int d\lambda\left[\frac{x'_c}{\ell}+(\xi-2)\frac{\ell'}{\ell}\right]\,,
\ee
which  for $\xi=1,2$ indeed agrees with (\ref{e1continuum}) and (\ref{e2continuum}), respectively. For $\xi=3$  one recovers the definition used in the main body of the paper
\be
E^{(3)}\to L\int d\lambda\,\frac{x'_+}{\ell}=E~,
\ee
as seen in (\ref{e2}). Notice that $E^{(3)}$ can be obtained from $E^{(1)}$ (up to a boundary term) using integration by parts, thus interchanging the role of $x_+$ and $x_-$. Such a boundary term was neglected in the previous literature, since the focus there was on closed curves, but it is actually very important when considering open curves as we do in this paper.

It is worth emphasizing that the boundary terms needed in the definition (\ref{ecal}) of renormalized differential entropy,
\be\label{ecalrepeated}
\mathcal{E}\equiv E-f(\lambda_f)+f(\lambda_i)~,
\ee
depend on the definition of $E$ that we start with. From the definition (\ref{e}) used in this paper,
\be
E=\int d\lambda\,\frac{\partial S(x_L(\lambda),x_R(\bar{\lambda}))}{\partial \bar{\lambda}}\bigg|_{\bar{\lambda}=\lambda}\,,
\ee
we obtained in (\ref{f})
\be
f(\lambda)=L \ln\left(\frac{2|\ell|}{\epsilon}\right)+L\sinh^{-1}\left(\frac{z'}{|x'|}\right)\,.
\ee
As explained in Section~\ref{opensubsec}, these two pieces correspond respectively to the length of $i)$ half of the geodesic labeled by $\lambda$ and $ii)$ the arc of the geodesic labeled by $\lambda$, running from $x(\lambda)$ to $x_c(\lambda)$. The sum of the two, then, is minus the length of the arc of this same geodesic, running from $x(\lambda)$ all the way to the right endpoint $x_+(\lambda)$ at the boundary (or, more precisely, at the regularized endpoint (\ref{xpepsilon})). This was illustrated in Fig.~\ref{ffig}.


\begin{thebibliography}{99}

\bibitem{malda}
  J.~M.~Maldacena,
  ``The large $N$ limit of superconformal field theories and supergravity,''
  Adv.\ Theor.\ Math.\ Phys.\  {\bf 2}, 231 (1998)
  [Int.\ J.\ Theor.\ Phys.\  {\bf 38}, 1113 (1999)]
  [arXiv:hep-th/9711200].
  %%CITATION = HEP-TH 9711200;%%

\bibitem{gkp}
  S.~S.~Gubser, I.~R.~Klebanov and A.~M.~Polyakov,
  ``Gauge theory correlators from non-critical string theory,''
  Phys.\ Lett.\ B {\bf 428}, 105 (1998)
  [arXiv:hep-th/9802109].
  %%CITATION = HEP-TH 9802109;%%%\cite{Witten:1998qj}

\bibitem{w}
  E.~Witten,
  ``Anti-de Sitter space and holography,''
  Adv.\ Theor.\ Math.\ Phys.\  {\bf 2}, 253 (1998)
  [arXiv:hep-th/9802150].
  %%CITATION = HEP-TH 9802150;%%

\bibitem{rt}
  S.~Ryu and T.~Takayanagi,
  ``Holographic derivation of entanglement entropy from AdS/CFT,''
  Phys.\ Rev.\ Lett.\  {\bf 96} (2006) 181602
%  doi:10.1103/PhysRevLett.96.181602
  [hep-th/0603001].
  %%CITATION = doi:10.1103/PhysRevLett.96.181602;%%

\bibitem{hrt}
  V.~E.~Hubeny, M.~Rangamani and T.~Takayanagi,
  ``A Covariant holographic entanglement entropy proposal,''
  JHEP {\bf 0707} (2007) 062
 % doi:10.1088/1126-6708/2007/07/062
  [arXiv:0705.0016 [hep-th]].
  %%CITATION = doi:10.1088/1126-6708/2007/07/062;%%

\bibitem{lm}
  A.~Lewkowycz and J.~Maldacena,
  ``Generalized gravitational entropy,''
  JHEP {\bf 1308} (2013) 090
 % doi:10.1007/JHEP08(2013)090
  [arXiv:1304.4926 [hep-th]].
  %%CITATION = doi:10.1007/JHEP08(2013)090;%%

 \bibitem{dlr}
  X.~Dong, A.~Lewkowycz and M.~Rangamani,
``Deriving covariant holographic entanglement,''
  JHEP {\bf 1611} (2016) 028
%  doi:10.1007/JHEP11(2016)028
  [arXiv:1607.07506 [hep-th]].
  %%CITATION = doi:10.1007/JHEP11(2016)028;%%

\bibitem{hms}
  L.~Y.~Hung, R.~C.~Myers and M.~Smolkin,
  ``On Holographic Entanglement Entropy and Higher Curvature Gravity,''
  JHEP {\bf 1104} (2011) 025
 % doi:10.1007/JHEP04(2011)025
  [arXiv:1101.5813 [hep-th]].
  %%CITATION = doi:10.1007/JHEP04(2011)025;%%

\bibitem{dong}
  X.~Dong,
  ``Holographic Entanglement Entropy for General Higher Derivative Gravity,''
  JHEP {\bf 1401} (2014) 044
 % doi:10.1007/JHEP01(2014)044
  [arXiv:1310.5713 [hep-th]].
  %%CITATION = doi:10.1007/JHEP01(2014)044;%%

\bibitem{camps}
  J.~Camps,
  ``Generalized entropy and higher derivative Gravity,''
  JHEP {\bf 1403} (2014) 070
 % doi:10.1007/JHEP03(2014)070
  [arXiv:1310.6659 [hep-th]].
  %%CITATION = doi:10.1007/JHEP03(2014)070;%%

 \bibitem{castro}
  M.~Ammon, A.~Castro and N.~Iqbal,
  ``Wilson Lines and Entanglement Entropy in Higher Spin Gravity,''
  JHEP {\bf 1310} (2013) 110
 % doi:10.1007/JHEP10(2013)110
  [arXiv:1306.4338 [hep-th]].
  %%CITATION = doi:10.1007/JHEP10(2013)110;%%


 \bibitem{bdhm}
  T.~Barrella, X.~Dong, S.~A.~Hartnoll and V.~L.~Martin,
  ``Holographic entanglement beyond classical gravity,''
  JHEP {\bf 1309} (2013) 109
%  doi:10.1007/JHEP09(2013)109
  [arXiv:1306.4682 [hep-th]].
  %%CITATION = doi:10.1007/JHEP09(2013)109;%%

\bibitem{flm}
  T.~Faulkner, A.~Lewkowycz and J.~Maldacena,
  ``Quantum corrections to holographic entanglement entropy,''
  JHEP {\bf 1311} (2013) 074
 % doi:10.1007/JHEP11(2013)074
  [arXiv:1307.2892 [hep-th]].
  %%CITATION = doi:10.1007/JHEP11(2013)074;%%

\bibitem{ew}
  N.~Engelhardt and A.~C.~Wall,
  ``Quantum Extremal Surfaces: Holographic Entanglement Entropy beyond the Classical Regime,''
  JHEP {\bf 1501} (2015) 073
 % doi:10.1007/JHEP01(2015)073
  [arXiv:1408.3203 [hep-th]].
  %%CITATION = doi:10.1007/JHEP01(2015)073;%%

 \bibitem{deboer}
  J.~de Boer, A.~Castro, E.~Hijano, J.~I.~Jottar and P.~Kraus,
  ``Higher spin entanglement and $ {\mathcal{W}}_{\mathrm{N}} $ conformal blocks,''
  JHEP {\bf 1507} (2015) 168
 % doi:10.1007/JHEP07(2015)168
  [arXiv:1412.7520 [hep-th]].
  %%CITATION = doi:10.1007/JHEP07(2015)168;%%

 \bibitem{hofman}
  A.~Castro, D.~M.~Hofman and N.~Iqbal,
  ``Entanglement Entropy in Warped Conformal Field Theories,''
  JHEP {\bf 1602} (2016) 033
%  doi:10.1007/JHEP02(2016)033
  [arXiv:1511.00707 [hep-th]].
  %%CITATION = doi:10.1007/JHEP02(2016)033;%%

 \bibitem{elena}
  E.~C\'aceres, R.~Mohan and P.~H.~Nguyen,
  ``On holographic entanglement entropy of Horndeski black holes,''
  arXiv:1707.06322 [hep-th].
  %%CITATION = ARXIV:1707.06322;%%

 \bibitem{janiszewski}
  S.~Janiszewski,
  ``Non-relativistic entanglement entropy from Horava gravity,''
  arXiv:1707.08231 [hep-th].
   %%CITATION = ARXIV:1707.08231;%%

 \bibitem{klebanov}
  I.~R.~Klebanov, D.~Kutasov and A.~Murugan,
``Entanglement as a probe of confinement,''
  Nucl.\ Phys.\ B {\bf 796} (2008) 274
  %  doi:10.1016/j.nuclphysb.2007.12.017
  [arXiv:0709.2140 [hep-th]].
  %%CITATION = doi:10.1016/j.nuclphysb.2007.12.017;%%

 \bibitem{vr}
  M.~Van Raamsdonk,
  ``Building up spacetime with quantum entanglement,''
  Gen.\ Rel.\ Grav.\  {\bf 42} (2010) 2323
   [Int.\ J.\ Mod.\ Phys.\ D {\bf 19} (2010) 2429]
%  doi:10.1007/s10714-010-1034-0, 10.1142/S0218271810018529
  [arXiv:1005.3035 [hep-th]].
  %%CITATION = doi:10.1007/s10714-010-1034-0, 10.1142/S0218271810018529;%%

\bibitem{headrick}
  M.~Headrick,
  ``Entanglement Renyi entropies in holographic theories,''
  Phys.\ Rev.\ D {\bf 82} (2010) 126010
%  doi:10.1103/PhysRevD.82.126010
  [arXiv:1006.0047 [hep-th]].
  %%CITATION = doi:10.1103/PhysRevD.82.126010;%%

\bibitem{myerssinha}
  R.~C.~Myers and A.~Sinha,
  ``Holographic c-theorems in arbitrary dimensions,''
  JHEP {\bf 1101} (2011) 125
  doi:10.1007/JHEP01(2011)125
  [arXiv:1011.5819 [hep-th]].
  %%CITATION = doi:10.1007/JHEP01(2011)125;%%

\bibitem{chm}
  H.~Casini, M.~Huerta and R.~C.~Myers,
  ``Towards a derivation of holographic entanglement entropy,''
  JHEP {\bf 1105}, 036 (2011)
  %doi:10.1007/JHEP05(2011)036
  [arXiv:1102.0440 [hep-th]].

\bibitem{hhm}
  P.~Hayden, M.~Headrick and A.~Maloney,
  ``Holographic Mutual Information is Monogamous,''
  Phys.\ Rev.\ D {\bf 87} (2013) no.4,  046003
 % doi:10.1103/PhysRevD.87.046003
  [arXiv:1107.2940 [hep-th]].
  %%CITATION = doi:10.1103/PhysRevD.87.046003;%%

\bibitem{hm}
  T.~Hartman and J.~Maldacena,
``Time Evolution of Entanglement Entropy from Black Hole Interiors,''
  JHEP {\bf 1305} (2013) 014
  %  doi:10.1007/JHEP05(2013)014
  [arXiv:1303.1080 [hep-th]].
  %%CITATION = doi:10.1007/JHEP05(2013)014;%%

 \bibitem{tsunami}
  H.~Liu and S.~J.~Suh,
  ``Entanglement Tsunami: Universal Scaling in Holographic Thermalization,''
  Phys.\ Rev.\ Lett.\  {\bf 112} (2014) 011601
 % doi:10.1103/PhysRevLett.112.011601
  [arXiv:1305.7244 [hep-th]].
  %%CITATION = doi:10.1103/PhysRevLett.112.011601;%%

 \bibitem{veronikaplateaux}
  V.~E.~Hubeny, H.~Maxfield, M.~Rangamani and E.~Tonni,
  ``Holographic entanglement plateaux,''
  JHEP {\bf 1308} (2013) 092
 % doi:10.1007/JHEP08(2013)092
  [arXiv:1306.4004 [hep-th]].
  %%CITATION = doi:10.1007/JHEP08(2013)092;%%

\bibitem{fghmvr}
  T.~Faulkner, M.~Guica, T.~Hartman, R.~C.~Myers and M.~Van Raamsdonk,
  ``Gravitation from Entanglement in Holographic CFTs,''
  JHEP {\bf 1403} (2014) 051
%  doi:10.1007/JHEP03(2014)051
  [arXiv:1312.7856 [hep-th]].
  %%CITATION = doi:10.1007/JHEP03(2014)051;%%

\bibitem{ooguri}
  N.~Bao, S.~Nezami, H.~Ooguri, B.~Stoica, J.~Sully and M.~Walter,
  ``The Holographic Entropy Cone,''
  JHEP {\bf 1509} (2015) 130
%  doi:10.1007/JHEP09(2015)130
  [arXiv:1505.07839 [hep-th]].
  %%CITATION = doi:10.1007/JHEP09(2015)130;%%

\bibitem{dongrenyi}
  X.~Dong,
  ``The Gravity Dual of Renyi Entropy,''
  Nature Commun.\  {\bf 7} (2016) 12472
 % doi:10.1038/ncomms12472
  [arXiv:1601.06788 [hep-th]].
  %%CITATION = doi:10.1038/ncomms12472;%%

 \bibitem{fh}
  M.~Freedman and M.~Headrick,
  ``Bit threads and holographic entanglement,''
  Commun.\ Math.\ Phys.\  {\bf 352} (2017) no.1,  407
  doi:10.1007/s00220-016-2796-3
  [arXiv:1604.00354 [hep-th]].
  %%CITATION = doi:10.1007/s00220-016-2796-3;%%

\bibitem{taylor}
  M.~Taylor and W.~Woodhead,
  ``Renormalized entanglement entropy,''
  JHEP {\bf 1608} (2016) 165
%  doi:10.1007/JHEP08(2016)165
  [arXiv:1604.06808 [hep-th]].
  %%CITATION = doi:10.1007/JHEP08(2016)165;%%

 \bibitem{fhhprvr}
  T.~Faulkner, F.~M.~Haehl, E.~Hijano, O.~Parrikar, C.~Rabideau and M.~Van Raamsdonk,
 ``Nonlinear Gravity from Entanglement in Conformal Field Theories,''
  arXiv:1705.03026 [hep-th].
  %%CITATION = ARXIV:1705.03026;%%

\bibitem{nrt}
  T.~Nishioka, S.~Ryu and T.~Takayanagi,
  ``Holographic Entanglement Entropy: An Overview,''
  J.\ Phys.\ A {\bf 42} (2009) 504008
 % doi:10.1088/1751-8113/42/50/504008
  [arXiv:0905.0932 [hep-th]].
  %%CITATION = doi:10.1088/1751-8113/42/50/504008;%%

\bibitem{vrlectures}
  M.~Van Raamsdonk,
  ``Lectures on Gravity and Entanglement,''
%  doi:10.1142/9789813149441_0005
  arXiv:1609.00026 [hep-th].
  %%CITATION = doi:10.1142/9789813149441_0005;%%

\bibitem{mukund}
  M.~Rangamani and T.~Takayanagi,
  ``Holographic Entanglement Entropy,''
  Lect.\ Notes Phys.\  {\bf 931} (2017) pp.
  doi:10.1007/978-3-319-52573-0
  [arXiv:1609.01287 [hep-th]].
  %%CITATION = doi:10.1007/978-3-319-52573-0;%%

\bibitem{hole-ography}
  V.~Balasubramanian, B.~D.~Chowdhury, B.~Czech, J.~de Boer and M.~P.~Heller,
  ``Bulk curves from boundary data in holography,''
  Phys.\ Rev.\ D {\bf 89} (2014) no.8,  086004
%  doi:10.1103/PhysRevD.89.086004
  [arXiv:1310.4204 [hep-th]].
  %%CITATION = doi:10.1103/PhysRevD.89.086004;%%

\bibitem{uvir}
  L.~Susskind and E.~Witten,
  ``The Holographic bound in anti-de Sitter space,''
  hep-th/9805114.
  %%CITATION = HEP-TH/9805114;%%

\bibitem{pp}
  A.~W.~Peet and J.~Polchinski,
  ``UV / IR relations in AdS dynamics,''
  Phys.\ Rev.\ D {\bf 59} (1999) 065011
 % doi:10.1103/PhysRevD.59.065011
  [hep-th/9809022].
  %%CITATION = doi:10.1103/PhysRevD.59.065011;%%

\bibitem{ewshadows}
  N.~Engelhardt and A.~C.~Wall,
  ``Extremal Surface Barriers,''
  JHEP {\bf 1403} (2014) 068
  doi:10.1007/JHEP03(2014)068
  [arXiv:1312.3699 [hep-th]].
  %%CITATION = doi:10.1007/JHEP03(2014)068;%%

\bibitem{myers}
  R.~C.~Myers, J.~Rao and S.~Sugishita,
  ``Holographic Holes in Higher Dimensions,''
  JHEP {\bf 1406} (2014) 044
%  doi:10.1007/JHEP06(2014)044
  [arXiv:1403.3416 [hep-th]].
  %%CITATION = doi:10.1007/JHEP06(2014)044;%%

\bibitem{veronikaresidual}
  V.~E.~Hubeny,
  ``Covariant Residual Entropy,''
  JHEP {\bf 1409} (2014) 156
%  doi:10.1007/JHEP09(2014)156
  [arXiv:1406.4611 [hep-th]].
  %%CITATION = doi:10.1007/JHEP09(2014)156;%%

\bibitem{cds}
  B.~Czech, X.~Dong and J.~Sully,
  ``Holographic Reconstruction of General Bulk Surfaces,''
  JHEP {\bf 1411} (2014) 015
 % doi:10.1007/JHEP11(2014)015
  [arXiv:1406.4889 [hep-th]].
  %%CITATION = doi:10.1007/JHEP11(2014)015;%%

\bibitem{hmw}
  M.~Headrick, R.~C.~Myers and J.~Wien,
  ``Holographic Holes and Differential Entropy,''
  JHEP {\bf 1410} (2014) 149
%  doi:10.1007/JHEP10(2014)149
  [arXiv:1408.4770 [hep-th]].
  %%CITATION = doi:10.1007/JHEP10(2014)149;%%

\bibitem{nutsandbolts}
  B.~Czech and L.~Lamprou,
  ``Holographic definition of points and distances,''
  Phys.\ Rev.\ D {\bf 90} (2014) 106005
%  doi:10.1103/PhysRevD.90.106005
  [arXiv:1409.4473 [hep-th]].
  %%CITATION = doi:10.1103/PhysRevD.90.106005;%%

\bibitem{bartekinformation}
  B.~Czech, P.~Hayden, N.~Lashkari and B.~Swingle,
  ``The Information Theoretic Interpretation of the Length of a Curve,''
  JHEP {\bf 1506} (2015) 157
 % doi:10.1007/JHEP06(2015)157, 10.1007/jhep06(2015)157
  [arXiv:1410.1540 [hep-th]].
  %%CITATION = doi:10.1007/JHEP06(2015)157, 10.1007/jhep06(2015)157;%%

\bibitem{freivogel}
  B.~Freivogel, R.~A.~Jefferson, L.~Kabir, B.~Mosk and I.~S.~Yang,
  ``Casting Shadows on Holographic Reconstruction,''
  Phys.\ Rev.\ D {\bf 91} (2015) no.8,  086013
 % doi:10.1103/PhysRevD.91.086013
  [arXiv:1412.5175 [hep-th]].
  %%CITATION = doi:10.1103/PhysRevD.91.086013;%%

\bibitem{integralgeometry}
  B.~Czech, L.~Lamprou, S.~McCandlish and J.~Sully,
``Integral Geometry and Holography,''
  JHEP {\bf 1510} (2015) 175
  %  doi:10.1007/JHEP10(2015)175
  [arXiv:1505.05515 [hep-th]].
  %%CITATION = doi:10.1007/JHEP10(2015)175;%%

\bibitem{taylorflavors}
  P.~A.~R.~Jones and M.~Taylor,
  ``Entanglement entropy and differential entropy for massive flavors,''
  JHEP {\bf 1508} (2015) 014
  doi:10.1007/JHEP08(2015)014
  [arXiv:1505.07697 [hep-th]].
  %%CITATION = doi:10.1007/JHEP08(2015)014;%%

 \bibitem{ef}
  N.~Engelhardt and S.~Fischetti,
  ``Covariant Constraints on Hole-ography,''
  Class.\ Quant.\ Grav.\  {\bf 32} (2015) no.19,  195021
%  doi:10.1088/0264-9381/32/19/195021
  [arXiv:1507.00354 [hep-th]].
  %%CITATION = doi:10.1088/0264-9381/32/19/195021;%%

\bibitem{keeler}
  S.~A.~Gentle and C.~Keeler,
  ``On the reconstruction of Lifshitz spacetimes,''
  JHEP {\bf 1603} (2016) 195
 % doi:10.1007/JHEP03(2016)195
  [arXiv:1512.04538 [hep-th]].
  %%CITATION = doi:10.1007/JHEP03(2016)195;%%

\bibitem{entropyanomalya}
  A.~Schwimmer and S.~Theisen,
  ``Entanglement Entropy, Trace Anomalies and Holography,''
  Nucl.\ Phys.\ B {\bf 801} (2008) 1
 % doi:10.1016/j.nuclphysb.2008.04.015
  [arXiv:0802.1017 [hep-th]].
  %%CITATION = doi:10.1016/j.nuclphysb.2008.04.015;%%

 \bibitem{entropyanomalyb}
  R.~X.~Miao,
  ``A Note on Holographic Weyl Anomaly and Entanglement Entropy,''
  Class.\ Quant.\ Grav.\  {\bf 31} (2014) 065009
%  doi:10.1088/0264-9381/31/6/065009
  [arXiv:1309.0211 [hep-th]].
  %%CITATION = doi:10.1088/0264-9381/31/6/065009;%%

  \bibitem{entropyanomalyc}
  V.~Rosenhaus and M.~Smolkin,
  ``Entanglement Entropy Flow and the Ward Identity,''
  Phys.\ Rev.\ Lett.\  {\bf 113} (2014) no.26,  261602
  %doi:10.1103/PhysRevLett.113.261602
  [arXiv:1406.2716 [hep-th]].
  %%CITATION = doi:10.1103/PhysRevLett.113.261602;%%

\bibitem{entropyanomalyd}
  S.~Sachan and D.~V.~Singh,
  ``Entanglement Entropy of BTZ Black Hole and Conformal Anomaly,''
  arXiv:1412.7170 [hep-th].
  %%CITATION = ARXIV:1412.7170;%%

\bibitem{entropyanomalye}
  A.~Allais and M.~Mezei,
  ``Some results on the shape dependence of entanglement and Rényi entropies,''
  Phys.\ Rev.\ D {\bf 91} (2015) no.4,  046002
 % doi:10.1103/PhysRevD.91.046002
  [arXiv:1407.7249 [hep-th]].
  %%CITATION = doi:10.1103/PhysRevD.91.046002;%%

 \bibitem{entropyanomalyf}
  D.~Carmi,
  ``On the Shape Dependence of Entanglement Entropy,''
  JHEP {\bf 1512} (2015) 043
 % doi:10.1007/JHEP12(2015)043
  [arXiv:1506.07528 [hep-th]].
  %%CITATION = doi:10.1007/JHEP12(2015)043;%%

 \bibitem{entropyanomalyg}
  P.~Fonda, D.~Seminara and E.~Tonni,
  ``On shape dependence of holographic entanglement entropy in AdS$_{4}$/CFT$_{3}$,''
  JHEP {\bf 1512} (2015) 037
 % doi:10.1007/JHEP12(2015)037
  [arXiv:1510.03664 [hep-th]].
  %%CITATION = doi:10.1007/JHEP12(2015)037;%%

\bibitem{bh}
  J.~D.~Brown and M.~Henneaux,
  ``Central Charges in the Canonical Realization of Asymptotic Symmetries: An Example from Three-Dimensional Gravity,''
  Commun.\ Math.\ Phys.\  {\bf 104} (1986) 207.
%  doi:10.1007/BF01211590
  %%CITATION = doi:10.1007/BF01211590;%%

\bibitem{cc}
  P.~Calabrese and J.~Cardy,
``Entanglement entropy and conformal field theory,''
  J.\ Phys.\ A {\bf 42} (2009) 504005
  %  doi:10.1088/1751-8113/42/50/504005
  [arXiv:0905.4013 [cond-mat.stat-mech]].
  %%CITATION = doi:10.1088/1751-8113/42/50/504005;%%

\bibitem{ch}
  H.~Casini and M.~Huerta,
  ``Entanglement entropy for the $n$-sphere,''
  Phys.\ Lett.\ B {\bf 694}, 167 (2011)
  %doi:10.1016/j.physletb.2010.09.054
  [arXiv:1007.1813 [hep-th]].


\bibitem{btz}
  M.~Ba\~nados, C.~Teitelboim and J.~Zanelli,
  ``The Black hole in three-dimensional space-time,''
  Phys.\ Rev.\ Lett.\  {\bf 69} (1992) 1849
 % doi:10.1103/PhysRevLett.69.1849
  [hep-th/9204099].
  %%CITATION = doi:10.1103/PhysRevLett.69.1849;%%

\bibitem{densitymatrix}
  B.~Czech, J.~L.~Karczmarek, F.~Nogueira and M.~Van Raamsdonk,
  ``The Gravity Dual of a Density Matrix,''
  Class.\ Quant.\ Grav.\  {\bf 29} (2012) 155009
 % doi:10.1088/0264-9381/29/15/155009
  [arXiv:1204.1330 [hep-th]].
  %%CITATION = doi:10.1088/0264-9381/29/15/155009;%%

\bibitem{wall}
  A.~C.~Wall,
  ``Maximin Surfaces, and the Strong Subadditivity of the Covariant Holographic Entanglement Entropy,''
  Class.\ Quant.\ Grav.\  {\bf 31} (2014) no.22,  225007
%  doi:10.1088/0264-9381/31/22/225007
  [arXiv:1211.3494 [hep-th]].
  %%CITATION = doi:10.1088/0264-9381/31/22/225007;%%

\bibitem{hhlr}
  M.~Headrick, V.~E.~Hubeny, A.~Lawrence and M.~Rangamani,
``Causality \& holographic entanglement entropy,''
  JHEP {\bf 1412} (2014) 162
  %  doi:10.1007/JHEP12(2014)162
  [arXiv:1408.6300 [hep-th]].
  %%CITATION = doi:10.1007/JHEP12(2014)162;%%

\bibitem{wedge}
  R.~Esp\'indola, A.~G\"uijosa and J.~F.~Pedraza, ``Living on the Wedge: Hole-ography and Reconstruction of the Entanglement Wedge'',
  in preparation.

\bibitem{entwinement}
  V.~Balasubramanian, B.~D.~Chowdhury, B.~Czech and J.~de Boer,
  ``Entwinement and the emergence of spacetime,''
  JHEP {\bf 1501} (2015) 048
 % doi:10.1007/JHEP01(2015)048
  [arXiv:1406.5859 [hep-th]].
  %%CITATION = doi:10.1007/JHEP01(2015)048;%%

\bibitem{lin}
  J.~Lin,
  ``A Toy Model of Entwinement,''
  arXiv:1608.02040 [hep-th].
  %%CITATION = ARXIV:1608.02040;%%

\bibitem{bbcdjg}
  V.~Balasubramanian, A.~Bernamonti, B.~Craps, T.~De Jonckheere and F.~Galli,
  ``Entwinement in discretely gauged theories,''
  JHEP {\bf 1612} (2016) 094
 % doi:10.1007/JHEP12(2016)094
  [arXiv:1609.03991 [hep-th]].
  %%CITATION = doi:10.1007/JHEP12(2016)094;%%

\bibitem{stereoscopy}
  B.~Czech, L.~Lamprou, S.~McCandlish, B.~Mosk and J.~Sully,
  ``A Stereoscopic Look into the Bulk,''
  JHEP {\bf 1607} (2016) 129
%  doi:10.1007/JHEP07(2016)129
  [arXiv:1604.03110 [hep-th]].
  %%CITATION = doi:10.1007/JHEP07(2016)129;%%

\bibitem{guica}
  B.~Carneiro da Cunha and M.~Guica,
  ``Exploring the BTZ bulk with boundary conformal blocks,''
  arXiv:1604.07383 [hep-th].
  %%CITATION = ARXIV:1604.07383;%%

\bibitem{diamondography}
  J.~de Boer, F.~M.~Haehl, M.~P.~Heller and R.~C.~Myers,
``Entanglement, holography and causal diamonds,''
  JHEP {\bf 1608} (2016) 162
  %  doi:10.1007/JHEP08(2016)162
  [arXiv:1606.03307 [hep-th]].
  %%CITATION = doi:10.1007/JHEP08(2016)162;%%

\bibitem{blmms2}
  B.~Czech, L.~Lamprou, S.~McCandlish, B.~Mosk and J.~Sully,
``Equivalent Equations of Motion for Gravity and Entropy,''
  JHEP {\bf 1702} (2017) 004
  %  doi:10.1007/JHEP02(2017)004
  [arXiv:1608.06282 [hep-th]].
  %%CITATION = doi:10.1007/JHEP02(2017)004;%%

\bibitem{guica2}
  M.~Guica,
  ``Bulk fields from the boundary OPE,''
  arXiv:1610.08952 [hep-th].
  %%CITATION = ARXIV:1610.08952;%%

\bibitem{ksuw}
  A.~Karch, J.~Sully, C.~F.~Uhlemann and D.~G.~E.~Walker,
  ``Boundary Kinematic Space,''
  arXiv:1703.02990 [hep-th].
  %%CITATION = ARXIV:1703.02990;%%

\bibitem{kl}
  D.~Kabat and G.~Lifschytz,
``Local bulk physics from intersecting modular Hamiltonians,''
  JHEP {\bf 1706} (2017) 120
  %  doi:10.1007/JHEP06(2017)120
  [arXiv:1703.06523 [hep-th]].
  %%CITATION = doi:10.1007/JHEP06(2017)120;%%

\bibitem{fl}
  T.~Faulkner and A.~Lewkowycz,
  ``Bulk locality from modular flow,''
  arXiv:1704.05464 [hep-th].
  %%CITATION = ARXIV:1704.05464;%%

\bibitem{insideout}
  A.~Almheiri, T.~Anous and A.~Lewkowycz,
  ``Inside Out: Meet The Operators Inside The Horizon,''
  arXiv:1707.06622 [hep-th].
  %%CITATION = ARXIV:1707.06622;%%

\bibitem{verlinde}
  H.~Verlinde,
``Poking Holes in AdS/CFT: Bulk Fields from Boundary States,''
  %  arXiv:1505.05069 [hep-th].
  %%CITATION = ARXIV:1505.05069;%%

\bibitem{mnstw}
  M.~Miyaji, T.~Numasawa, N.~Shiba, T.~Takayanagi and K.~Watanabe,
``Continuous Multiscale Entanglement Renormalization Ansatz as Holographic Surface-State Correspondence,''
  Phys.\ Rev.\ Lett.\  {\bf 115} (2015) no.17,  171602
  %  doi:10.1103/PhysRevLett.115.171602
  [arXiv:1506.01353 [hep-th]].
  %%CITATION = doi:10.1103/PhysRevLett.115.171602;%%

\bibitem{no}
  Y.~Nakayama and H.~Ooguri,
``Bulk Locality and Boundary Creating Operators,''
  JHEP {\bf 1510} (2015) 114
  %  doi:10.1007/JHEP10(2015)114
  [arXiv:1507.04130 [hep-th]].
  %%CITATION = doi:10.1007/JHEP10(2015)114;%%

\bibitem{no2}
  Y.~Nakayama and H.~Ooguri,
``Bulk Local States and Crosscaps in Holographic CFT,''
  JHEP {\bf 1610} (2016) 085
 % doi:10.1007/JHEP10(2016)085
 [arXiv:1605.00334 [hep-th]].
  %%CITATION = doi:10.1007/JHEP10(2016)085;%%

\bibitem{verlinde2}
  A.~Lewkowycz, G.~J.~Turiaci and H.~Verlinde,
``A CFT Perspective on Gravitational Dressing and Bulk Locality,''
  JHEP {\bf 1701} (2017) 004
  %  doi:10.1007/JHEP01(2017)004
  [arXiv:1608.08977 [hep-th]].
  %%CITATION = doi:10.1007/JHEP01(2017)004;%%

\bibitem{gt}
  K.~Goto and T.~Takayanagi,
``CFT descriptions of bulk local states in the AdS black holes,''
  arXiv:1704.00053 [hep-th].
  %%CITATION = ARXIV:1704.00053;%%


\end{thebibliography}
\end{document}